\documentclass[3p,twocolumn]{elsarticle}

\usepackage{lineno}
\usepackage{url}
\usepackage{here}
\usepackage{ascmac}
\usepackage{amsmath}
\usepackage{color}
\usepackage{mathtools}

\modulolinenumbers[1]
\journal{Journal of \LaTeX\ Templates}
\def\vector#1{\mbox{\boldmath $#1$}} 
\def\dd{\mathrm{d}}

\begin{document}

\begin{frontmatter}

\title{Reconstruction of multiple Compton scattering events in MeV gamma-ray Compton telescopes towards GRAMS: the physics-based probabilistic model}

\author[1]{Hiroki Yoneda \corref{corresponding}}
\cortext[corresponding]{Corresponding author}
\ead{hiroki.yoneda@riken.jp}

\author[2,3,4]{Hirokazu Odaka}

\author[5]{Yuto Ichinohe}

\author[2,1]{Satoshi Takashima}

\author[6]{Tsuguo Aramaki}


\author[7]{Kazutaka Aoyama}
\author[8]{Jonathan Asaadi}
\author[9]{Lorenzo Fabris}
\author[10,11,3]{Yoshiyuki Inoue}
\author[12]{Georgia Karagiorgi}
\author[13]{Dmitry Khangulyan}
\author[7]{Masato Kimura}
\author[6]{Jonathan Leyva}
\author[14]{Reshmi Mukherjee}
\author[7]{Taichi Nakasone}
\author[15]{Kerstin Perez}
\author[7]{Mayu Sakurai}
\author[12]{William Seligman}
\author[7]{Masashi Tanaka}
\author[16,11,5]{Naomi Tsuji}
\author[7]{Kohei Yorita}
\author[6]{Jiancheng Zeng}

\address[1]{RIKEN Nishina Center, 2-1 Hirosawa, Wako, Saitama 351-0198, Japan}
\address[2]{Department of Physics, The University of Tokyo, 7-3-1 Hongo, Bunkyo, Tokyo 113-0033, Japan}
\address[3]{Kavli Institute for the Physics and Mathematics of the Universe (WPI), The University of Tokyo, Kashiwa 277-8583, Japan}
\address[4]{Research Center for the Early Universe, The University of Tokyo, 7-3-1 Hongo, Bunkyo, Tokyo 113-0033, Japan}
\address[5]{Department of Physics, Rikkyo University, 3-34-1 Nishi Ikebukuro, Toshima, Tokyo 171-8501, Japan}
\address[6]{Northeastern University, 360 Huntington Ave, Boston, MA 02115, USA}
\address[7]{Waseda University, 3-4-1, Okubo, Shinjuku, Tokyo 169-8555, Japan}
\address[8]{University of Texas-Arlington, Arlington, TX 76019, USA}
\address[9]{Oak Ridge National Laboratory, Oak Ridge, TN 37830, USA}
\address[10]{Department of Earth and Space Science, Graduate School of Science, Osaka University, Toyonaka, Osaka 560-0043, Japan}
\address[11]{Interdisciplinary Theoretical \& Mathematical Science Program (iTHEMS), RIKEN, 2-1 Hirosawa, Wako, Saitama 351-0198, Japan}
\address[12]{Columbia University, 538 West 120th Street, New York, NY 10027, USA}
\address[13]{Graduate School of Artificial Intelligence and Science, Rikkyo University, 3-34-1 Nishi Ikebukuro, Toshima, Tokyo 171-8501, Japan}
\address[14]{Barnard College, 3009 Broadway, New York, NY 10027, USA}
\address[15]{Massachusetts Institute of Technology, 77 Massachusetts Ave, Cambridge, MA 02139, USA}
\address[16]{Fucalty of Science, Kanagawa University, 2946 Tsuchiya, Hiratsuka-shi, Kanagawa 259-1293, Japan}

\begin{abstract}
Aimed at progress in mega-electron volt (MeV) gamma-ray astronomy, which has not yet been well-explored,
Compton telescope missions with a variety of detector concepts have been proposed so far.
One of the key techniques for these future missions is an event reconstruction algorithm that is able to determine the scattering orders of multiple Compton scattering events and to identify events in which gamma rays escape from the detectors before they deposit all of their energies.
We revisit previous event reconstruction methods and propose a modified algorithm based on a probabilistic method.
First, we present a general formalism of the probabilistic model of Compton scattering describing physical interactions inside the detector and measurement processes.
Then, we also introduce several approximations in the calculation of the probability functions for efficient computation.
For validation, the developed algorithm has been applied to simulation data of a Compton telescope using a liquid argon time projection chamber, which is a new type of Compton telescope proposed for the GRAMS project.
We have confirmed that it works successfully for up to 8-hit events, including correction of incoming gamma-ray energies for escape events.
The proposed algorithm can be used for next-generation MeV gamma-ray missions featured by large-volume detectors, e.g., GRAMS.
\end{abstract}


\end{frontmatter}


\section{Introduction}
\label{sec_introduction}
Astronomical observations of gamma rays from a few 100 keV to a few 10 MeV remain to be explored in modern astronomy.
COMPTEL, aboard the Compton Gamma-Ray Observatory, observed the MeV gamma-ray universe in 1991-2000 \cite{schonfelder1991Imaging} and found more than 30 sources in the energy band from 0.75 to 30 MeV \cite{schoenfelder2000First}.
After this mission, a few space satellite missions, e.g., INTEGRAL/SPI \cite{INTEGRAL2003} have succeeded, but the number of the detected sources in the MeV band is still limited.
These days, stimulated by the dawn of multi-messenger astronomy, including gravitational wave \cite{gw2017} and high-energy neutrino observations \cite{icecube2018},
this curtained window of electromagnetic waves is drawing increasing attention.
Towards high-sensitivity observations in the 2020s and 2030s,
several MeV gamma-ray missions have been proposed at this moment: 
COSI\footnote{COSI has been selected as a SMEX mission by NASA. It is expected to be launched in 2025.} \cite{tomsick2019Compton}, SMILE \cite{hamaguchi2019Spacebased}, e-ASTROGAM \cite{deangelis2018Science}, AMEGO \cite{mcenery2019Allsky}, GRAMS \cite{aramaki2019Dual}, etc.

All of the missions above utilize the Compton telescope, one of the most promising techniques for imaging gamma-ray sources in the sub-MeV/MeV bands \cite{Schonfelder1973,Herzo1975,Lockwood1979}.
A Compton telescope measures the position and deposited energy at each interaction,
and calculates the scattering angle by the kinematics of Compton scattering.
Then, the incoming gamma-ray direction is constrained to a ring in the sky, traditionally called an ``event circle'', and in the following referred to as a ``Compton circle''.
If recoiled electron trajectories can be measured additionally,
the gamma-ray direction is constrained on an arc-shaped region \cite{tanimori2004MeV,vetter2011First,yoneda2018Development}.
After an accumulation of many events,
the incoming gamma-ray direction can be identified as intersections of the constrained circles or arcs.
This basic principle of Compton imaging
\cite{Schoenfelder1982}, called a back-projection method, has been superseded by statistical methods for image reconstruction \cite{Ballmoos1989,Strong1995,Oberlack1996,Wilderman1998,Knodlseder1999,APRILE2008,ikeda2014Bin,Siegert2020}.

To calculate the scattering angle of a gamma-ray event,
it is required to accurately identify the scattering order of the detected signals and estimate the incident gamma-ray energy.
When a gamma ray deposits its total energy, it is sufficient to identify the first and second interactions.
If the third interaction is identified additionally, the incident energy can be estimated even if a gamma ray escapes outside the detector before they are absorbed, as discussed later.
The scattering order determination becomes complicated as the gamma-ray energy increases 
since gamma rays can easily be scattered multiple times in a detector and can escape from a detector.
These multiple scattering events are considered to be dominant in multi-layered or large-volume detectors.
GRAMS is an example since it utilizes a thick and wide-area liquid argon time projection chamber as a high-efficiency Compton telescope \cite{aramaki2019Dual}.
In this project, like many future missions, it is difficult to measure the time-of-flight between interactions of scattered gamma rays considering its timing resolution because the length between interactions gets shorter than a detector with two separated layers like COMPTEL. In many future missions, the scattering order is difficult to determine directly, and thus, one has to determine it based on the detected energies and positions.
This paper focuses on this order determination problem of the multiple scattering events.

Several approaches have been studied in a variety of fields, e.g., astronomy \citep{kamae1987New,boggs2000Event,kurfess2000Considerations,oberlack2000Compton,Kroeger2001,zoglauer2007Recognition,zoglauer2007Application,SGD2016}, 
homeland security \citep{ShyPhd, ChuPhd, JaworskiPhd, WahlPhd, XuPhd, LehnerPhd, Xu4pi, Shy2020},
and nuclear physics \citep{TANGO2010,Tashenov2011,Korichi2019}.
A widely used method is the {\it mean squared difference (MSD) method} \citep{boggs2000Event,Xu4pi}.
For events with $n~(\geq 3)$ interactions,
the scattering angles of all the interaction sites except for the first and last interaction points, i.e., $i$-th interactions with $2\le i \le n-1$, can be calculated in two ways, namely--one is based on kinematics and the other on geometrical information.
Then, by comparing the calculated angles at each site,
the most plausible scattering order is estimated using a figure-of-merit based on the MSD of the calculated scattering angles.
Another similar and well-formulated method is to calculate the probability of Compton scattering, assuming the scattering order,
and choose the one with the highest probability ({\it deterministic method} \citep{Xu4pi}).
A Bayesian method \cite{zoglauer2007Recognition} and a neural network method \cite{zoglauer2007Application} 
are also proposed, and sometimes they outperform the above approaches.
However, these approaches require large simulation data sets and a lot of computer resources.
One should be careful that when the detector properties change, it may be required to re-compute the training process in a neutral network method.

Usually, events such that gamma rays escape from the detector ({\it escape events}) are rejected in the analysis.
However, as gamma-ray energy increases, photoabsorption becomes less likely,
and the escape events become non-negligible to achieve high detection efficiency.
Then, reconstruction methods should also estimate the escape energy and distinguish the events from those that deposit all of their gamma-ray energies in the detector ({\it fully-absorbed events}).
If the number of interactions is three or more,
the gamma-ray energy of escape events can be estimated by combining the position and energy information (see, e.g., \citep{kurfess2000Considerations,Kroeger2001}).
For example, using the MSD method,
\citep{TANGO2010} proposed an algorithm for germanium spectrometers in nuclear physics experiments \cite{Korichi2019}.
They demonstrated that the gamma-ray detection efficiency can be increased by identifying escape events and estimating their escape energies correctly.
This consideration is also essential for astronomical observations.

Towards the future missions, especially GRAMS,
in this work, we revisit the classical approaches and formulate an event reconstruction algorithm that performs full treatment of both the fully absorbed events and the escape events based on a probabilistic method.
To derive the quantities for the event reconstruction deductively,
first, we formulate the probability functions of the event occurrence by considering physical and measurement processes in Compton telescopes.
In Section 2, we briefly review the problem and describe the basic concept of the present algorithm.
The mathematical formulas of the probability functions are shown in Section~\ref{sec_def_likelihood_function}.
In Section~\ref{sec_implementation},
we implement the algorithm by introducing several approximations to calculate the derived functions' values efficiently.
Here we consider a Compton telescope that measures 
the interaction positions and deposited energies, in which information on the recoil electron trajectories is not contained.
A summary of the developed algorithm is given in Section~\ref{sec_algorithm}.
As a numerical test, we apply it to simulation data sets of a liquid argon detector, regarded as a ideal concept model of the GRAMS project, in Section~\ref{sec_pv}.
Finally, we summarize our results and discuss further improvements in Section~\ref{sec_discussion}.

We have developed the present algorithm to conduct a proof-of-concept study of the GRAMS project, which newly proposes a large-volume liquid argon Compton telescope.
The algorithm will be considered as a standard event reconstruction algorithm in the project's data analysis pipeline. Noted that we are also developing another type of a reconstruction algorithm using a multi-task deep neural network as an alternative approach. 
A detailed description of the algorithm will be presented by a forthcoming publication, which will also show the performance comparison between the neural network method and the physics-based probabilistic method in this paper.

\section{Basic Concept}
\label{sec_concept}

In a Compton telescope, ideally an incident gamma ray deposits its energy at interaction sites via Compton scattering or photoabsorption.
In general, the measured values at each site are expressed by a tuple of several physical quantities, which is denoted by $\vector{D}_I$.
The index $I$ is the label of each tuple in an event that corresponds to a gamma-ray incidence into the telescope,
and the scattering order is unknown at this moment.
We define a hit as a measured interaction with $\vector{D}_I$.
When a Compton telescope measures deposited energies and interaction positions,
$\vector{D}_I$ is a pair of them:
\begin{align}
\vector{D}_I = \left( \vector{r}_{I}, \varepsilon_{I} \right)~,
\end{align}
where $\vector{r}_{I}$ and $\varepsilon_{I}$ are the measured position and energy, respectively.
Since a gamma ray is scattered multiple times or absorbed by a detector,
we obtain a list of $\vector{D}_I$:
\begin{align}
(\vector{D}_I) = (\vector{D}_1, \vector{D}_2, ...)~.
\end{align}
When the number of hits in an event is $n$,
we refer to the event as an $n$-hit event.
Note that we do not consider events including pair creation in this work, though they are not negligible at gamma-ray energies above $\sim 5$ MeV, as discussed in \S\ref{sec_discussion}.

\begin{figure}[!htbp]
\begin{center}
\includegraphics[width = 8.0 cm]{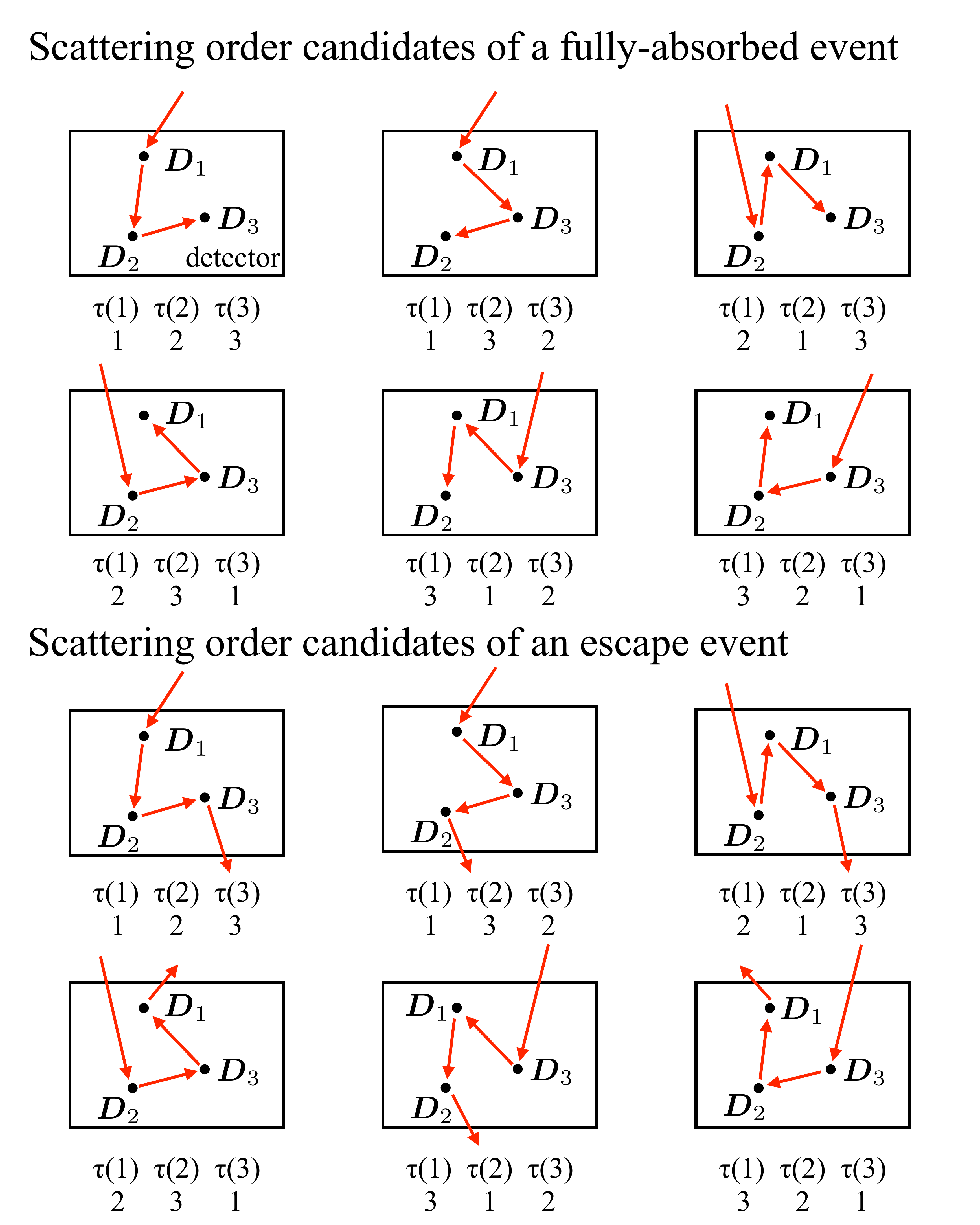}
\end{center}
\caption{Scattering order candidates of a 3-hit event and their corresponding map functions $\tau(\cdot)$.}
\label{fig_scattering_pattern}
\end{figure}

In the reconstruction of Compton scattering events,
the task is divided into the following:
\begin{enumerate}
\item  to determine the event type, i.e., a fully-absorbed event or an escape event, for a given event.
\item  to determine the scattering order of the detected hits:
\begin{align}
(\vector{D}_I)_{\mathrm{ordered}} = (\vector{D}_{\tau(1)}, \vector{D}_{\tau(2)}, ..., \vector{D}_{\tau(n)})_{\mathrm{ordered}}~,
\end{align}
where $(\vector{D}_I)_{\mathrm{ordered}}$ is a re-ordered list of $(\vector{D}_I)$ by determining or assuming the scattering order.
Here we define the function $\tau(\cdot)$ which maps the scattering order to the data label ($I$), i.e.,
$i$-th interaction corresponds to $\vector{D}_{\tau(i)}$.
Figure~\ref{fig_scattering_pattern} shows the scattering order candidates and their corresponding map functions $\tau(\cdot)$ for a 3-hit event.
\item  to estimate the incoming gamma-ray energy.
\item  to estimate the incoming gamma-ray direction.
\end{enumerate}

Related to the second task,
the MSD method \cite{kamae1987New,boggs2000Event,oberlack2000Compton,Xu4pi,TANGO2010}
determines the scattering order by calculating the scattering angles redundantly from kinematics and geometrical information.
For example, \cite{boggs2000Event} defines the following quantity:
\begin{align}
\chi^2_c = \frac{1}{N-2} \sum_{i=2}^{N-1} \frac{(\cos \vartheta_{i}^{\mathrm{kin}} - \cos \vartheta_{i}^{\mathrm{geo}} )^2}{\left(\Delta \cos \vartheta_{i}^{\mathrm{kin}} \right)^2 + \left( \Delta \cos \vartheta_{i}^{\mathrm{geo}}\right)^2 }~(N \geq 3)~,
\end{align}
where $\vartheta_{i}^{\mathrm{geo}}$ and $\vartheta_{i}^{\mathrm{kin}}$ are $i$-th scattering angles calculated by kinematics and geometrically, respectively;
$\Delta \cos \vartheta_{i}$ is the measurement uncertainty, and $N$ is the number of hits.
This quantity is interpreted as a modified chi-squared value.
For all $N!$ scattering order candidates,
$\chi^2_c$ are calculated, and 
the best scattering order is determined as one that yields the smallest $\chi^2_c$.

Related to the energy reconstruction,
the so-called {\it three-Compton} method is usually used to estimate escape gamma-ray energy \citep{kurfess2000Considerations,Kroeger2001}.
If the energy loss at the first and second sites ($\varepsilon_{1}, \varepsilon_{2}$) and the scattering angle at the second interaction ($\vartheta_{2}$) are known,
then the incident gamma-ray energy ($\hat{E}_0$) can be calculated as
\begin{align}
\hat{E}_0 = \varepsilon_{1} + \frac{\varepsilon_{2}}{2} + \sqrt{\frac{\varepsilon_{2}^2}{4} + \frac{\varepsilon_{2} m_e c^2}{1 - \cos \vartheta_{2}}}~,
\end{align}
where $m_e$ is the electron rest mass and $c$ is the speed of light.
For example, \cite{TANGO2010} calculates the incident gamma-ray energy for escape events using the three-Compton method and defines figure-of-merits for both fully-absorbed and escape events when the gamma-ray generation position is known.

In this work, aiming at MeV gamma-ray telescopes in astronomy,
we formulate a probabilistic model of Compton scattering in a detector instead of the MSD methods mentioned above.
The most plausible event type and scattering order can be determined as those that yield the maximum probability.
While the basic concept is similar to the deterministic methods (e.g., \citep{Xu4pi}),
we derive a probabilistic model that describes both fully-absorbed and escape events in a unified framework.
Then, our approach can also identify and reconstruct escape events which have been usually rejected.
In our probabilistic model, the physical processes of gamma rays and the measurement processes of the detector are explicitly considered.

The main advantage of this approach is that the probabilistic model derives the figure-of-merit for the event reconstruction deductively. The relative ratio between the figure-of-merits for the fully-absorbed and escape events can be determined based on the physics-based modeling. It is expected that this approach makes the event type determination more accurate compared to heuristic approaches in the MSD methods. 
We will examine this point in \S\ref{sec_comparsion}. Additionally, when more information is obtained, the proposed algorithm can be naturally extensible by including probability functions corresponding to the new information (see Section \ref{sec_discussion_etcc}).

\section{Probabilistic model of an event type and scattering order}
\label{sec_def_likelihood_function}

In this section, we formulate a probabilistic model of a sequence of Compton scattering and photoabsorption in a detector, given an event type and scattering order.
Here we assume that an incoming gamma ray is Compton-scattered in the detector $n-1$ times and photo-absorbed at last, or is scattered $n$ times and escapes from the detector, i.e., a $n$-hit event.
In addition, we assume that all interactions are measured and 
a list of the measurement values 
$(\vector{D}_I) = (\vector{D}_1, \vector{D}_2, ..., \vector{D}_n )$ are obtained.
Note that in reality, there is a case that some interactions are not detected due to interactions in passive materials, the detection threshold, and multiple scattering within the same spatial resolution element of the detector.
Although the probability of such events strongly depends on the actual detector configuration, we ignore these possibilities, and as a result, our algorithm treats those events as fully-absorbed or escape events.
Especially, the effect of passive materials on the algorithm performance is important for realistic situations, and thus we will discuss it later in the discussion section (see \S\ref{sec_passive_material}).

\begin{figure*}[!htbp]
\begin{center}
\includegraphics[width = 12.0 cm]{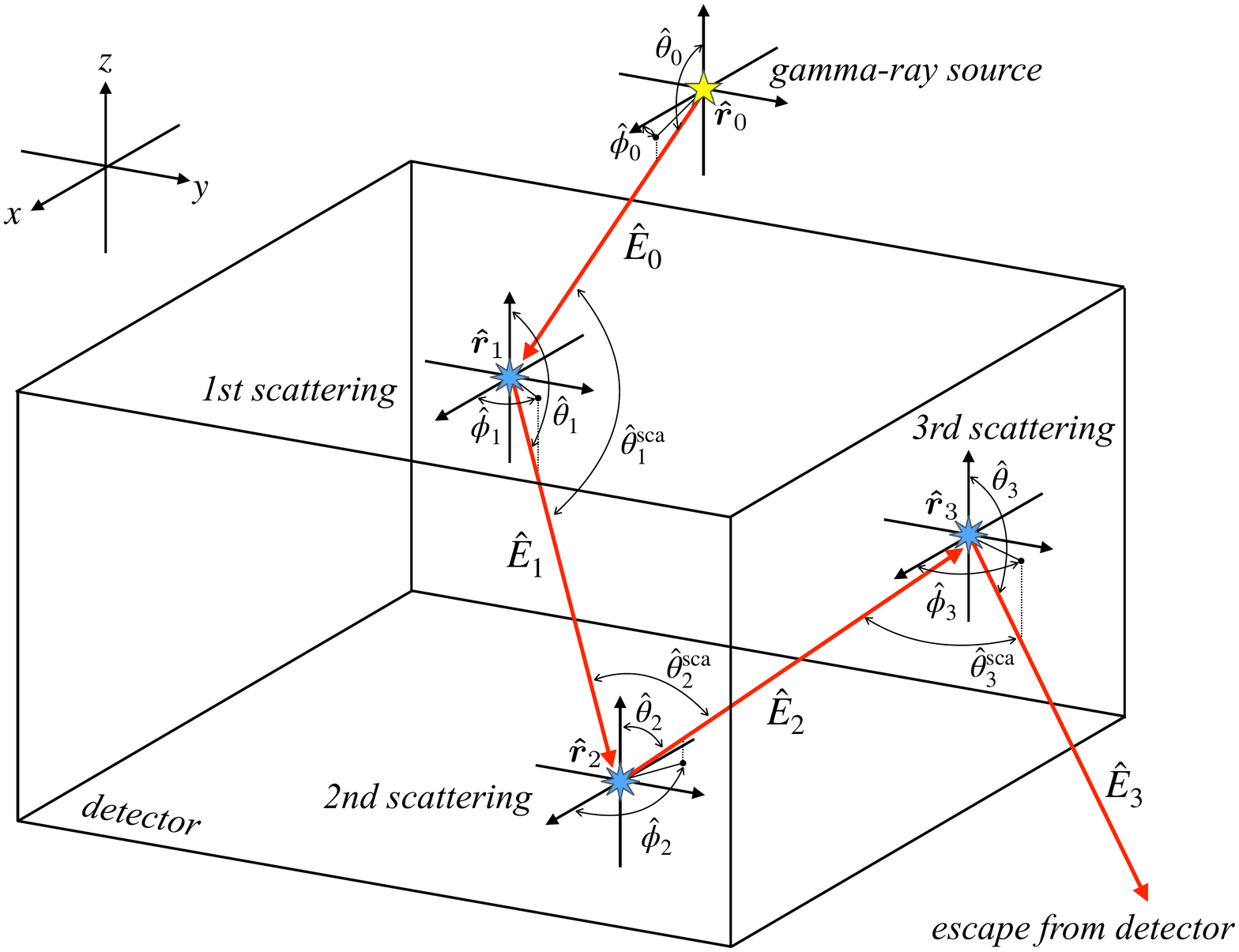}
\end{center}
\caption{A schematic of a multiple Compton scattering event in a detector. The red arrows represent the path of the gamma ray.
Note that the initial gamma-ray position and energy are unknown, and $|\hat{\vector{r}}_0| \rightarrow \infty$ in astronomical situations. 
In the current algorithm, some approximations are introduced for use in astronomy (see \S~\ref{subsec_approximation}).
}
\label{fig_schematic_scattering}
\end{figure*}

As shown in Figure~\ref{fig_schematic_scattering},
the gamma ray changes its energy and its direction of travel every time it interacts with the detector.
Then, the gamma ray after the $i$-th interaction in the detector can be described with
the quantities $\hat{\vector{q}}_{i}$ defined as
\begin{align}
\begin{aligned}
\hat{\vector{q}}_{i} &= \left(\hat{\vector{r}}_{i}, \hat{E}_{i}, \hat{\theta}_i, \hat{\phi}_i \right)~,\\
\hat{\vector{r}}_{i} &= \left(
    \begin{array}{c}
    x_i \\
    y_i \\
    z_i
    \end{array}
  \right)~,\\
\hat{\vector{p}}_{i} &= \frac{\hat{E}_{i}}{c} \left(
    \begin{array}{c}
    \sin \hat{\theta}_i \cos \hat{\phi}_i \\
    \sin \hat{\theta}_i \sin \hat{\phi}_i \\
    \cos \hat{\theta}_i
    \end{array}
  \right)~,
\end{aligned}
\end{align}
where $\hat{\vector{r}}_{i}$ represents the $i$-th interaction position;
$\hat{E}_{i}$ and $\hat{\vector{p}}_{i}$ represent the energy and momentum vector, respectively, of the gamma ray after the $i$-th interaction (see Figure~\ref{fig_schematic_scattering});
$\hat{\theta}_i$ and $\hat{\phi}_i$ describe the direction of travel of the gamma ray.
We refer to the quantity $\hat{\vector{q}}_{i}$ as {\it gamma-ray state} in this work.
To distinguish explicitly the parameters in the gamma-ray state and those measured by experiments,
we put hats on the former ones.
Note that $\hat{\vector{q}}_{0}$ represents the initial gamma-ray state, and the polarization state of incoming and scattered gamma-rays is neglected in this study.

In astronomical observations, 
$\hat{\vector{q}}_{0}$ is described by just three parameters:
\begin{align}
\hat{\vector{q}}_{0} &= \left(\hat{E}_{0}, \hat{\theta}_{0}, \hat{\phi}_{0} \right)~,
\end{align}
since incoming gamma rays originate from distant sources.
In astronomical cases, the incoming gamma-ray direction and energy are usually unknown, and rather an important goal is the reconstruction of the scattering order. 
In this section, we formulate the probabilistic model assuming the initial gamma-ray state, and in \S\ref{subsec_approximation}
we explain the approximations implemented in the current algorithm considering that the initial gamma-ray state is unknown in reality.

The gamma-ray state $\hat{\vector{q}}_{i}$ ($i \geq 1$) can be determined when we assume the initial gamma ray state $\hat{\vector{q}}_{0}$ and the position of each interaction ($\hat{\vector{r}}_{i(\geq 1)}$).
Except for $i=n$,
$\hat{\theta}_i$ and $\hat{\phi}_i$ are determined 
because the momentum of a scattered gamma ray is parallel to $\hat{\vector{r}}_{i+1} - \hat{\vector{r}}_{i}$:
\begin{align}
\label{eq_phi_theta_scattering}
\frac{\hat{\vector{p}}_{i}}{|\hat{\vector{p}}_{i}|} = \frac{\hat{\vector{r}}_{i+1} - \hat{\vector{r}}_{i}}{|\hat{\vector{r}}_{i+1} - \hat{\vector{r}}_{i}|}~.
\end{align}
The gamma-ray energy $\hat{E}_{i}$ is determined by the kinematics of Compton scattering:
\begin{align}
\label{eq_ene_scattering}
\hat{E}_{i} = \frac{\hat{E}_{i-1}}{1 + \dfrac{\hat{E}_{i-1}}{m_e c^2}\left(1 - \cos \hat{\vartheta}^\mathrm{scat}_i \right)}~,
\end{align}
where $\hat{\vartheta}^\mathrm{scat}_i$ is the $i$-th scattering angle which is calculated as
\begin{align}
\label{eq_ene_scattering_2}
\cos \hat{\vartheta}^\mathrm{scat}_i = \frac{ \hat{\vector{p}}_{i-1} \cdot \hat{\vector{p}}_{i} }{ |\hat{\vector{p}}_{i-1}| |\hat{\vector{p}}_{i}| }~.
\end{align}
Note that Eq.~\ref{eq_ene_scattering} does not take account of the uncertainty of $\hat{E}_{i}$ due to the finite momentum fluctuation of the target electrons in the detector material, which is also known as the Doppler broadening effect \cite{zoglauer2003Doppler}.
We will discuss a possible treatment to include this effect in Section~\ref{sec_discussion}.

\begin{figure}[!htbp]
\begin{center}
\includegraphics[width = 8 cm]{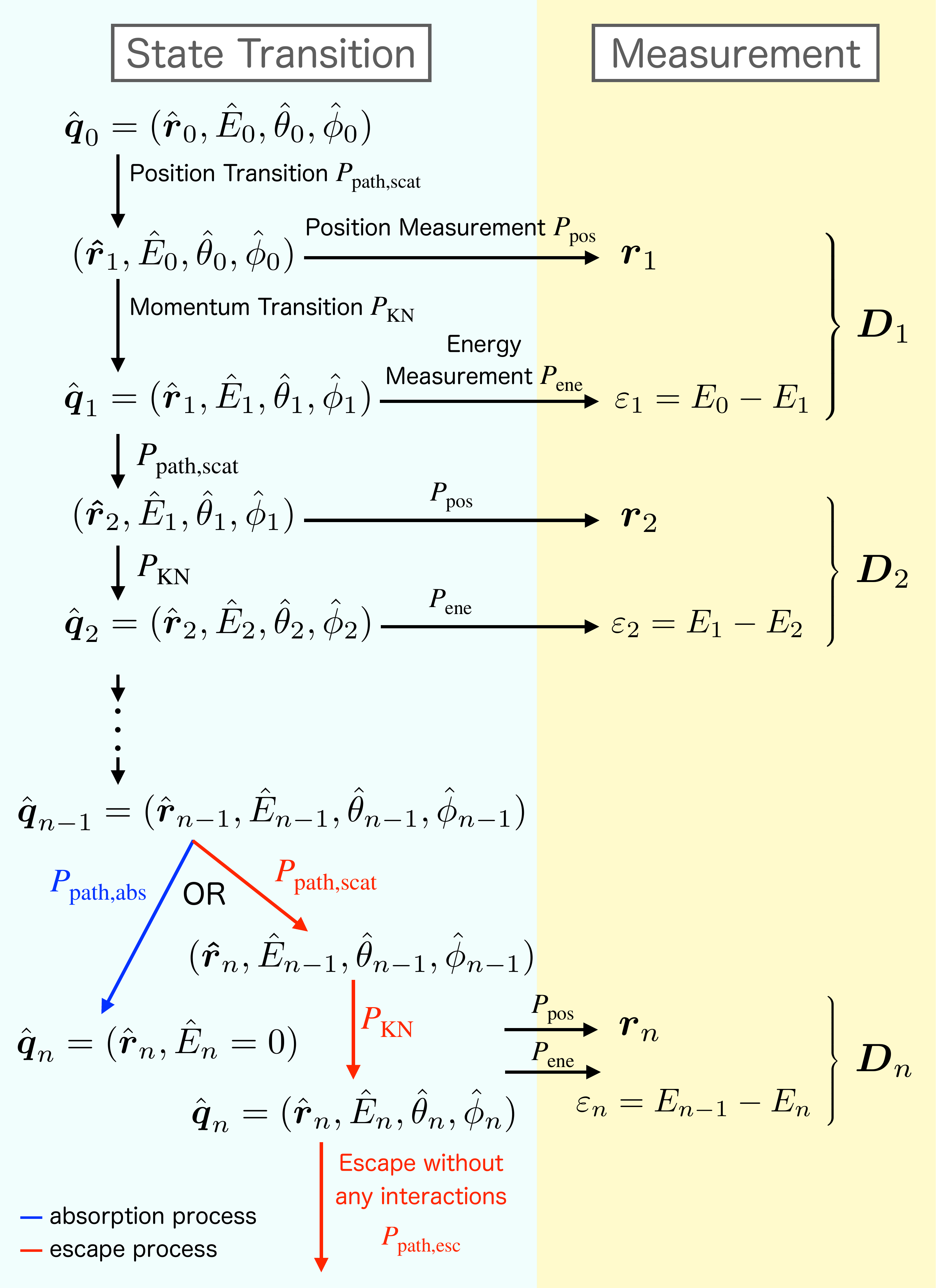}
\end{center}
\caption{A graphical representation of Compton scattering in a detector and measurement in a Compton telescope. The red arrows correspond to probabilistic processes that take place only in escape events while the blue arrow corresponds to that for fully-absorbed events.
Note again that $\hat{\vector{q}}_0$ is unknown and $|\hat{\vector{r}}_0| \rightarrow \infty$ in astronomical situations.
The approximations implemented for astronomy are explained in \S~\ref{subsec_approximation}.
}
\label{fig_model_picture}
\end{figure}

To describe the probabilistic model,
we show the graphical representation of a Compton scattering event in Figure~\ref{fig_model_picture}.
The change of the state $\vector{\hat{q}}_i$ is a probabilistic process determined by the physics of Compton scattering.
Then, the parameters of the state are related to the measured values $(\vector{D}_I)$ through measurement processes.
For example, $\hat{E}_{i-1} - \hat{E}_{i}$ is measured as the deposited energy at $i$-th interaction.

As a general expression, given a fully-absorbed event, its initial state $\hat{\vector{q}}_{0}$ and the scattering order map $\tau(\cdot)$,
the conditional probability to produce its interaction positions $\hat{\vector{r}}_i$ ($1 \leq i \leq n-1)$ 
and the data $(\vector{D}_I)$
is described as
\begin{align}
\label{eq_likelihood_fulldep}
\begin{split}
\mathcal{P}_\mathrm{fullabs} & \left((\vector{D}_I), (\hat{\vector{r}}_{i(\geq 1)}) \mid \hat{\vector{q}}_{0},  \tau(\cdot)\right)\\
& = P_\mathrm{abs} \left( \hat{\vector{q}}_{n}, \vector{D}_{\tau(n)} \mid \hat{\vector{q}}_{n-1}\right) \\
&\times \prod_{i=1}^{n-1} P_\mathrm{scat} \left( \hat{\vector{q}}_{i}, \vector{D}_{\tau(i)} \mid \hat{\vector{q}}_{i-1}\right)~(n \geq 2)~.
\end{split}
\end{align}
The gamma-ray energy $\hat{E}_{i}$ and the direction of travel $(\hat{\theta}_i$, $\hat{\phi}_i)$ after the $i(\geq 1)$-th interaction 
are determined from the model parameters (see Eqs.~\ref{eq_phi_theta_scattering} and \ref{eq_ene_scattering}).

Here $P_\mathrm{scat} \left( \hat{\vector{q}}_{i}, \vector{D}_{\tau(i)} \mid \hat{\vector{q}}_{i-1}\right)$
is the probability that a gamma ray with a given state $\hat{\vector{q}}_{i-1}$ is scattered with changing its state to $\hat{\vector{q}}_{i}$,
and the interaction is measured as $\vector{D}_{\tau(i)}$.
It is described as the product of three functions: 
\begin{align}
\begin{aligned}
P_\mathrm{scat} & \left( \hat{\vector{q}}_{i}, \vector{D}_{\tau(i)} \mid \hat{\vector{q}}_{i-1}\right) = P_\mathrm{path, scat} \left( \hat{\vector{r}}_{i} \mid \hat{\vector{q}}_{i-1}\right) \\
& \times P_\mathrm{KN} \left(\hat{\theta}_{i}, \hat{\phi}_{i} \mid \hat{\vector{q}}_{i-1}\right) 
P_\mathrm{det} \left( \vector{D}_{\tau(i)} \mid \hat{\vector{q}}_{i-1}, \hat{\vector{q}}_{i} \right)~,
\end{aligned}
\end{align}
where $P_\mathrm{path, scat}(\cdot)$ 
is the probability
that a gamma ray of $\hat{\vector{q}}_{i-1}$ is scattered 
at $\hat{\vector{r}}_{i}$;
$P_\mathrm{KN}(\cdot)$ is the probability that 
the scattering direction is $(\hat{\theta}_{i}, \hat{\phi}_{i})$ and it is determined by the Klein-Nishina formula \cite{klein1929Ueber};
$P_\mathrm{det}(\cdot)$ is a probability function related to the detector response.
We will define it in \S\ref{sec_Pdet}.
The first and second functions correspond to the probability to produce $\hat{\vector{q}}_{i}$ and the last one calculates the probability to obtain $\vector{D}_{\tau(i)}$ given $\hat{\vector{q}}_{i-1}$ and $\hat{\vector{q}}_{i}$.
Also, $P_\mathrm{abs} \left( \hat{\vector{q}}_{n}, \vector{D}_{\tau(n)} \mid  \hat{\vector{q}}_{n-1} \right)$ is the probability that a gamma ray of $\hat{\vector{q}}_{n-1}$ is absorbed at $\hat{\vector{r}}_{n}$,
and the interaction is measured as $\vector{D}_{\tau(n)}$:
\begin{align}
\begin{aligned}
P_\mathrm{abs} & \left( \hat{\vector{q}}_{n}, \vector{D}_{\tau(n)} \mid  \hat{\vector{q}}_{n-1}\right) \\
= P&_\mathrm{path, abs} \left( \hat{\vector{r}}_{n} \mid \hat{\vector{q}}_{n-1}\right) P_\mathrm{det} \left(\vector{D}_{\tau(n)} \mid \hat{\vector{q}}_{n-1}, \hat{\vector{q}}_{n} \right)~,
\end{aligned}
\end{align}
where $P_\mathrm{path, abs}(\cdot)$ is 
the probability that a gamma ray with $\hat{\vector{q}}_{i-1}$ is absorbed at $\hat{\vector{r}}_{i}$.
In the following subsections,
we explain these functions in detail.

Besides, the conditional probability corresponding to the escape event is described as
\begin{align}
\label{eq_likelihood_escape}
\begin{split}
\mathcal{P}_\mathrm{escape} & \left((\vector{D}_I), (\hat{\vector{r}}_{i(\geq 1)}) \mid \hat{\vector{q}}_{0},  \tau(\cdot)\right)\\
&= P_\mathrm{esc} \left( \hat{\vector{q}}_{n}, \vector{D}_{\tau(n)} \mid  \hat{\vector{q}}_{n-1}\right) \\
&\times \prod_{i=1}^{n-1} P_\mathrm{scat} \left( \hat{\vector{q}}_{i}, \vector{D}_{\tau(i)} \mid  \hat{\vector{q}}_{i-1}\right)~(n \geq 2)~.
\end{split}
\end{align}
The only difference between Eq.~\ref{eq_likelihood_fulldep} and Eq.~\ref{eq_likelihood_escape} is the treatment of the last interaction.
Here we introduce $P_\mathrm{esc} \left( \hat{\vector{q}}_{n}, \vector{D}_{\tau(n)} \mid \hat{\vector{q}}_{n-1}\right)$,
which is the probability that a gamma ray with a given state $\hat{\vector{q}}_{n-1}$ is scattered with changing its state to $\hat{\vector{q}}_{n}$
and escape from the detector,
and the interaction is measured as $\vector{D}_{\tau(n)}$.
It is described as
\begin{align}
\label{eq_p_esc}
\begin{aligned}
P_\mathrm{esc} \left( \hat{\vector{q}}_{n}, \vector{D}_{\tau(n)} \mid \hat{\vector{q}}_{n-1} \right) = 
P_\mathrm{path,scat} \left(  \hat{\vector{r}}_{n} \mid \hat{\vector{q}}_{n-1}\right) &\\
\times \int \dd(\cos\hat{\theta}_n) \dd\hat{\phi}_n P_\mathrm{KN} \left( \hat{\theta}_n, \hat{\phi}_n \mid  \hat{\vector{q}}_{n-1}\right) &\\
\times P_\mathrm{det}  \left( \vector{D}_{\tau(n)}  \mid \hat{\vector{q}}_{n-1}, \hat{\vector{q}}_{n}\right)  P_\mathrm{path, esc} \left( \hat{\vector{q}}_{n} \right)&~,
\end{aligned}
\end{align}
where $P_\mathrm{path, esc}(\hat{\vector{q}}_{n})$ is the probability that a gamma ray with a state of $\hat{\vector{q}}_{n}$ escapes from the detector.
Since we cannot know the momentum direction of the escape gamma ray, 
we integrate the functions over $\hat{\phi}_n$ and $\cos\hat{\theta}_n$.

\subsection{Probability functions related to physical processes}

In the above equations,
$P_\mathrm{path,scat}( \hat{\vector{r}}_{i} \mid \hat{\vector{q}}_{i-1} )$ and $P_\mathrm{path,abs}(\hat{\vector{r}}_{i} \mid \hat{\vector{q}}_{i-1})$ describe the probability that a gamma ray of $\hat{\vector{q}}_{i-1}$ is scattered or absorbed at $\hat{\vector{r}}_{i}$.
They are defined as
\begin{align}
&\begin{aligned}
\label{eq_P_mov}
&P_\mathrm{path,scat} \left(  \hat{\vector{r}}_{i} \mid \hat{\vector{q}}_{i-1}\right) \dd^3 \hat{\vector{r}}_{i} = \\
&~~~~~\frac{\rho \sigma_\mathrm{scat}}{|\hat{\vector{r}}_{i} - \hat{\vector{r}}_{i-1}|^2} \exp\left(-\rho \sigma_\mathrm{all} |\hat{\vector{r}}_{i} - \hat{\vector{r}}_{i-1}|\right) \dd^3 \hat{\vector{r}}_{i}~,
\end{aligned}\\
&\begin{aligned}
\label{eq_P_mov_abs}
&P_\mathrm{path,abs} \left(  \hat{\vector{r}}_{i} \mid \hat{\vector{q}}_{i-1} \right) \dd^3 \hat{\vector{r}}_{i} = \\
&~~~~~\frac{\rho \sigma_\mathrm{abs}}{|\hat{\vector{r}}_{i} - \hat{\vector{r}}_{i-1}|^2} \exp\left(-\rho \sigma_\mathrm{all} |\hat{\vector{r}}_{i} - \hat{\vector{r}}_{i-1}|\right) \dd^3 \hat{\vector{r}}_{i}~,
\end{aligned}
\end{align}
where $\sigma_\mathrm{abs}$ and $\sigma_\mathrm{scat}$ are the cross sections of photoabsorption and Compton scattering at an energy of $\hat{E}_{i-1}$, respectively;
$\rho$ is the number density of the detector material;
$\sigma_\mathrm{all}$ is the sum of the cross sections of the photon interactions.
The term $1/|\hat{\vector{r}}_{i} - \hat{\vector{r}}_{i-1}|^2$ corresponds to the solid angle of the volume element at $\hat{\vector{r}}_{i}$ seen from $\hat{\vector{r}}_{i-1}$.
When incoming gamma rays are considered to be parallel light as in astronomical observations,
the term with $i = 1$ must be removed, and Eq.~\ref{eq_P_mov} is described as
\begin{align}
\label{eq_P_mov_para}
\begin{aligned}
&P_\mathrm{path,scat} \left( \hat{\vector{r}}_{1} \mid \hat{\vector{q}}_{0} \right) \dd^3 \hat{\vector{r}}_{1} \\
&~~~~~\propto
\rho \sigma_\mathrm{scat} \exp\left(-\rho \sigma_\mathrm{all} \hat{l}_{\mathrm{first}}\right) \dd^3 \hat{\vector{r}}_{1}~,
\end{aligned}
\end{align}
where $\hat{l}_{\mathrm{first}}$ is the length of the gamma-ray path inside the detector before it arrives at $\hat{\vector{r}}_{1}$, and depends on the unknown incoming direction ($\hat{\theta}_0$, $\hat{\phi}_0$) in astronomical cases.
Later, we will introduce an approximation of this value (see Eq.~\ref{eq_l_first}. Additional position dependence from edge effects is neglected).
Note that here it is assumed that the detector consists of a single material.
If a Compton telescope consists of several detectors with different materials, e.g., semiconductor detectors and scintillators,
then $\sigma_\mathrm{abs/scat/pair}$ and $\rho$ also depend on the position.
In this case, Eq.~\ref{eq_P_mov} is modified as
\begin{align}
\label{eq_P_mov_integral}
\begin{aligned}
&P_\mathrm{path,scat} \left( \hat{\vector{r}}_{i} \mid \hat{\vector{q}}_{i-1} \right) \dd^3 \hat{\vector{r}}_{i} = \\
&\frac{\rho (\hat{\vector{r}}_{i}) \sigma_\mathrm{scat} (\hat{\vector{r}}_{i}) }{|\hat{\vector{r}}_{i} - \hat{\vector{r}}_{i-1}|^2} \exp\left(-
\int_{C} \rho (\hat{\vector{r}}) \sigma_\mathrm{all} (\hat{\vector{r}}) \dd |\hat{\vector{r}}| \right) \dd^3 \hat{\vector{r}}_{i}~,
\end{aligned}
\end{align}
where $C$ is the straight line from $\hat{\vector{r}}_{i-1}$ to $\hat{\vector{r}}_{i}$.
The same applies to Eqs.~\ref{eq_P_mov_abs} and \ref{eq_P_mov_para}.

The function $P_\mathrm{KN} \left(\hat{\theta}_{i}, \hat{\phi}_{i} \mid \hat{\vector{q}}_{i-1}\right)$ corresponds to the probability that
a gamma ray of $\hat{\vector{q}}_{i-1}$ is scattered to the direction described by $(\hat{\theta}_{i}, \hat{\phi}_{i})$.
Namely it is the normalized differential cross section of Compton scattering.
Following Klein-Nishina's formula \cite{klein1929Ueber},
it is defined as
\begin{align}
\begin{aligned}
P_\mathrm{KN} & \left( \hat{\theta}_i, \hat{\phi}_i \mid \hat{\vector{q}}_{i-1} \right) \dd \hat{\Omega}_i = \frac{1}{\sigma_\mathrm{scat}} \frac{\dd \sigma_\mathrm{scat}}{\dd \hat{\Omega}_i} \dd \hat{\Omega}_i = \frac{1}{\sigma_\mathrm{scat}} \frac{r_e^2}{2} \\
\times & \left(\frac{\hat{E}_{i}}{\hat{E}_{i-1}}\right)^2 \left( \frac{\hat{E}_{i}}{\hat{E}_{i-1}} + \frac{\hat{E}_{i-1}}{\hat{E}_{i}} - \sin^2\hat{\vartheta}^\mathrm{scat}_i\right) \dd \hat{\Omega}_i~,
\end{aligned}
\end{align}
where $r_e$ is the classical electron radius and
$\hat{E}_{i}$ is the energy of the scattered gamma ray calculated by Eq.~\ref{eq_ene_scattering} and
$\hat{\vartheta}^\mathrm{scat}_i$ is the $i$-th scattering angle defined in Eq.~\ref{eq_ene_scattering_2} and
$\dd \hat{\Omega}_i$ is equal to $\dd(\cos \hat{\theta}_i) \dd \hat{\phi}_i$.
Note that the polarization is not considered in this process to reduce the amount of calculation because the target position is unknown.
The treatment of the polarization with known target position is discussed in \cite{Fernandez1995,TANGO2010}.

Finally we define $P_\mathrm{path,esc} \left( \cdot \right)$ as the probability that a gamma ray escapes from the detector without any interaction.
It is described as
\begin{align}
\label{eq_p_mov_esc}
P_\mathrm{path,esc} \left( \hat{\vector{q}}_{n} \right) =  \exp\left(-\rho \sigma_\mathrm{all} \hat{l}_n \right)~,
\end{align}
where $\hat{l}_n$ is the length between $\vector{\hat{r}}_n$ and the position at which the gamma ray escaped from the detector.

\subsection{Probability functions related to measurements}
\label{sec_Pdet}
The function $P_\mathrm{det}(\cdot)$ corresponds to the measurement process, and then it is determined by the detector response and what kind of quantities a Compton telescope can measure.
Here we assume that a Compton telescope measures the deposited energy and position of each interaction independently.
In this case, $P_\mathrm{det}(\cdot)$ is usually expressed as the product of two detector response functions defined as
\begin{align}
\label{eq_det}
\begin{aligned}
P_\mathrm{det} \left( \vector{D}_{\tau(i)} \mid \hat{\vector{q}}_{i-1}, \hat{\vector{q}}_{i}\right) = & P_\mathrm{ene}(\varepsilon_{\tau(i)} \mid \hat{\varepsilon}_{i}) \\
&\times P_\mathrm{pos}(\vector{r}_{\tau(i)} \mid \hat{\vector{r}}_i)~,
\end{aligned}
\end{align}
where $\hat{\varepsilon}_{i}$ is the true deposited energy at $i$-th interaction, which is equal to $\hat{E}_{i-1} - \hat{E}_{i}$.
Here $P_\mathrm{ene}(\varepsilon_{\tau(i)} \mid \hat{\varepsilon}_{i})$
is the probability that the true deposited energy of $\hat{\varepsilon}_{i}$ is detected as $\varepsilon_{\tau(i)}$, and 
$P_\mathrm{pos}(\vector{r}_{\tau(i)} \mid \hat{\vector{r}}_i)$
is one that the interaction position of $\hat{\vector{r}}_i$ is detected as $\vector{r}_{\tau(i)}$.
In general, these two can have any functional forms and they should be assumed so as to model the detector response adequately using experiments with calibration sources, simulations or simplifications.
We will introduce specific formulas for these functions in the following section.

\section{Implementation}
\label{sec_implementation}

The task of this algorithm is to determine the event type, the scattering order and the incident energy and to constrain the incoming direction (see \S\ref{sec_concept}).
In our approach, these parameters are estimated as those that yield the maximum probability defined as Eq.~\ref{eq_likelihood_fulldep} or Eq.~\ref{eq_likelihood_escape}.
Hereafter we notate the estimated incident energy and incoming direction as $\hat{E}_0^{\ast}$ and ($\hat{\theta}_0^{\ast}, \hat{\phi}_0^{\ast}$), respectively.
Ideally, this task is achieved by sweeping the parameter space of the incoming gamma-ray energy and direction, and marginalizing out the true interaction positions ($\vector{r}_{\tau(i)}$).
However, it is not practical in terms of computational time because the parameter space becomes of high dimensions, i.e., $(3n+3)$ dimensions when the number of interactions is $n$.

In this section,
we implement the order determination algorithm by focusing on a Compton telescope that measures the interaction positions and deposited energies and does not obtain the trajectories of recoiled electrons.
First, we introduce specific forms for the detector response terms.
Then, we introduce several approximations to eliminate integral calculations and parameter space search.

\subsection{Detector response}

To describe the detector response terms $P_\mathrm{ene}(\varepsilon \mid \hat{\varepsilon})$ and $P_\mathrm{pos}(\vector{r} \mid \hat{\vector{r}})$,
we adopt the Gaussian function as a simple model.
Namely, these functions are formulated as
\begin{align}
\label{eq_energy_resolution}
P_\mathrm{ene}(\varepsilon \mid \hat{\varepsilon}) \dd \varepsilon = \frac{1}{\sqrt{2 \pi \sigma^2_{\hat{\varepsilon}}}} \exp\left( - \frac{(\hat{\varepsilon} - \varepsilon)^2}{2 \sigma^2_{\hat{\varepsilon}}} \right) \dd \varepsilon~,
\end{align}
\begin{align}
\label{eq_position_resolution}
\begin{aligned}
P_\mathrm{pos}(\vector{r} \mid \hat{\vector{r}}) \dd^3 \vector{r}
&= \frac{1}{\sqrt{2 \pi \sigma^2_{\hat{x}}}} 
\exp\left( - \frac{(\hat{x} - x)^2}{2 \sigma^2_{\hat{x}}} \right) \\
&\times \frac{1}{\sqrt{2 \pi \sigma^2_{\hat{y}}}} 
\exp\left( - \frac{(\hat{y} - y)^2}{2 \sigma^2_{\hat{y}}} \right) \\
&\times \frac{1}{\sqrt{2 \pi \sigma^2_{\hat{z}}}} 
\exp\left( - \frac{(\hat{z} - z)^2}{2 \sigma^2_{\hat{z}}} \right) \dd^3 \vector{r}~,
\end{aligned}
\end{align}
where $\sigma_{\hat{\varepsilon}}$, $\sigma_{\hat{x}}$, $\sigma_{\hat{y}}$, and $\sigma_{\hat{z}}$ are the energy and positional resolutions of the detector, respectively.

The response functions in Eqs.~\ref{eq_energy_resolution} and \ref{eq_position_resolution} are very simple,
e.g., Eqs.~\ref{eq_energy_resolution} considers only the energy resolution of the detector.
We note that one may adopt more complex models in principle.
For example, the positional resolution often depends on the interaction position due to the charge sharing effect, pixelization, e.t.c.
These effects may be modeled by including the dependence of $\sigma_{\hat{x}/\hat{y}/\hat{z}}$ on the interaction position.
While these effects vary with detector systems, and we ignore them here,
it may improve the algorithm performance by using more realistic positional/energy response functions.

\subsection{Approximations in the calculation of the probability functions}
\label{subsec_approximation}

\subsubsection{Fully-absorbed events}

For the fully-absorbed events,
$\hat{E}_0^{\ast}$ is estimated by a good approximation as the sum of the measured energies:
\begin{align}
\label{eq_e_ini_full_dep}
&\hat{E}_0^\ast \simeq \sum_{i=1}^{n} \varepsilon_i~.
\end{align}
The true interaction positions are approximated as the measured positions:
\begin{align}
\label{eq_pos_approx}
&\hat{\vector{r}}_i^{\ast} \simeq \vector{r}_{\tau(i)}~(i \geq 1)~,
\end{align}
i.e., the positional response function is approximated as a delta function:
\begin{align}
P_\mathrm{pos}(\vector{r} \mid \hat{\vector{r}}) \dd^3 \vector{r}
= \delta \left(\hat{x} - x\right) \delta \left(\hat{y} - y\right) \delta \left(\hat{z} - z\right) \dd^3 \vector{r}~.
\end{align}
However, with this approximation, the positional errors in Eq.~\ref{eq_position_resolution} become ignored.
To compensate for this,
we modify the energy resolution in Eq.~\ref{eq_energy_resolution}.
At $i$-th interaction ($2 \leq i \leq n-1$),
the scattering angle can be calculated geometrically,
and the positional errors ($\sigma_{\hat{x}}$, $\sigma_{\hat{y}}$, and $\sigma_{\hat{z}}$) produce the uncertainty on it, to which we refer as $\Delta (\cos \vartheta^\mathrm{scat}_{i})_{\mathrm{pos}}$.
Considering the propagation of the positional errors as discussed in the classical approach \cite{boggs2000Event},
$\Delta (\cos \vartheta^\mathrm{scat}_{i})_{\mathrm{pos}}$ can be calculated.
Then, the partial derivative of Eq.~\ref{eq_ene_scattering} with respect to $\cos \vartheta^\mathrm{scat}_{i}$ yields
\begin{align}
\Delta \hat{\varepsilon}_{i} = \frac{\hat{E}^2_{i}}{m_e c^2} \Delta (\cos \vartheta^\mathrm{scat}_{i})~,
\end{align}
which means that the positional errors produce uncertainty on the estimation of the gamma-ray energy through the scattering angle calculation.
In our algorithm, for $2 \leq i \leq n-1$,
we modify the energy resolution as
\begin{align}
\label{eq_modification_of_energy_resolution}
\sigma^2_{\hat{\varepsilon}} \rightarrow \sigma^2_{\hat{\varepsilon}} + \left(\frac{\hat{E}^2_{i}}{m_e c^2} \Delta (\cos \vartheta^\mathrm{scat}_{i})_{\mathrm{pos}}\right)^2~.
\end{align}
Thus, though the approximation by Eq.~\ref{eq_pos_approx} ignores the positional errors described in Eq.~\ref{eq_position_resolution}, we include them in the energy response function effectively.

The incoming direction ($\hat{\theta}_0^{\ast}, \hat{\phi}_0^{\ast}$) should be also determined.
In Compton telescopes,
the incoming gamma-ray direction is constrained on a circle in the sky.
We calculate the first scattering angle under assumption of 
the incident energy ($\hat{E}_{0}^\ast$), the deposited energy ($\varepsilon_{\tau(1)}$), and the first and second interaction positions ($\vector{r}_{\tau(1)}, \vector{r}_{\tau(2)}$),
and then the circle can be obtained 
as the solutions of ($\hat{\theta}_0^{\ast}, \hat{\phi}_0^{\ast}$) that satisfy the Compton kinematics:
\begin{align}
\label{eq_ini_theta_phi}
\hat{E}_{0}^\ast = \hat{E}_{1} (\hat{E}_{0}^\ast, \hat{\theta}_{0}^\ast, \hat{\phi}_{0}^\ast, \vector{r}_{\tau(1)}, \vector{r}_{\tau(2)}) + \varepsilon_{\tau(1)}~.
\end{align}
Here $\hat{E}_{1}$ is calculated by Eqs.~\ref{eq_ene_scattering} and \ref{eq_ene_scattering_2}.

The function $P_\mathrm{path,scat} \left( \hat{\vector{r}}_{1} \mid \hat{\vector{q}}_{0} \right)$ also depends on $\hat{\theta}_0^{\ast}$ and $\hat{\phi}_0^{\ast}$.
Here we assume a constant value $L_0$ for the path length in Eq.~\ref{eq_P_mov_para}:
\begin{align}
\label{eq_l_first}
\hat{l}_\mathrm{first} = L_0~.
\end{align}
This should be comparable to the size scale of the detector.
Then, the probabilities are the same for the parameters ($\hat{\theta}_0^{\ast}, \hat{\phi}_0^{\ast}$) that satisfy Eq.~\ref{eq_ini_theta_phi}, i.e., on a Compton circle,
and they are considered to be the approximation of the maximum probability for the assumed scattering order.
It should be noted that this approximation ignores the dependence of $\hat{l}_\mathrm{first}$ on the first interaction position, i.e., detector geometry effect. We examined how the choice of $L_0$ affects the algorithm performance and found that it does not significantly depend on $L_0$ (see \ref{sec_effect_length_scale}).

\subsubsection{Escape events}

For escape events,
the incident gamma-ray energy and escape energy can be
estimated using the scattering angle measured from geometrical information \cite{kamae1987New}.
Here we calculate the incident gamma-ray energy at each $i$-th interaction site ($2 \leq i \leq n-1$),
and use the averaged value as the estimation.
Based on the three-Compton method \cite{kurfess2000Considerations,Kroeger2001}, $\hat{E}_0^\ast$ is estimated as
\begin{align}
\label{eq_esc_energy}
\hat{E}_0^\ast = \frac{1}{n-2} \sum_{m=2}^{n-1}  \hat{E}_0^\mathrm{\ast, m}~(n \geq 3)~,
\end{align}
where,
\begin{align}
& \hat{E}_0^\mathrm{\ast, m} = \sum_{i=1}^{m} \varepsilon_{\tau(i)} + E'_{m}~, \\
& E'_m = -\frac{\varepsilon_{\tau(m)}}{2} + \sqrt{\frac{\varepsilon_{\tau(m)}^2}{4} + \frac{\varepsilon_{\tau(m)} m_e c^2}{1 - \cos \vartheta^{\mathrm{scat},G}_{m}}}~, \\
& \cos \vartheta^{\mathrm{scat},G}_{m} =  \frac{(\vector{r}_{\tau(m)} - \vector{r}_{\tau(m-1)})\cdot(\vector{r}_{\tau(m+1)} - \vector{r}_{\tau(m)})}{|\vector{r}_{\tau(m)} - \vector{r}_{\tau(m-1)}| |\vector{r}_{\tau(m+1)} - \vector{r}_{\tau(m)}|}~.
\end{align}

In the calculation of the probability function for escape events,
the term $P_\mathrm{esc} \left( \hat{\vector{q}}_{n}, \vector{D}_{\tau(n)} \mid \hat{\vector{q}}_{n-1} \right)$ needs the integration over the direction of escape (see Eq.~\ref{eq_p_esc}).
To calculate it approximately without integral computation,
here we assume that
\begin{align}
&P_\mathrm{ene} (\varepsilon_{\tau(n)} \mid \hat{\varepsilon}_{n})
\simeq 
\delta\left( \varepsilon_{\tau(n)} - \hat{\varepsilon}_{n}\right)~.
\end{align}
We also approximate $\hat{l}_n$ in Eq.~\ref{eq_p_mov_esc} as a constant value:
\begin{align}
\label{eq_l_esc}
\hat{l}_\mathrm{esc} = L_\mathrm{esc}~,
\end{align}
where $L_\mathrm{esc}$ should be also comparable to the detector size scale like $L_0$.
Again this implementation ignores the detector geometry effect.
We also found that the algorithm performance is not significantly affected by the choice of $L_\mathrm{esc}$ in \ref{sec_effect_length_scale}.
Then these approximations yield
\begin{align}
\label{eq_last_ene_esc}
\begin{aligned}
&P_\mathrm{esc} \left( \hat{\vector{q}}_{n}, \vector{D}_{\tau(n)} \mid \hat{\vector{q}}_{n-1}\right) \\
&= P_\mathrm{path,scat} \left(  \vector{r}_{\tau(n)} \mid \hat{\vector{q}}_{n-1}\right) \\
&~~~~\times \int  \dd(\cos\hat{\vartheta}^\mathrm{scat}_n) \dd\hat{\varphi}^{\mathrm{scat}}_n P_\mathrm{KN} \left( 
\hat{\theta}_{n}, \hat{\phi}_{n} \mid \hat{\vector{q}}_{n-1} \right) \\
&~~~~~~~~\times P_\mathrm{det} \left( \vector{D}_{\tau(n)} \mid \hat{\vector{q}}_{n-1}, \hat{\vector{q}}_{n}\right)  P_\mathrm{path,esc} \left( \hat{\vector{q}}_{n} \right) \\
&= P_\mathrm{path,scat} \left(  \vector{r}_{\tau(n)} \mid \hat{\vector{q}}_{n-1}\right) 
\frac{P_{\mathrm{pos}} (\vector{r}_{\tau(n)}, \vector{r}_{\tau(n)}) }{\sigma_\mathrm{scat}} \\
&~~~~\times \int \dd(\cos\hat{\vartheta}^\mathrm{scat}_n) \dd\hat{\varphi}^{\mathrm{scat}}_n \frac{\dd \sigma_\mathrm{scat}}{\dd \Omega_n}\\
&~~~~~~~~\times \delta\left( \varepsilon_{\tau(n)} - \hat{\varepsilon}_{n} \right)
\exp\left(-\rho \sigma_\mathrm{all} L_\mathrm{esc} \right)\\
&= P_\mathrm{path,scat} \left(  \vector{r}_{\tau(n)} \mid \hat{\vector{q}}_{n-1}\right) 
\frac{P_{\mathrm{pos}} (\vector{r}_{\tau(n)}, \vector{r}_{\tau(n)}) }{\sigma_\mathrm{scat}} \\
&~~~~\times \frac{2 \pi m_e c^2}{(\hat{E}_{n-1} - \varepsilon_{\tau(n)})^2} 
\frac{\dd \sigma_\mathrm{scat}}{\dd \Omega}
\exp\left(-\rho \sigma_\mathrm{all} L_\mathrm{esc}\right)~.
\end{aligned}
\end{align}
In the last equation, $\displaystyle \frac{\dd \sigma_\mathrm{scat}}{\dd \Omega}$ is a fixed value because after the integration the scattering angle $\vartheta^\mathrm{scat}_n$ is determined from $\hat{E}_{n-1}$ and $\varepsilon_{\tau(n)}$ with Compton kinematics.
Note that to derive the last equation we used change of variables $\cos\hat{\vartheta}^\mathrm{scat}_n \rightarrow \hat{\varepsilon}_{n}$ using Eq.~\ref{eq_ene_scattering}.
The term $\frac{m_e c^2}{(\hat{E}_{n-1} - \varepsilon_{\tau(n)})^2}$ is due to this change of variables.
Here the direction of escape is described with the scattering angle $\hat{\vartheta}^\mathrm{scat}_n$ at the last interaction and the azimuth angle $\hat{\varphi}^{\mathrm{scat}}_n$ along $\vector{r}_{\tau(n)} - \vector{r}_{\tau(n-1)}$, not $\hat{\theta}_n$ and $\hat{\phi}_n$,
because it makes the calculation simpler, e.g., $\dd\cos \vartheta^\mathrm{scat}_n / \dd\hat{\varepsilon}_{n}$.

\section{Algorithm Summary}
\label{sec_algorithm}

By calculating the probability functions as described in the previous section,
the scattering order and the event type (fully-absorbed or escape) can be determined as follows.
\begin{enumerate}
\item One selects a scattering order from all candidates. 
Here the selected order is labeled with $k$:
\begin{align}
( \vector{D}_I)_{\mathrm{ordered}}^{k} = (\vector{D}_{\tau_{k} (1)}, \vector{D}_{\tau_{k} (2)}, ..., \vector{D}_{\tau_{k} (n)} )_{\mathrm{ordered}}~.
\end{align}
When the number of the hits is $n$,
the number of the candidates is $n!$, i.e., $1 \leq k \leq n!$.
\item One calculates the conditional probabilities given the event type and scattering order described in Eqs.~\ref{eq_likelihood_fulldep} and \ref{eq_likelihood_escape}, and uses the approximations introduced in Section~\ref{subsec_approximation}.
Then, for each scattering candidate, one obtains the two probabilities $\mathcal{P}_\mathrm{fullabs}^{k}$ and $\mathcal{P}_\mathrm{escape}^{k}$, which correspond to the fully-absorbed and escape events, respectively.
The calculation procedure of $\mathcal{P}_\mathrm{fullabs}^{k}$ and $\mathcal{P}_\mathrm{escape}^{k}$ is described in the following subsections.
\item One determines the scattering order and the event type as a set of them that yield the maximum value in all of the calculated 
$\mathcal{P}_\mathrm{fullabs}^{k}$ and $\mathcal{P}_\mathrm{escape}^{k}$.
\end{enumerate}

\subsection{Calculation of $\mathcal{P}_\mathrm{fullabs}^{k}$}
\begin{enumerate}
\item The incoming gamma-ray energy is calculated as the sum of the detected energy (Eq.~\ref{eq_e_ini_full_dep}), and the interaction positions are assumed to be the same as the detected ones (Eq.~\ref{eq_pos_approx}).
\item The first scattering angle is calculated by Eq.~\ref{eq_ini_theta_phi} and the incoming gamma-ray direction is constrained on a Compton circle in the sky.
\item The probability is calculated using Eq.~\ref{eq_likelihood_fulldep} by applying the approximations described in Eqs.~\ref{eq_modification_of_energy_resolution},~\ref{eq_l_first} 
\end{enumerate}

\subsection{Calculation of $\mathcal{P}_\mathrm{escape}^{k}$}
\begin{enumerate}
\item The incoming gamma-ray energy is calculated by Eq.~\ref{eq_esc_energy}, and the interaction positions are assumed to be the same as the detected ones (Eq.~\ref{eq_pos_approx}).
\item The first scattering angle is calculated by Eq.~\ref{eq_ini_theta_phi} and incoming gamma-ray direction is constrained on a Compton circle in the sky.
\item The probability is calculated using Eq.~\ref{eq_likelihood_escape} by applying the approximations described in Eqs.~\ref{eq_modification_of_energy_resolution},~\ref{eq_l_first},~\ref{eq_l_esc},~\ref{eq_last_ene_esc}.
\end{enumerate}

As mentioned in \S\ref{sec_concept}, when the number of hits is two or fewer,
the escape gamma-ray energy cannot be estimated 
since Eq.~\ref{eq_esc_energy} requires the detection of three or more hits to calculate the second and subsequent scattering angles.
Thus, this algorithm can work for 3 or more hit events.

The derived algorithm differs from the MSD methods in that 
it compares the deposited energy rather than the scattering angle.
Effectively, this algorithm calculates the deposited energy at each site assuming the incident gamma-ray energy and interaction positions, and compares it with the measured value directly through the energy response function.
Moreover, by adopting the probabilistic method, the physical processes are naturally taken into account in the scattering order determination.

\section{Numerical Experiments}
\label{sec_pv}

In order to test the reconstruction algorithm developed in this work,
we apply it to a simulation data set generated by {\tt ComptonSoft}, a software package of Geant4-based simulation and data analysis for Compton telescopes \cite{ODAKA2010303,ODAKA2018,AGOSTINELLI2003250}.
The GRAMS experiment utilizes a large volume detector using a liquid argon time projection chamber, and MeV gamma-ray events are considered to be dominated by multiple Compton scattering events\cite{aramaki2019Dual}.
We choose this type of Compton telescope for a numerical demonstration of the reconstruction algorithm, and evaluate the performance of the algorithm, i.e.,
the accuracy of the event classification (fully-absorbed or escape events), the prediction accuracy of the incident gamma-ray energy,
the scattering order, and the angular resolution.

\subsection{Simulation setup}

\begin{figure}[!htbp]
\begin{center}
\includegraphics[width = 7.5 cm]{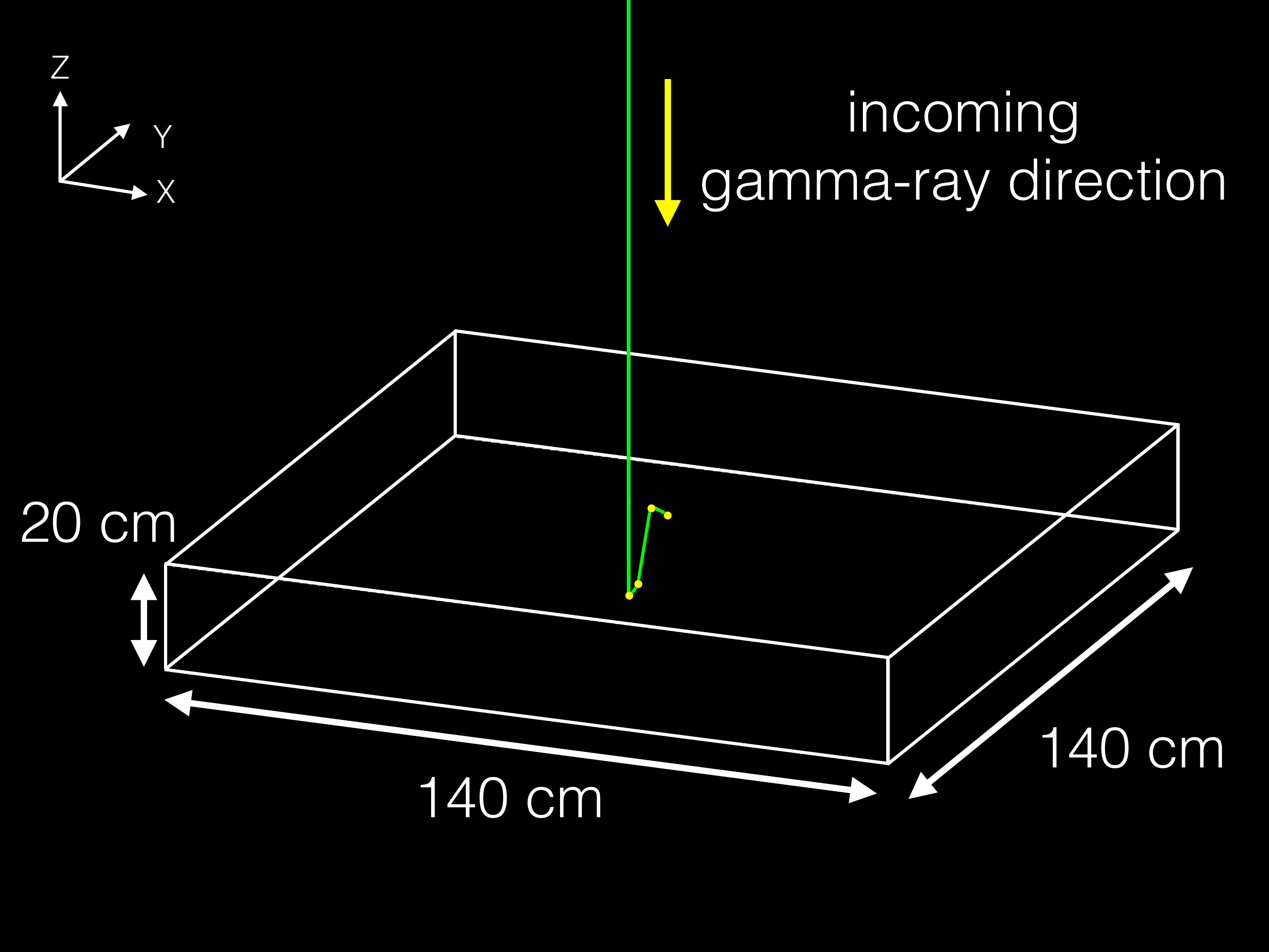}
\end{center}
\caption{The geometry of the Geant4 simulation.}
\label{fig_sim_setup}
\end{figure}

We assume a detector with a size of $140\times140\times20$ cm$^3$ filled with liquid argon as shown in Figure~\ref{fig_sim_setup}.
It is the same as proposed in \cite{aramaki2019Dual}.
The energy resolution $\sigma_{\hat{\varepsilon}}$ of the detector is set to as follows \cite{aramaki2019Dual}:
\begin{align}
\label{eq_sigma_dete}
\sigma_{\hat{\varepsilon}}^2 = (5~\mathrm{keV})^2 + 0.25~\mathrm{keV}^2 \times (\hat{\varepsilon}/\mathrm{keV})~.
\end{align}
Eq.~\ref{eq_sigma_dete} is also adopted in the probability calculation in Eq.~\ref{eq_energy_resolution}. 
We note that while the energy resolution of liquid argon time projection chambers in the MeV range has not yet been measured, the energy resolution of $\sim 1$\% ($1\sigma$) at 1 MeV is measured with a liquid argon scintillation detector \cite{Kimura2020}. Thus, the argon detector can have the assumed energy resolution in principle.

The X-Y positions of signals are pixelized with a size of 2 mm,
and we set $\sigma_{\hat{x}}$ and $\sigma_{\hat{y}}$ to be $2/\sqrt{12}$ mm in Eq.~\ref{eq_position_resolution}.
This is the standard deviation of X or Y positions of events distributed uniformly in a single pixel.
The direction of electron drift in the time projection chamber is assumed to be parallel to the Z-axis,
and the resolution of Z position is assumed to $\sigma_{\hat{z}} = $ 1 mm.
Note that X and Y positions are determined by the pixel positions, but Z position is estimated from the measured electron drift time. Thus $\sigma_{\hat{z}}$ can be different from $\sigma_{\hat{x}/\hat{y}}$.
In this simulation, the signals from adjacent pixels are merged into a single signal. The signal position after merging is set to the center of gravity (energy-weighted average) of the pixels before merging.
Note that the electron diffusion in the liquid argon is small; in this case, it is about 300 $\mu$m at the most, much smaller than the pixel size \cite{CENNINI1994,AMORUSO2004}. Thus we ignored the effect of electron diffusion in this simulation.
The energy threshold of each pixel is set to 25 keV.

In this demonstration, we set $L_\mathrm{first}$ in Eq.~\ref{eq_l_first} and 
$L_\mathrm{esc}$ in Eq.~\ref{eq_l_esc} to be 10 and 20 cm respectively (see \ref{sec_effect_length_scale} for this optimization).
The computational time for different numbers of hits is described in Table~\ref{tab_computational_time}.
For this table, we used 1 MeV gamma-ray events in the following subsection.

\begin{table}[!htbp]
\caption{The computational time for the event reconstruction of 1 MeV gamma-ray events. Here Mac mini (2018) with 3.2 GHz 6-Core Intel Core i7 and 32GB memery is used.}
\label{tab_computational_time}
\centering
\begin{tabular}{cc}
the number of hits & 
\begin{tabular}{c}
the number of processed \\ events per a second 
\end{tabular} \\ \hline
3 & $2.0\times10^{4}$ \\
4 & $9.8\times10^{3}$ \\
5 & $3.3\times10^{3}$ \\
6 & $7.5\times10^{2}$ \\
7 & $1.4\times10^{2}$ \\
8 & $2.1\times10^{1}$ \\ \hline
\end{tabular}
\end{table}

\subsection{Results of event classification and energy reconstruction}
\label{sec_event_classification}
\begin{figure}[!htbp]
\begin{center}
\includegraphics[width = 7.5 cm]{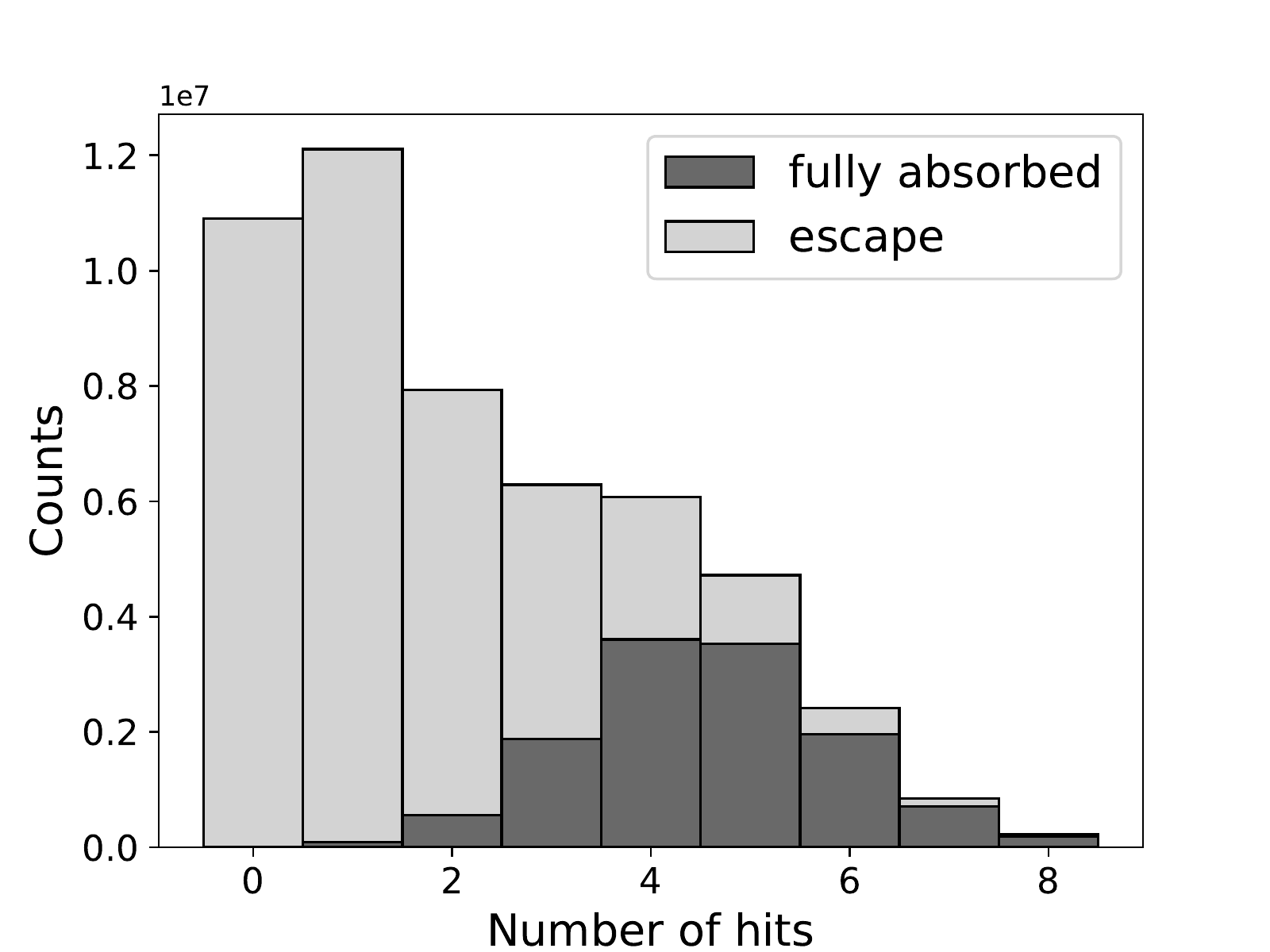}
\end{center}
\caption{The count of the detected events with the number of hits. Here $10^{8}$ events of 1 MeV gamma rays are simulated.
Note that 0 hit represents the event that gamma ray escapes from the detector without any interaction or all the produced signals are lower than the threshold.}
\label{fig_num_hist}
\end{figure}

\begin{figure}[!htbp]
\begin{center}
\includegraphics[width = 7.5 cm]{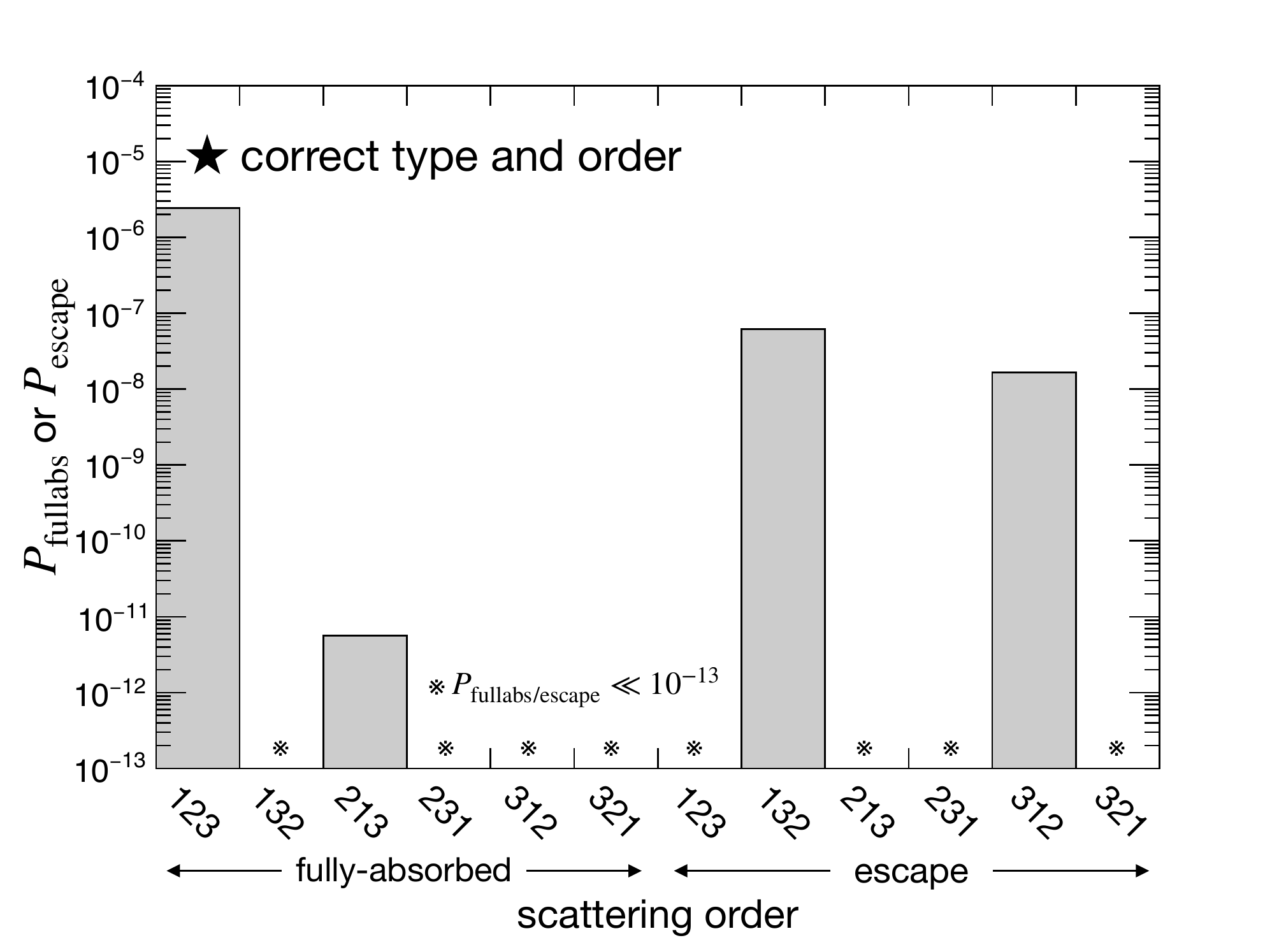}
\end{center}
\caption{A worked example of the algorithm to a 3-hit event, and the obtained conditional probabilities for each event type and scattering order.
Here we used an event of 
$\vector{D}_{1} = \left( \vector{r}_{1}, \varepsilon_{1} \right)
= (-5.3~\mathrm{cm}, -58.1~\mathrm{cm}, 3.64~\mathrm{cm}, 767.0~\mathrm{keV})$, 
$\vector{D}_{2} = (12.7~\mathrm{cm}, -60.9~\mathrm{cm}, 8.59~\mathrm{cm}, 37.7~\mathrm{keV})$
and 
$\vector{D}_{3} = (14.9~\mathrm{cm}, -59.3~\mathrm{cm}, 7.98~\mathrm{cm}, 168.6~\mathrm{keV})$.
The coordinate origin is the center of gravity of the detector in Figure~\ref{fig_sim_setup}.
The scattering order labels in the X-axis are the same as Figure~\ref{fig_scattering_pattern}.
}
\label{fig_conditional_prob_eventID_22}
\end{figure}

We simulated $10^{8}$ events of 1 MeV gamma-ray beam incoming from the top of the detector,
i.e., $\hat{\theta}_0 = \pi$.
The beam has a radius of 110 cm that covers the whole volume of the detector, and it is co-aligned at the detector center.
Figure~\ref{fig_num_hist} shows the count of the detected events with the number of hits.
The number of counts of fully-absorbed events reaches a maximum at 4 hits.
Here we focus on the events with the number of hits from 3 to 8.
The ratio of the number of the events with more than 8 hits to the total detected events is only 0.13\% and they are negligible.
Figure~\ref{fig_conditional_prob_eventID_22} shows an example of applying the algorithm to a 3-hit event.
The algorithm outputs 12 probabilities corresponding to each scattering order and event type.
In this case, the largest value is the leftmost one, and this event is identified as a fully-absorbed event with a scattering order of ``123''.

\begin{figure*}[!htbp]
\begin{center}
\includegraphics[width = 16.0 cm]{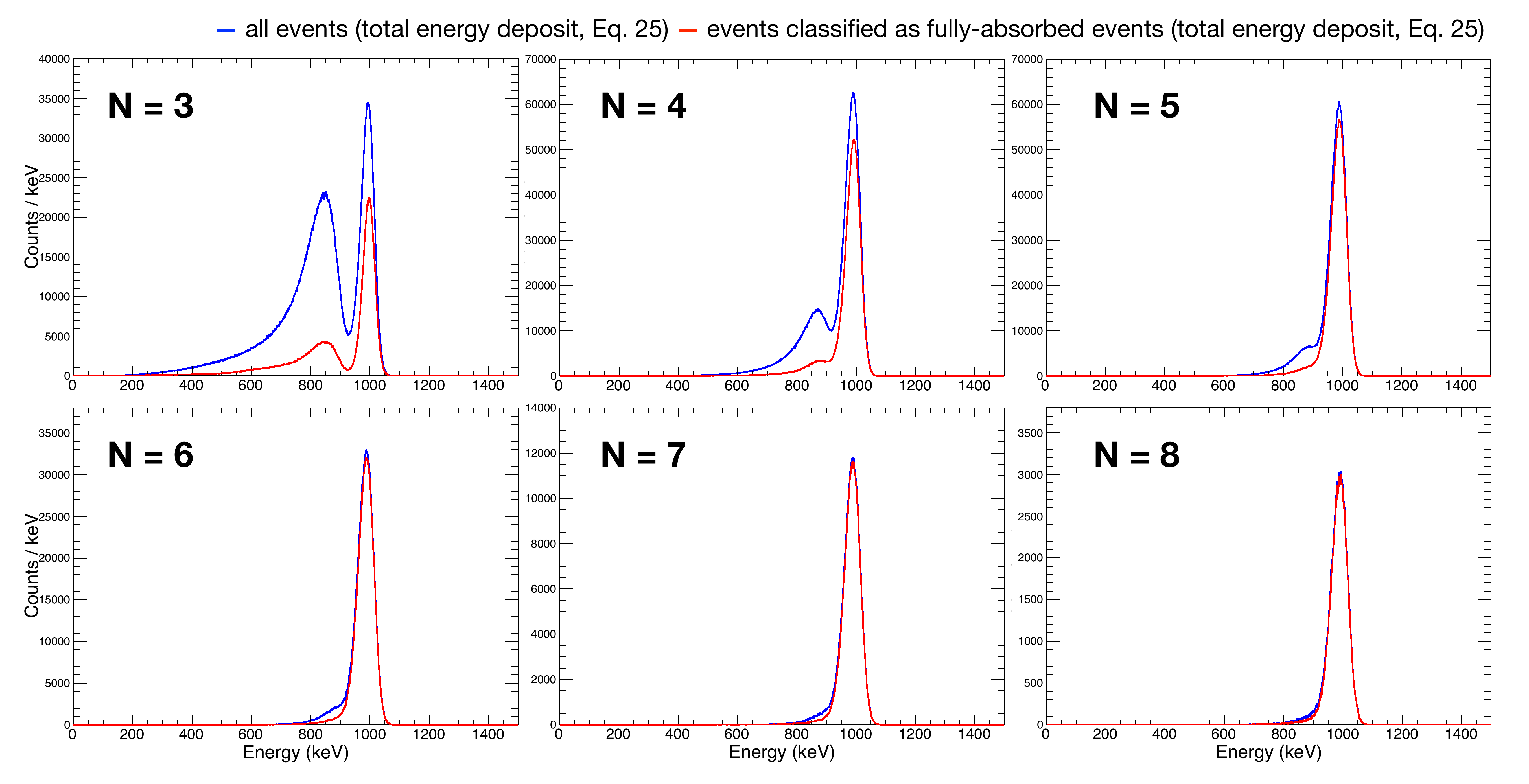}
\end{center}
\caption{The energy spectra of 1 MeV gamma-ray events classified as fully-absorbed events.
The blue and red lines represent the spectra of the total detected energies of all events and those of the events classified as fully-absorbed events by the algorithm, respectively.
}
\label{fig_spectrum_fulldep_1MeV}
\end{figure*}

First, we examine the algorithm performance against fully-absorbed events qualitatively.
Figure~\ref{fig_spectrum_fulldep_1MeV} shows the energy spectra of all detected events (blue, solid) and of those classified as fully-absorbed events after applying the algorithm (red, solid).
We confirmed that the events peaking around 1 MeV is correctly classified as fully-absorbed events and the component below 1 MeV, which corresponds to escape events, is reduced successfully after the reconstruction algorithm is applied.
As the number of hits is increased,
the events around 1 MeV are classified as fully-absorbed more accurately.
This is because events with more hits have more physical information to constrain the Compton scattering sequence.

\begin{figure*}[!htbp]
\begin{center}
\includegraphics[width = 16.0 cm]{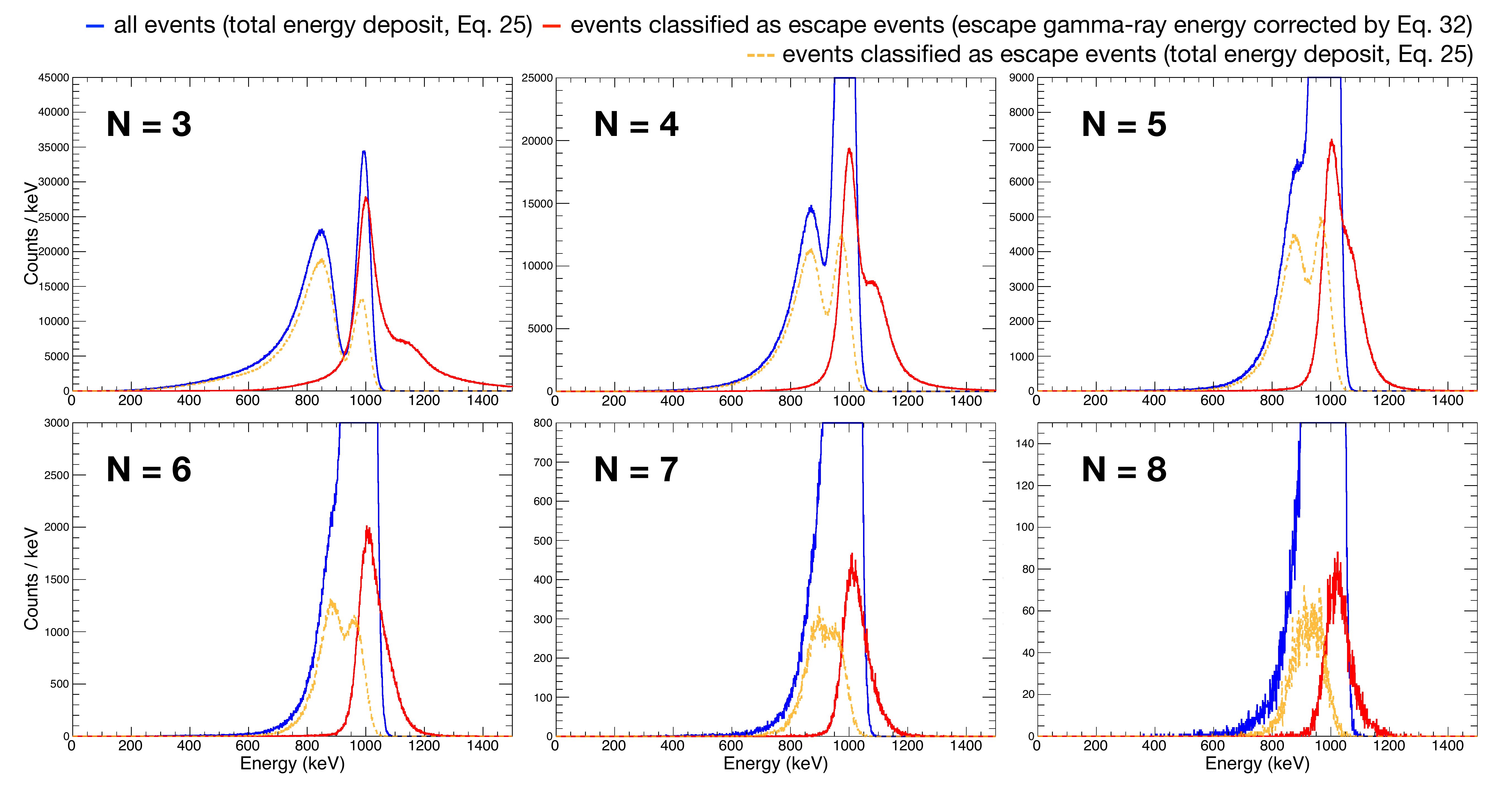}
\end{center}
\caption{The energy spectra of 1 MeV gamma-ray events classified as escape events.
The blue lines are the same as in Figure~\ref{fig_spectrum_fulldep_1MeV}.
The orange lines represent the spectra of the total detected energy of events classified as escape events by the algorithm.
The red ones represent the spectra of the incident energy of those events, estimated by the algorithm using Eq.~\ref{eq_esc_energy}.
Note that the events for the orange and red lines are the same, but the energies shown here are different (one is the total energy deposit, and the other is the reconstructed energy).
}
\label{fig_spectrum_escape_1MeV}
\end{figure*}

Next, the algorithm performance against escape events is checked.
Figure~\ref{fig_spectrum_escape_1MeV} shows energy spectra of events classified as escape events.
The orange lines are spectra of the sum of detected energies of events classified as escape events.
Note that the blue ones are the same as Figure~\ref{fig_spectrum_fulldep_1MeV}.
By comparing the blue and orange lines, we can see how accurately the algorithm identifies the escape events.
Considering that the events with total deposited energies less than 1 MeV are escape events, and most of these events are included in the orange line,
we confirmed that the algorithm successfully identifies the escape events.
Furthermore, we can also check whether the energy correction by Eq.~\ref{eq_esc_energy} works for the escape events correctly.
The red lines in Figure~\ref{fig_spectrum_escape_1MeV} are the spectra of estimated incident gamma-ray energy by applying Eq.~\ref{eq_esc_energy} to the events classified as escape events, i.e., the events in the orange lines.
The peak at 1 MeV is clearly reconstructed, confirming that the algorithm correctly estimates the escape energy.
Note that the reconstructed spectra of the escape events are broader than those of fully-absorbed events.
It is because the positional information is also used for estimating gamma-ray escape energy. Then the reconstructed energy is affected by both the errors of position and energy of each detected hit.

\begin{figure}[!htbp]
\begin{center}
\includegraphics[width = 7.5 cm]{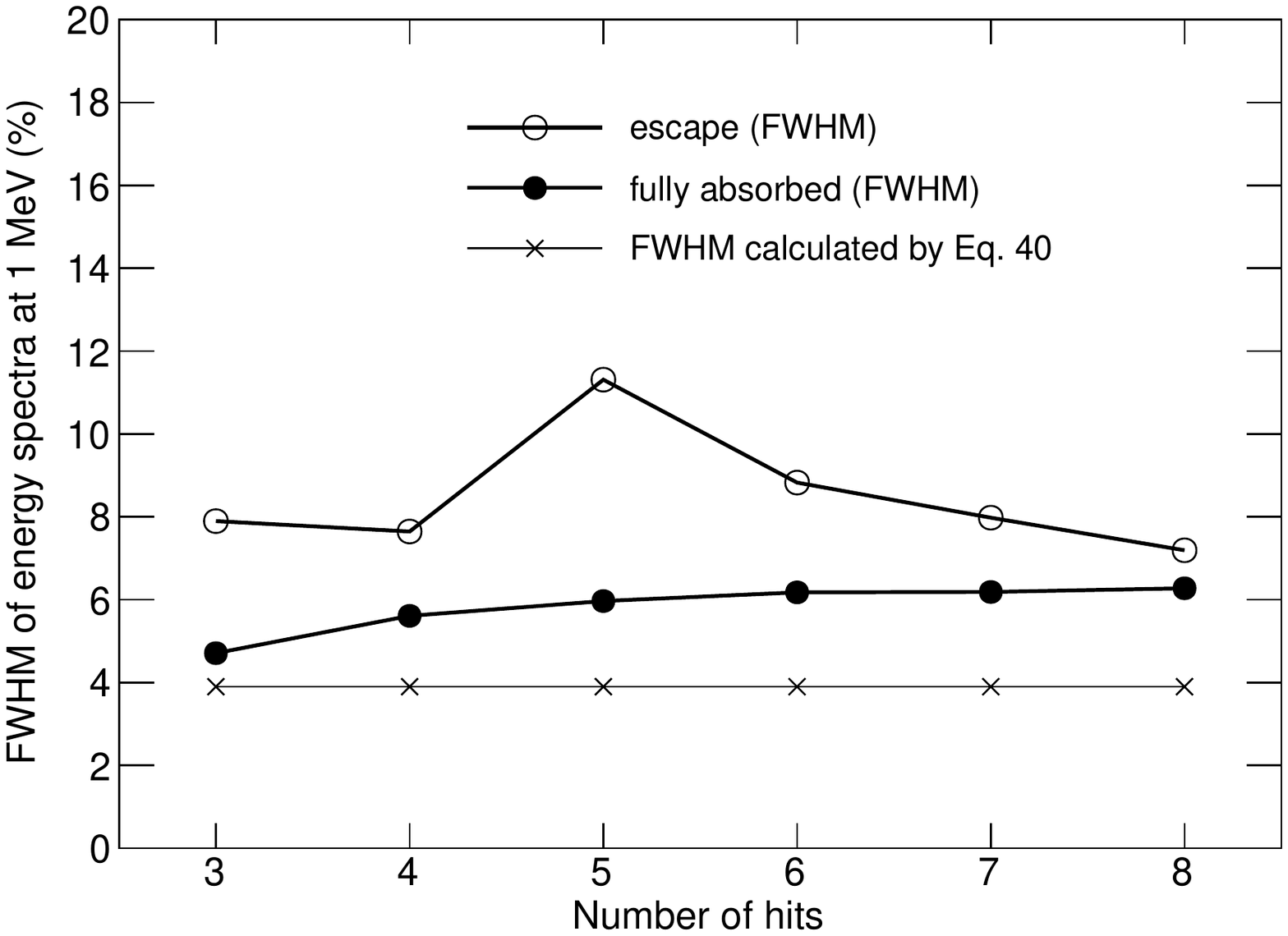}
\includegraphics[width = 7.5 cm]{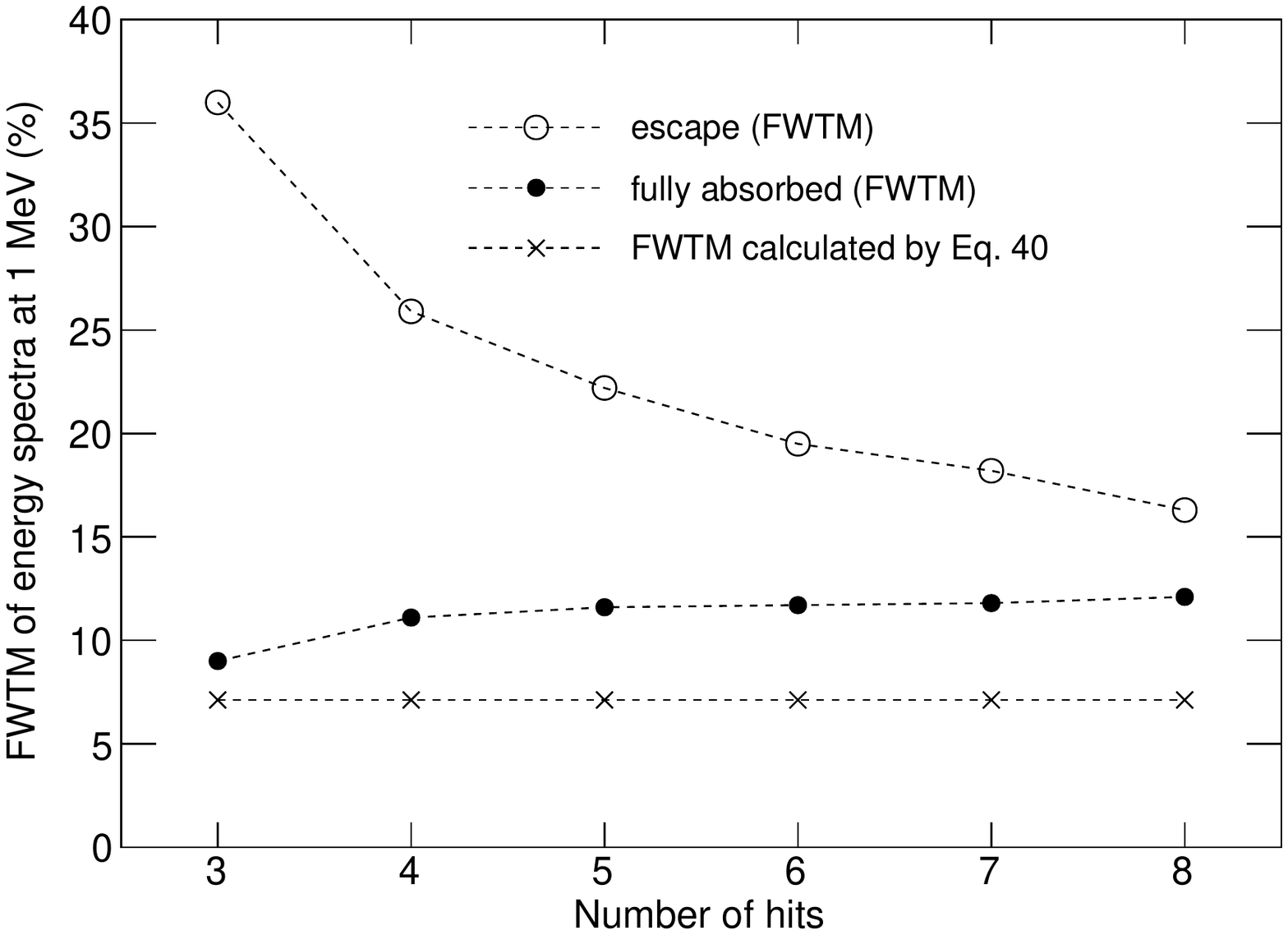}
\end{center}
\caption{The energy resolution of the reconstructed spectra for 1 MeV gamma-ray events. The top and bottom show the FWHM and FWTM of the reconstructed energy spectra, respectively. The line with ``$\times$'' represents the energy resolution of the detector (see Eq.~\ref{eq_sigma_dete})}
\label{fig_energy_resolution_at_1MeV}
\end{figure}

The dependence of the spectra on the event type and the number of hits is investigated quantitatively.
Figure~\ref{fig_energy_resolution_at_1MeV} shows the full width at half maximum (FWHM) of the reconstructed energy spectra with different numbers of hits.
Note that the detector has an energy resolution of 16.6 keV at 1 MeV as FWHM in this simulation.
For fully-absorbed events, the energy resolution does not depend on the number of hits so much.
On the other hand, for escape events, the FWHM for the escape events shows a peak at 5 hits.
When fully-absorbed events are misidentified as escape events, their incident energies are incorrectly reconstructed as above 1 MeV by the energy correction. Then, this misidentification makes a high-energy tail or hump in the reconstructed energy spectra, and from 5 hits, the FWHM catches it.
As mentioned before, the positional uncertainty can affect the energy spectra of the escape events, which also contributes to their tail-like structures in both low- and high-energy bands. 
To evaluate it, we also show the full width at tenth maximum (FWTM) of the spectra in Figure~\ref{fig_energy_resolution_at_1MeV}.
We can see that the FWTM of the escape events becomes smaller as the number of hits increases, which corresponds to the spectrum having shorter tails.
We want to note that though the absolute value of the energy resolution varies by the assumed detector response, in general, the energy resolutions of the fully-absorbed and escape events are different, and they have different dependence on the number of hits.

\subsection{Angular resolution and its dependence on the number of hits}

We also investigated the angular resolution of the reconstructed events and its dependence on the number of hits.
The angular resolution is evaluated by the angular resolution measure (ARM), which is defined as
\begin{align}
\mathrm{ARM} = \theta_K - \theta_G~,
\end{align}
where $\theta_K$ is the estimated first scattering angle:
\begin{align}
\theta_K = \arccos \left( 1 + m_e c^2 \left( \frac{1}{\hat{E}_0^\ast} - \frac{1}{\hat{E}_0^\ast - E_1} \right) \right)~,
\end{align}
and $\theta_G$ is the first scattering angle calculated from the incident direction and the reconstructed positions of the first and second hits.

\begin{figure*}[thbp]
\begin{center}
\includegraphics[width = 8.0 cm]{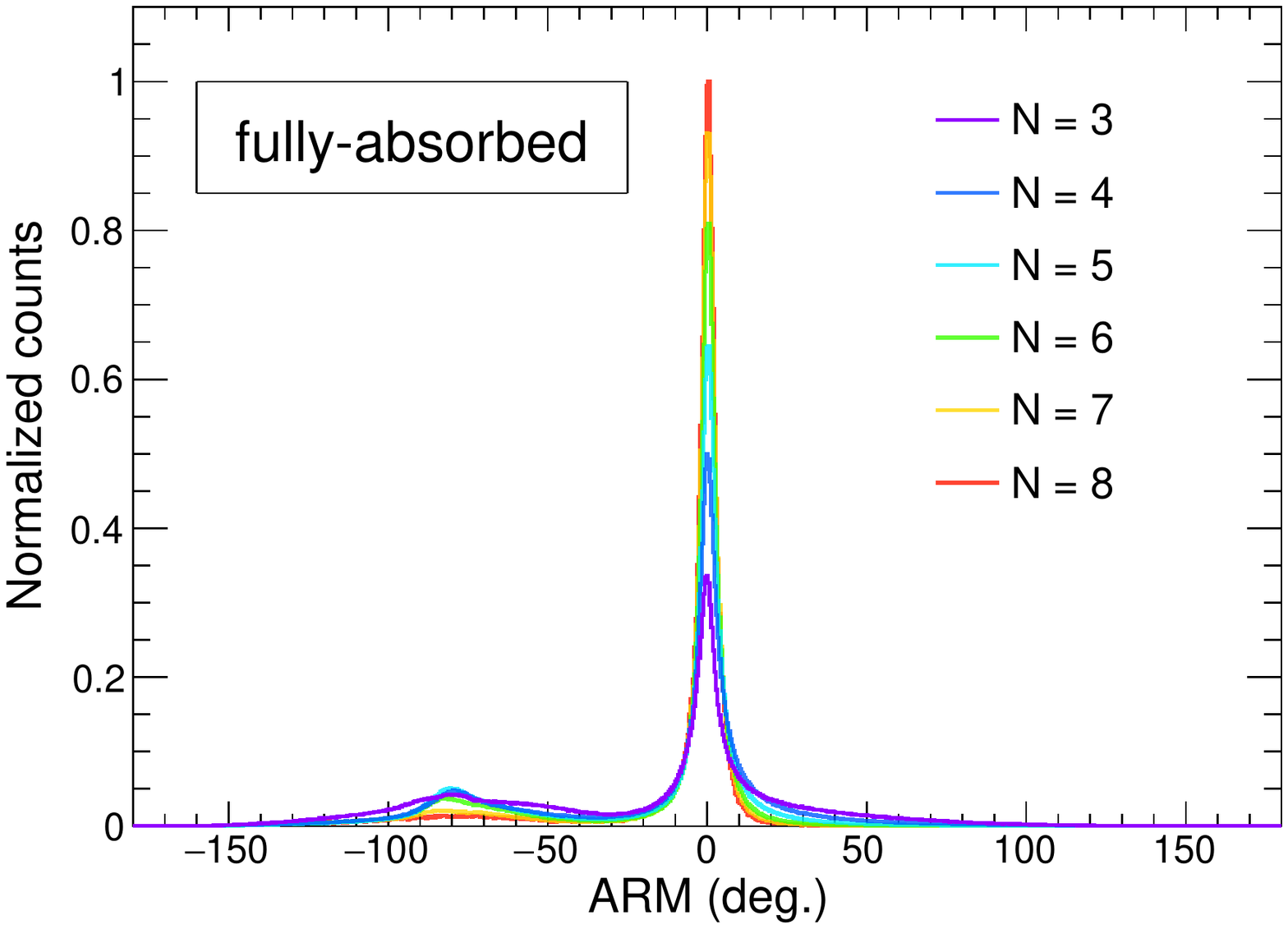}
\includegraphics[width = 8.0 cm]{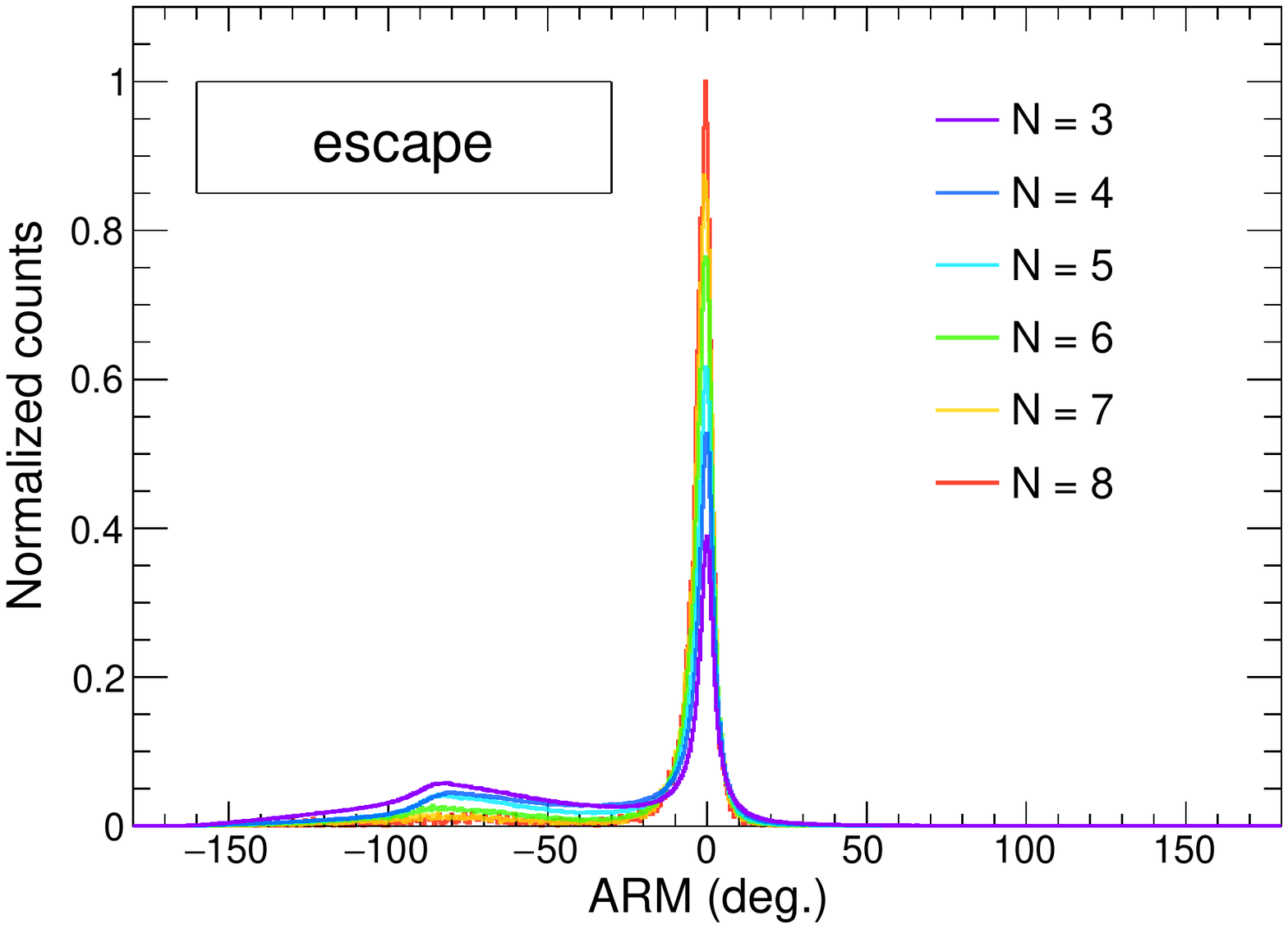}
\end{center}
\caption{ARM distributions of 1 MeV gamma rays with different numbers of hits.
The left and right panels correspond to fully-absorbed and escape events, respectively. The histograms are normalized by the total count in each.}
\label{fig_arm}
\end{figure*}

Figure~\ref{fig_arm} shows distributions of ARM for both fully-absorbed and escape events up to 8 hits, using the same data set in the previous subsection.
We normalized the histograms by the total count in each.
In both cases, as the number of hits increases, 
the main peak at 0 degrees becomes sharper,
and the tail components and a sub-peak at $\sim 90$ degrees are reduced.
It suggests that the scattering order is estimated more accurately with more hits.
We will examine this point quantitatively in the following subsection.

\begin{figure}[!htbp]
\begin{center}
\includegraphics[width = 7.5 cm]{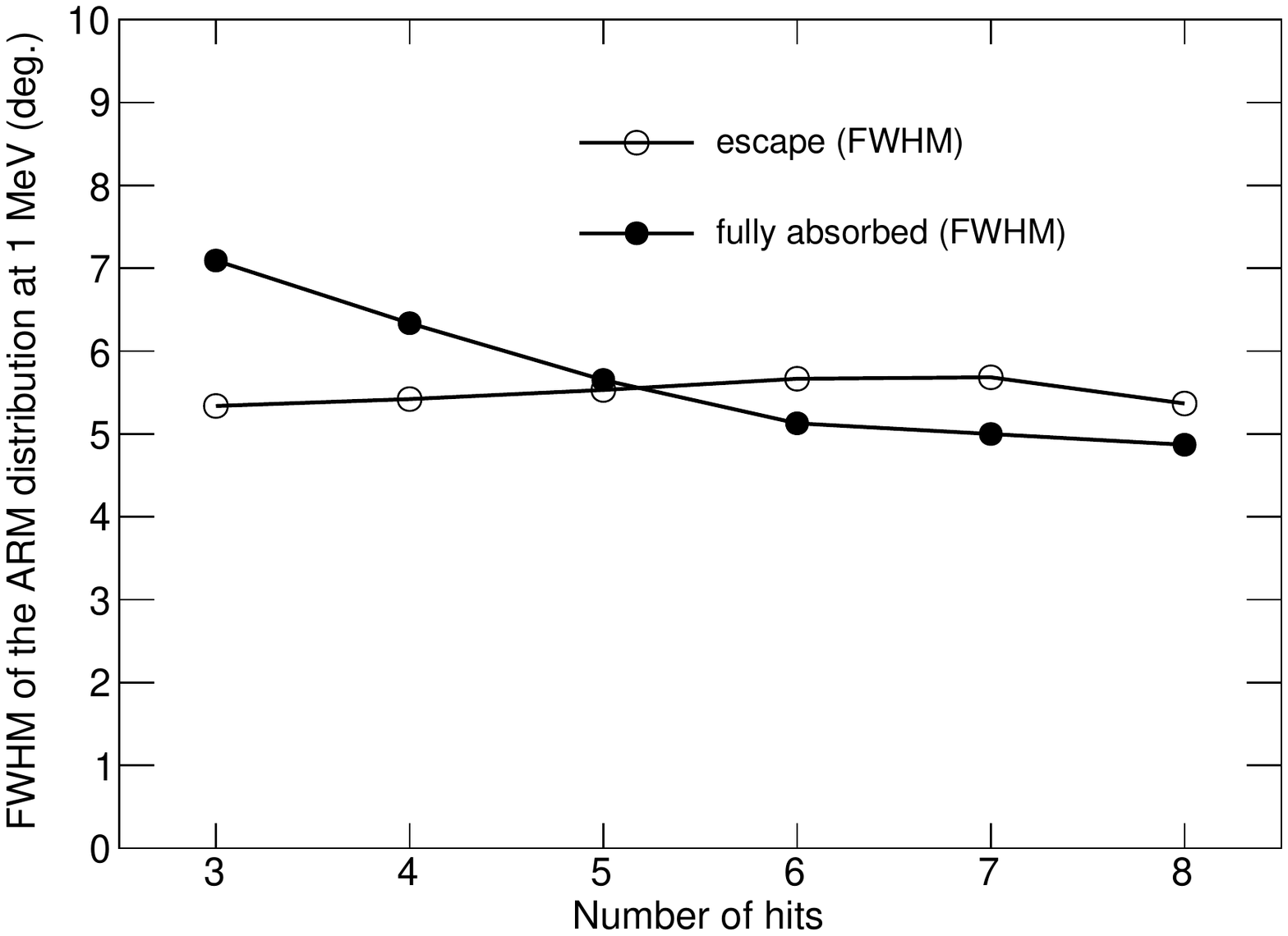}
\includegraphics[width = 7.5 cm]{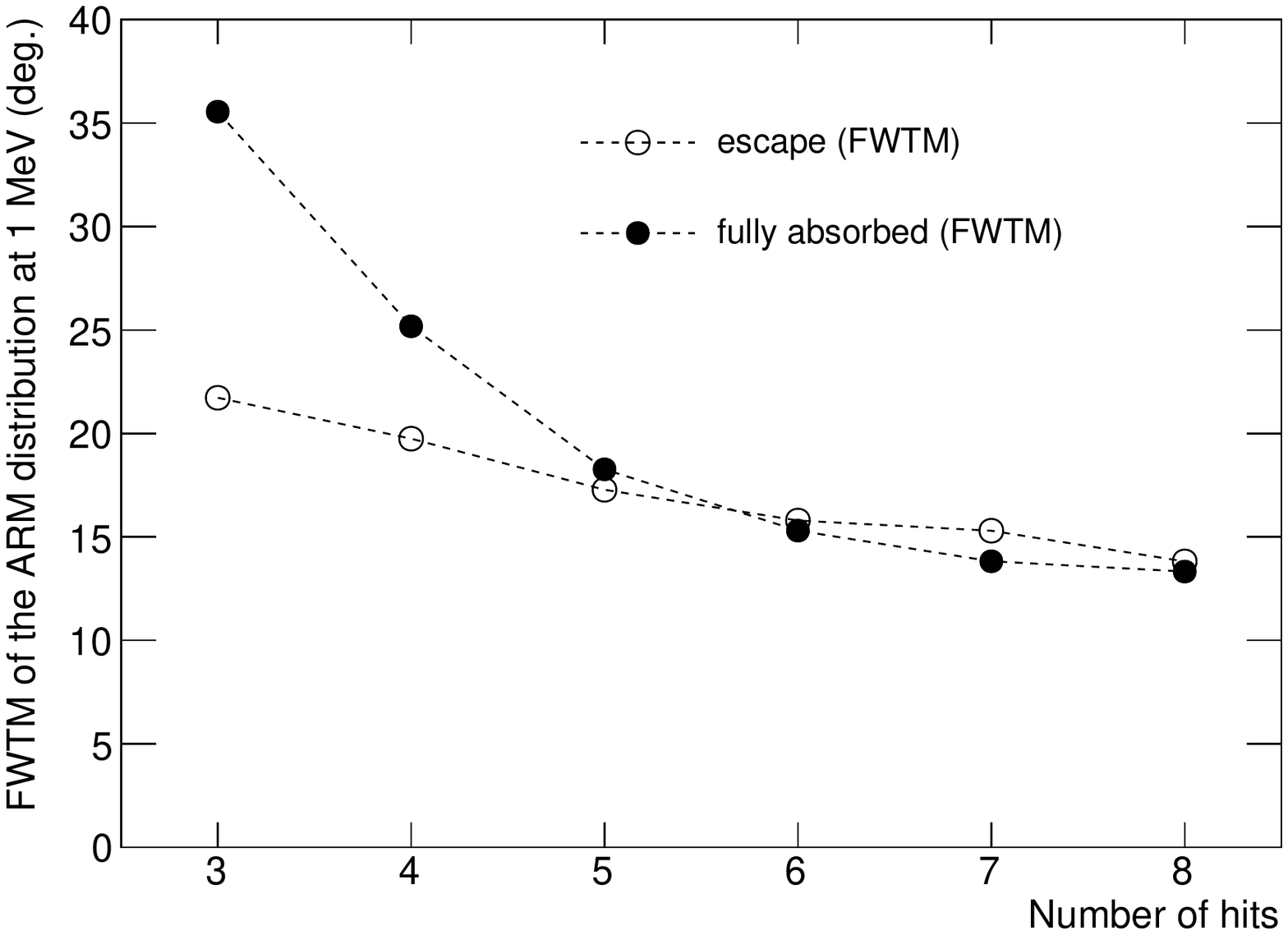}
\end{center}
\caption{The angular resolutions of 1 MeV gamma rays for fully-absorbed and escape events.
The FWHMs (top) and FHTMs (bottom) of the ARM distributions in Figure~\ref{fig_arm} are shown for different numbers of hits.
The filled and open markers correspond to the fully-absorbed and escape events, respectively.}
\label{fig_arm_number_of_hit}
\end{figure}

The FWHM of the obtained ARM distributions are shown in Figure~\ref{fig_arm_number_of_hit}.
It can be seen that they are from 4.9 to 7.1 degrees, depending on the number of hits and the event type.
In this case, the FWHM is larger for fully-absorbed events with a smaller number of hits, because these events contain a higher percentage of events with large scattering angles and large deposited energies at the first interaction, and the Doppler broadening effect limits the angular resolution. 
When 1 MeV gamma ray is back-scattered, the scattered gamma ray has an energy of 0.204 MeV. Then the gamma ray is difficult to escape from the detector since its interaction length is much smaller than that of 1 MeV gamma ray, and it is usually absorbed by the detector after few scattering.
Note that since the scattering angle at the first interaction is estimated, the angular resolution can be improved by selecting forward-scattering events. For example, when using events with scattering angles less than 60 degrees at the first interaction, the ARM FWHM of 3-hit fully-absorbed events is improved from 7.1 degrees to 4.1 degrees.
Figure~\ref{fig_arm_number_of_hit} also shows the FWTM of the ARM distributions.

\subsection{Accuracy of the event reconstruction}

In order to quantify the performance of the developed algorithm,
we calculate the fraction of the 1MeV gamma-ray events reconstructed with the correct event types (``accuracy (type)'' in the following figures),
the fraction of the events reconstructed with the correct scattering orders (``accuracy (order)''),
and the fraction of the events reconstructed correctly for both the event type and scattering order (``accuracy (type + order)'').

\begin{figure*}[thbp]
\begin{center}
\includegraphics[width = 8.0 cm]{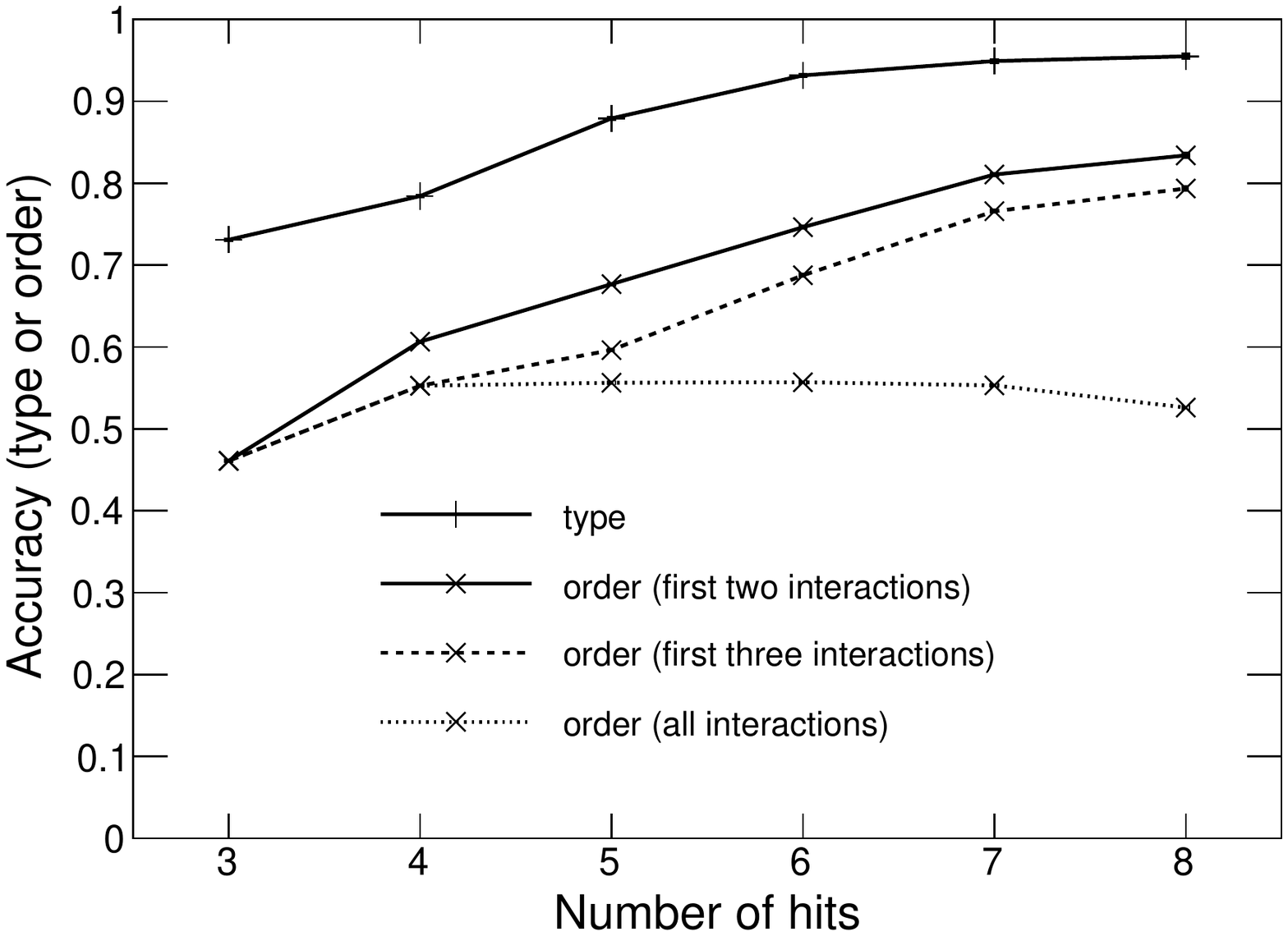}
\includegraphics[width = 8.0 cm]{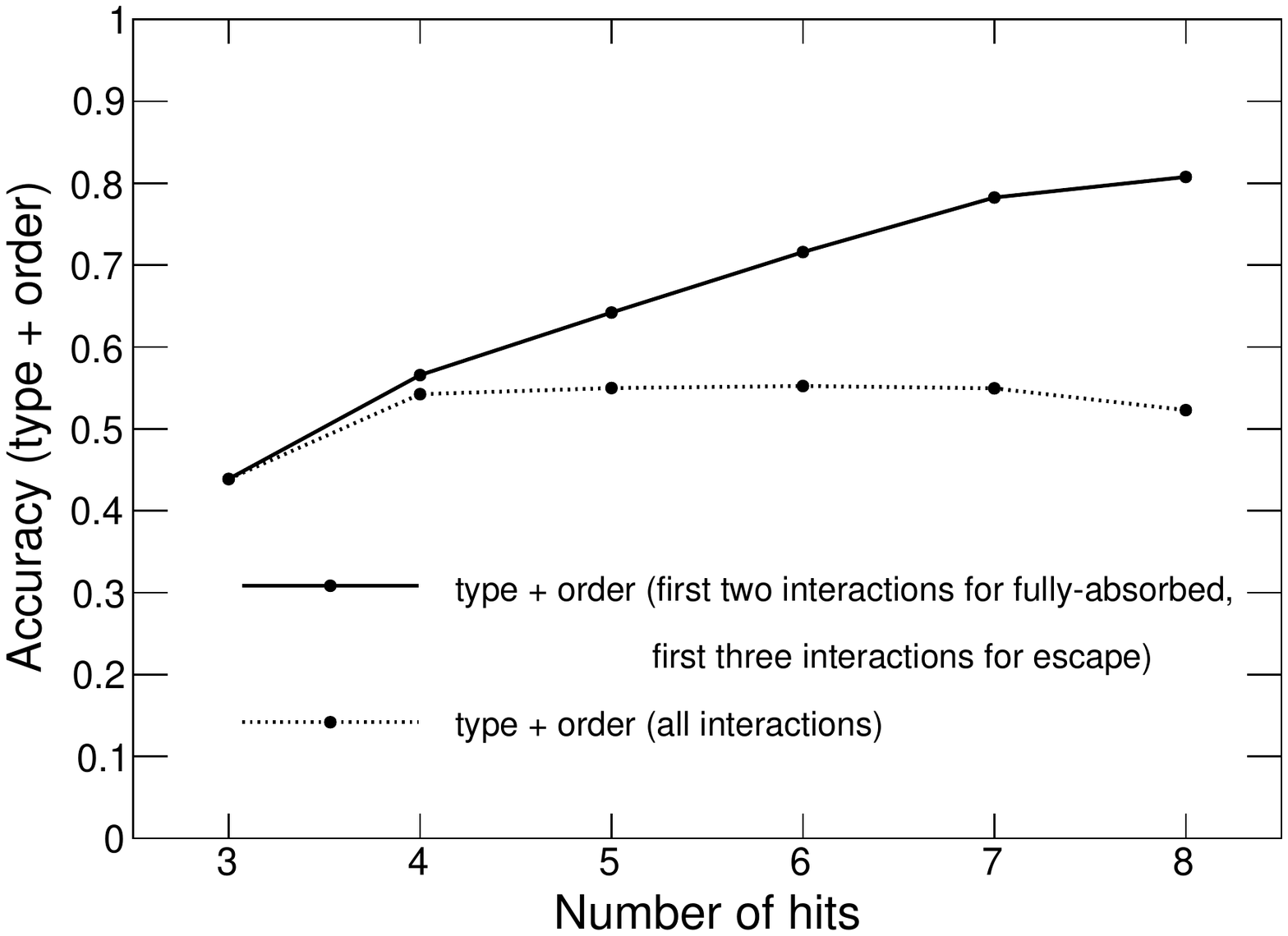}
\end{center}
\caption{The reconstruction accuracy of 1 MeV gamma-ray events. The left panel shows the accuracy for event type (``$+$'') or scattering order (``$\times$''). 
For the scattering order, the solid, dashed, and dotted lines represent the accuracy for the scattering order of the first two interactions, the first three interactions, and all interactions, respectively. 
The right panel shows the accuracy for both event type and scattering order. The solid line represents the accuracy when focusing on the scattering order of the first two interactions for fully-absorbed events and the first three for escape events, while the dotted line represents that of all interactions.}
\label{fig_accuray_at_1MeV}
\end{figure*}

We show these quantities for the 1 MeV gamma-ray events in Figure~\ref{fig_accuray_at_1MeV}.
The algorithm identifies the event type correctly with an accuracy of more than 73\%, and when the number of hits reaches 8, the accuracy is improved to 95\% (``$+$'' in the figure on the left).
On the other hand, the accuracy for the scattering order is about 55\%, and it reaches the peak at 4-5 hits (the dotted line with ``$\times$'' in the figure on the left).
This is because as the number of hits increases, 
the number of candidates for the scattering order increases with factorial. Then the correct order must be selected from a larger number of possibilities. 
However, if we focus on the scattering orders of the first two or three interactions (the solid and dashed lines, respectively), the accuracy of the scattering order improves with the number of hits, and it reaches 80\% for 8-hit events.
From a practical point of view, it is important to identify the first two or three interactions. In the case of fully-absorbed events, the scattering angle and scattered gamma-ray direction at the first interaction can be calculated correctly as long as the first two interactions are correctly identified.
Also, if the first three interactions are correctly identified for escape events, the escape energy can be calculated in principle \cite{Kroeger2001}.
The accuracy for both event type and scattering order also has a similar trend (the figure on the right). When only considering the first two interactions for fully-absorbed events and the first three for escape events, the algorithm reconstructs the gamma-ray events with an accuracy of 81\% for 8-hit events (see the solid line in the right).

\begin{figure*}[thbp]
\begin{center}
\includegraphics[width = 7.0 cm]{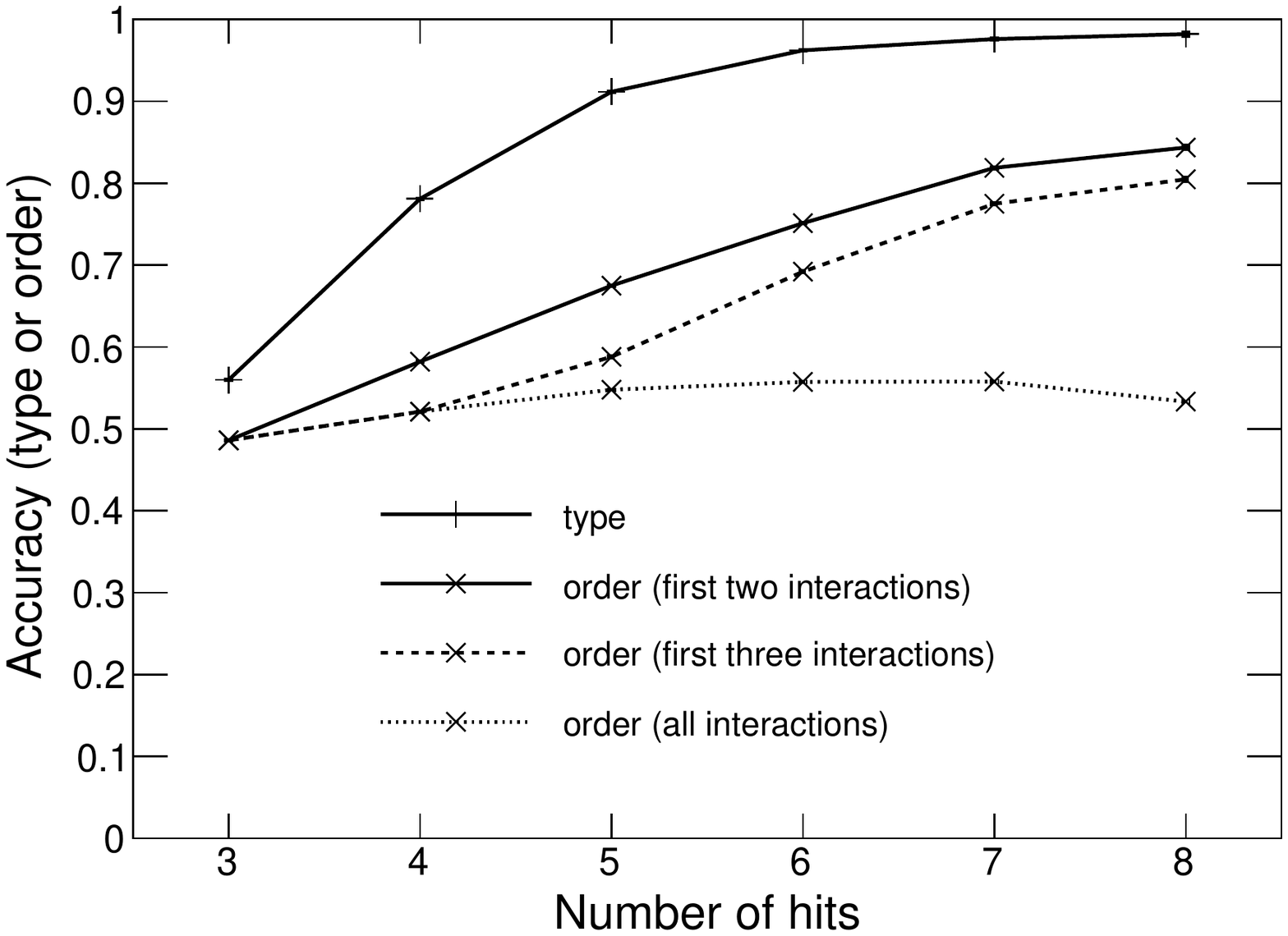}
\includegraphics[width = 7.0 cm]{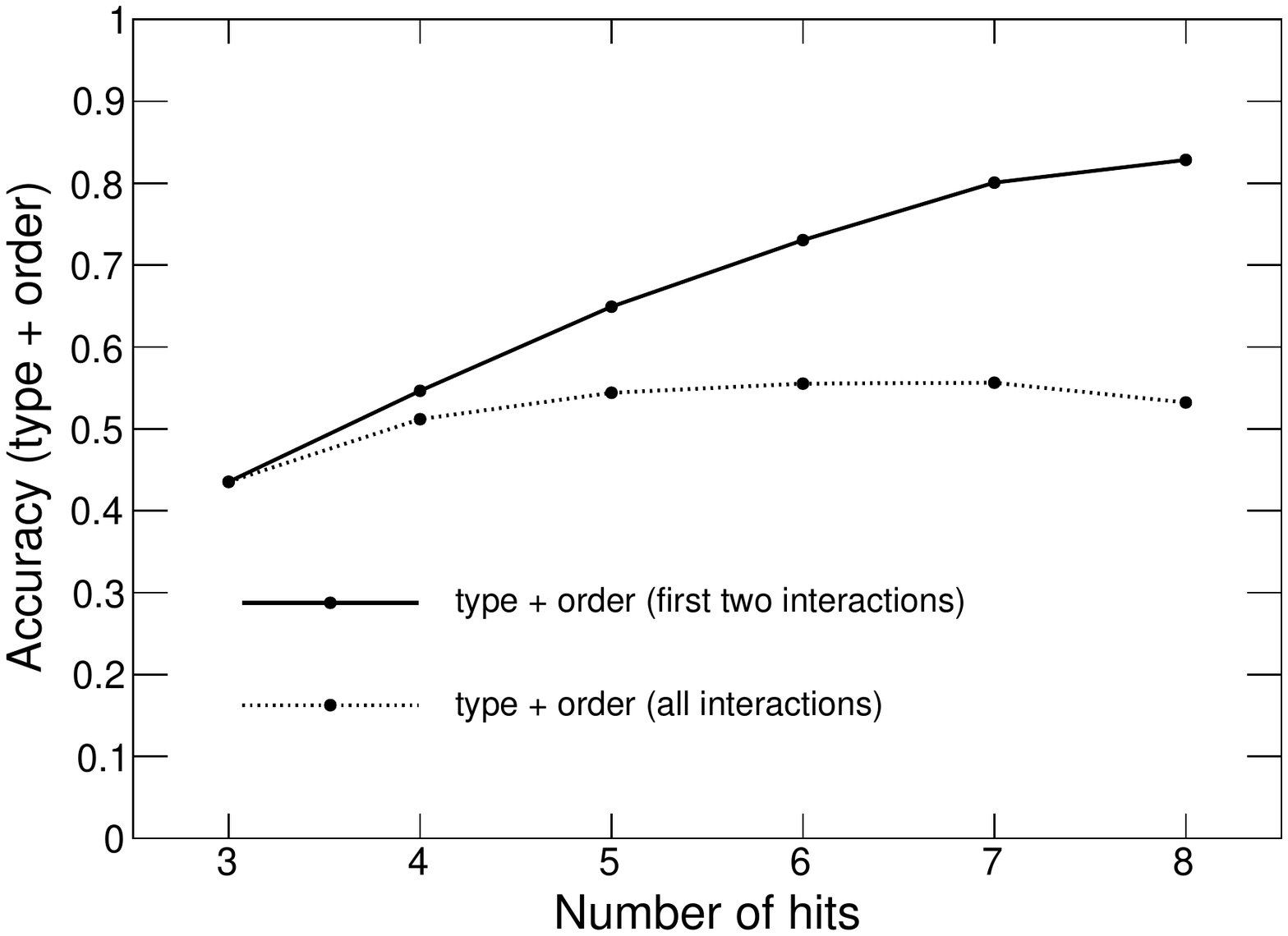}
\end{center}
\caption{The same as Figure~\ref{fig_accuray_at_1MeV}, but using truly fully-absorbed events.}
\label{fig_accuray_at_1MeV_fulldep}
\end{figure*}

\begin{figure*}[thbp]
\begin{center}
\includegraphics[width = 7.0 cm]{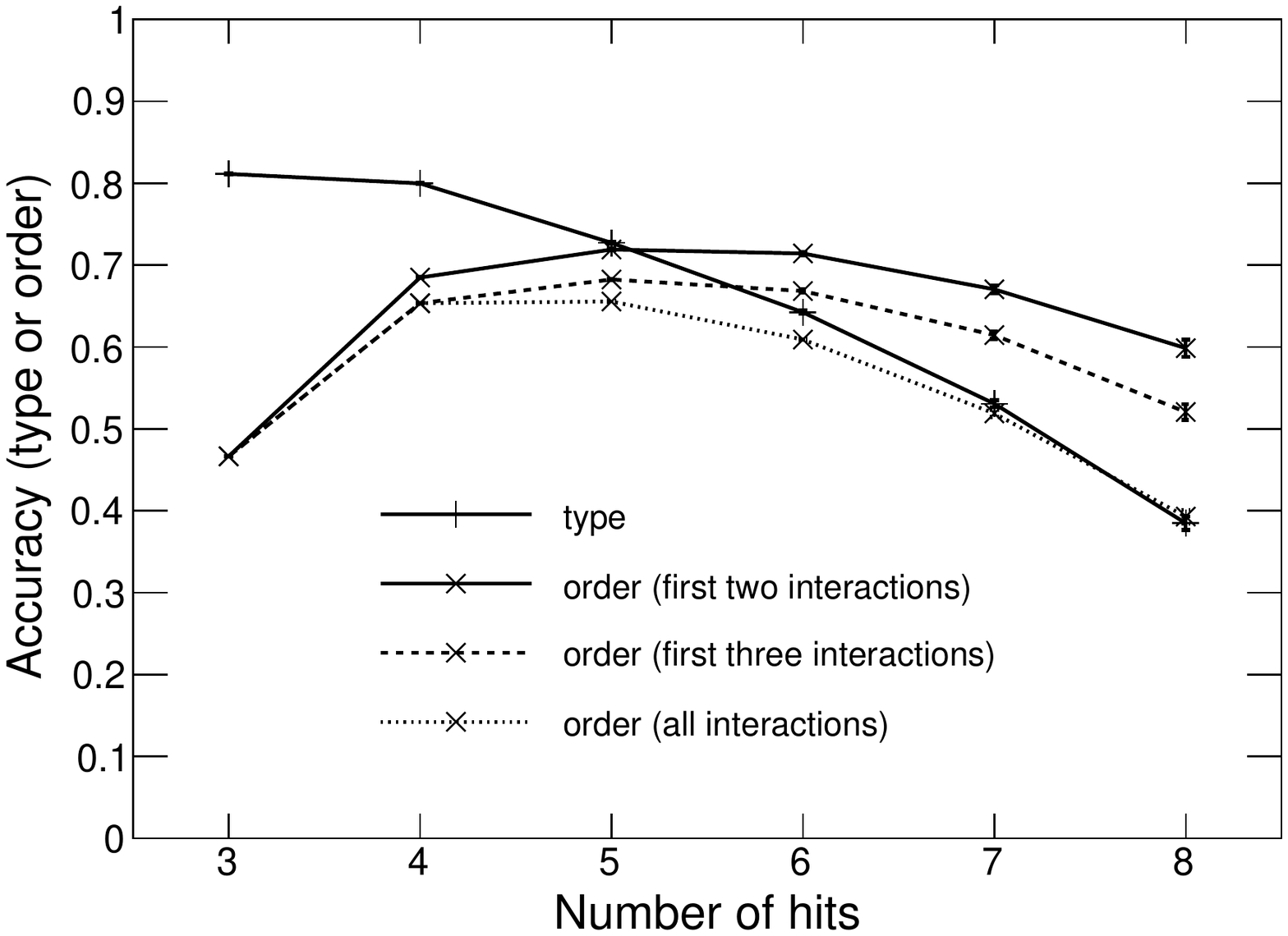}
\includegraphics[width = 7.0 cm]{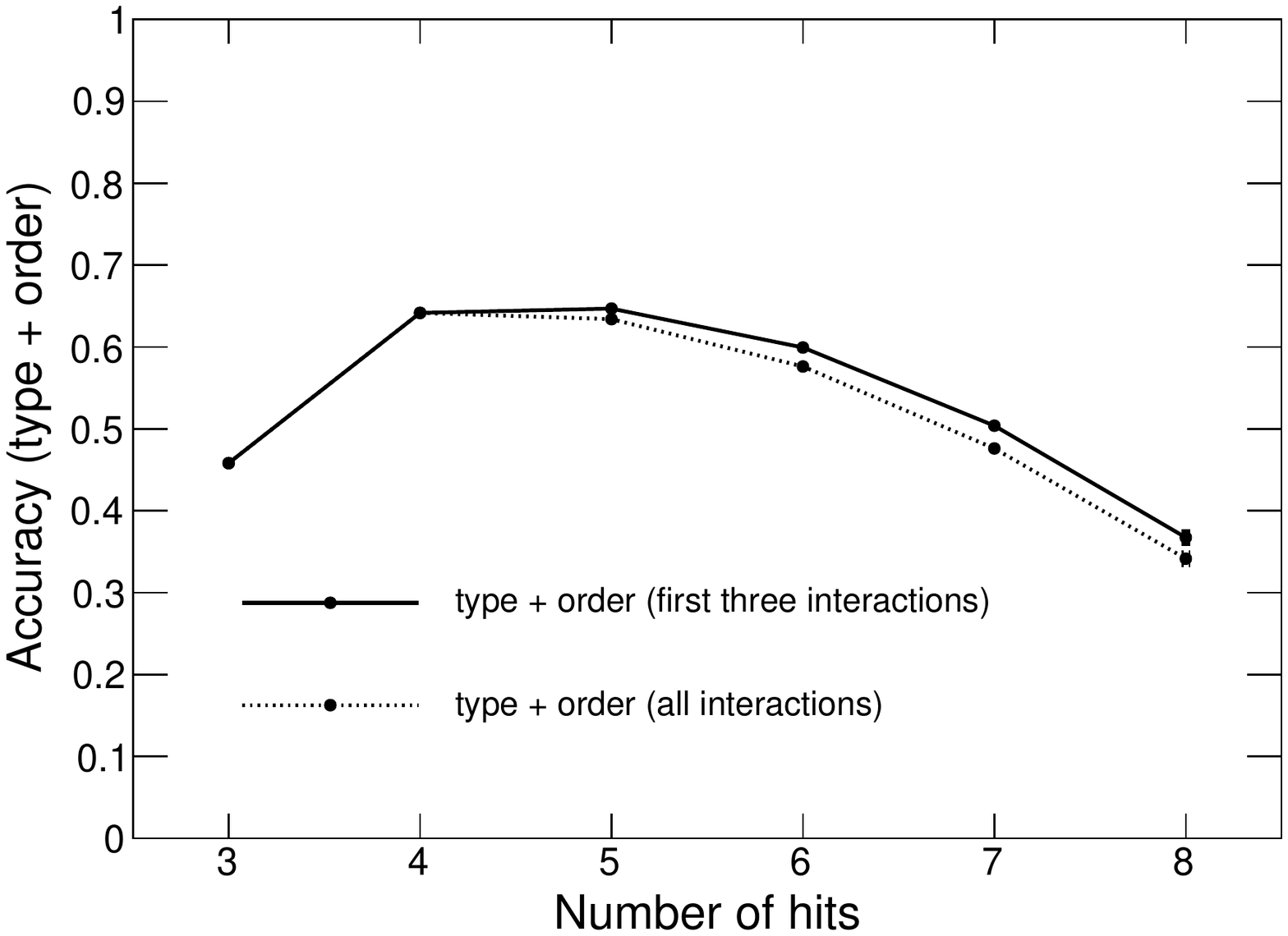}
\end{center}
\caption{The same as Figure~\ref{fig_accuray_at_1MeV}, but using escape events with truly deposited energies below 900 keV.}
\label{fig_accuray_at_1MeV_escape}
\end{figure*}

In Figures~\ref{fig_accuray_at_1MeV_fulldep} and \ref{fig_accuray_at_1MeV_escape}, we extract the truly fully-absorbed or escape events respectively, and show the fractions of correctly reconstructed events among them.
For the escape events,
we use events in which the sum of the truly deposited energies is less than 900 keV.
The accuracy of the fully-absorbed events shows a similar trend to Figure~\ref{fig_accuray_at_1MeV}. 
The accuracy for the event type is excellent for 8-hit events while this decreases to 56\% for 3-hits events.
However, for truly escape events with 3 hits, the event type is correctly identified with an accuracy of 81\%.
Note that the algorithm performance depends on the detector's energy resolution and position resolution. It is discussed in \ref{sec_effect_energy_position_resolution}.

\subsection{Comparison with other algorithms}
\label{sec_comparsion}

We compare the obtained performance with those of other two algorithms.
We adopt the TANGO algorithm \cite{TANGO2010} as reference, which can distinguish fully-absorbed and escape events based on the MSD method.
We slightly modified the reference algorithm to compare the cosine of the scattering angle directly rather than the scattering angle for the following reason. The assumed energy resolution here is not as good as the germanium detector in the original TANGO algorithm. Then the cosine of the scattering angle calculated from kinematics (see Eq.~\ref{eq_ene_scattering} and $|\cos \theta_{kin}|$ in \cite{TANGO2010}) is often larger than one even for a correct scattering order, mainly when the events include a Compton scattering interaction with a scattering angle close to 0 or 180 degrees. In the original implementation, the scattering angle, not the cosine of it, is used for the FoM, and the correct scattering order of such an event is rejected because the arc-cosine cannot be applied to them. We found that it degrades the algorithm’s performance significantly, and using the cosine of scattering angle results in much better performance because we do not have to use the arc-cosine and avoid rejecting the correct scattering order due to the mathematical reason.
Also, since the original implementation assumes the source position in their nuclear experiments, we tuned the TANGO algorithm as follows.
The comparison of the cosine of scattering angle is skipped at the first interaction and the polarization effect in the Klein-Nishina formula is ignored,
and the traveled distance in Eq.6 is fixed to 10 cm for the first interaction.
In addition, we also compare our algorithm with a classical algorithm \cite{oberlack2000Compton} that does not identify the escape events.

\begin{figure}[thbp]
\begin{center}
\includegraphics[width = 8.0 cm]{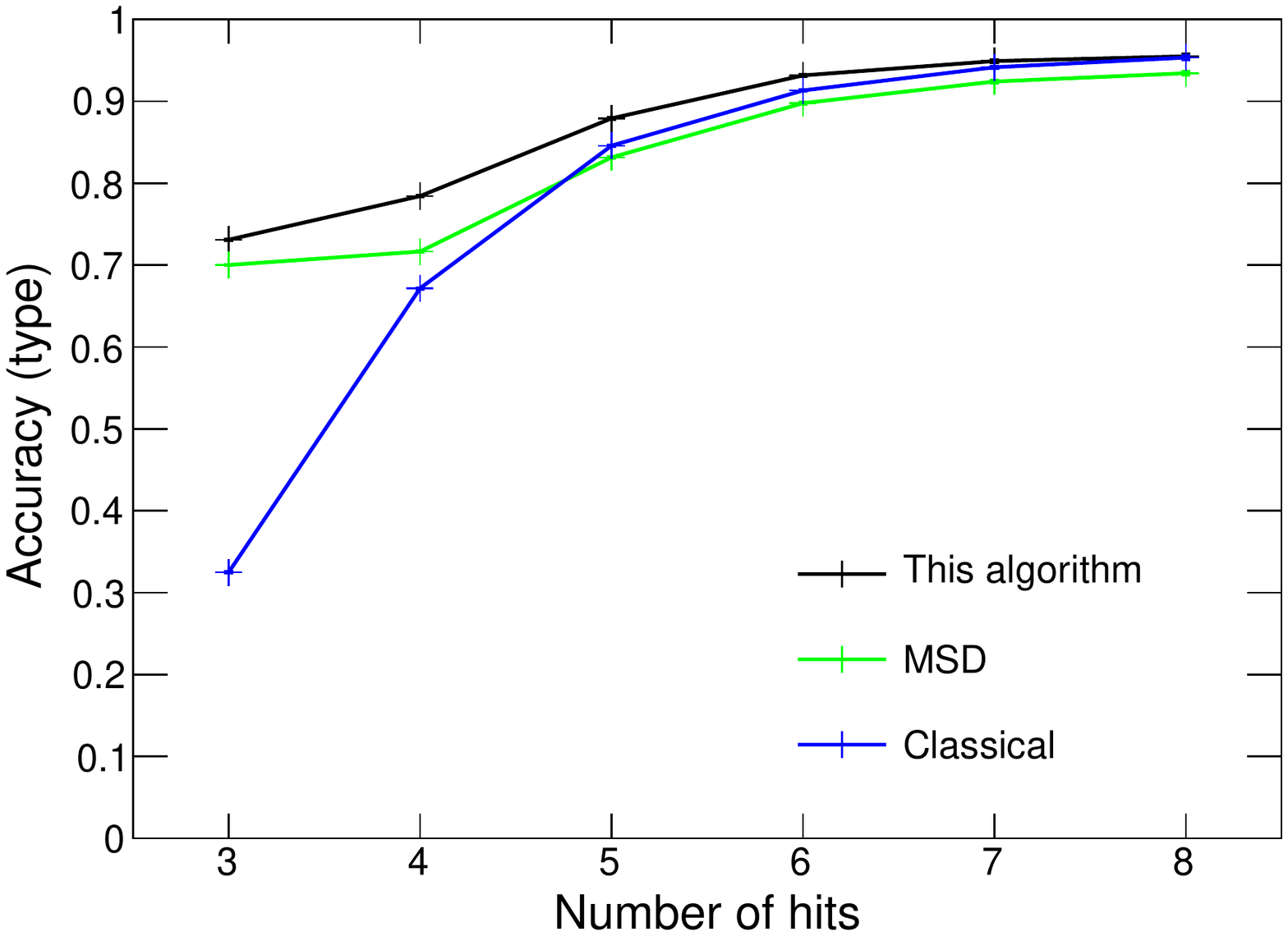}
\includegraphics[width = 8.0 cm]{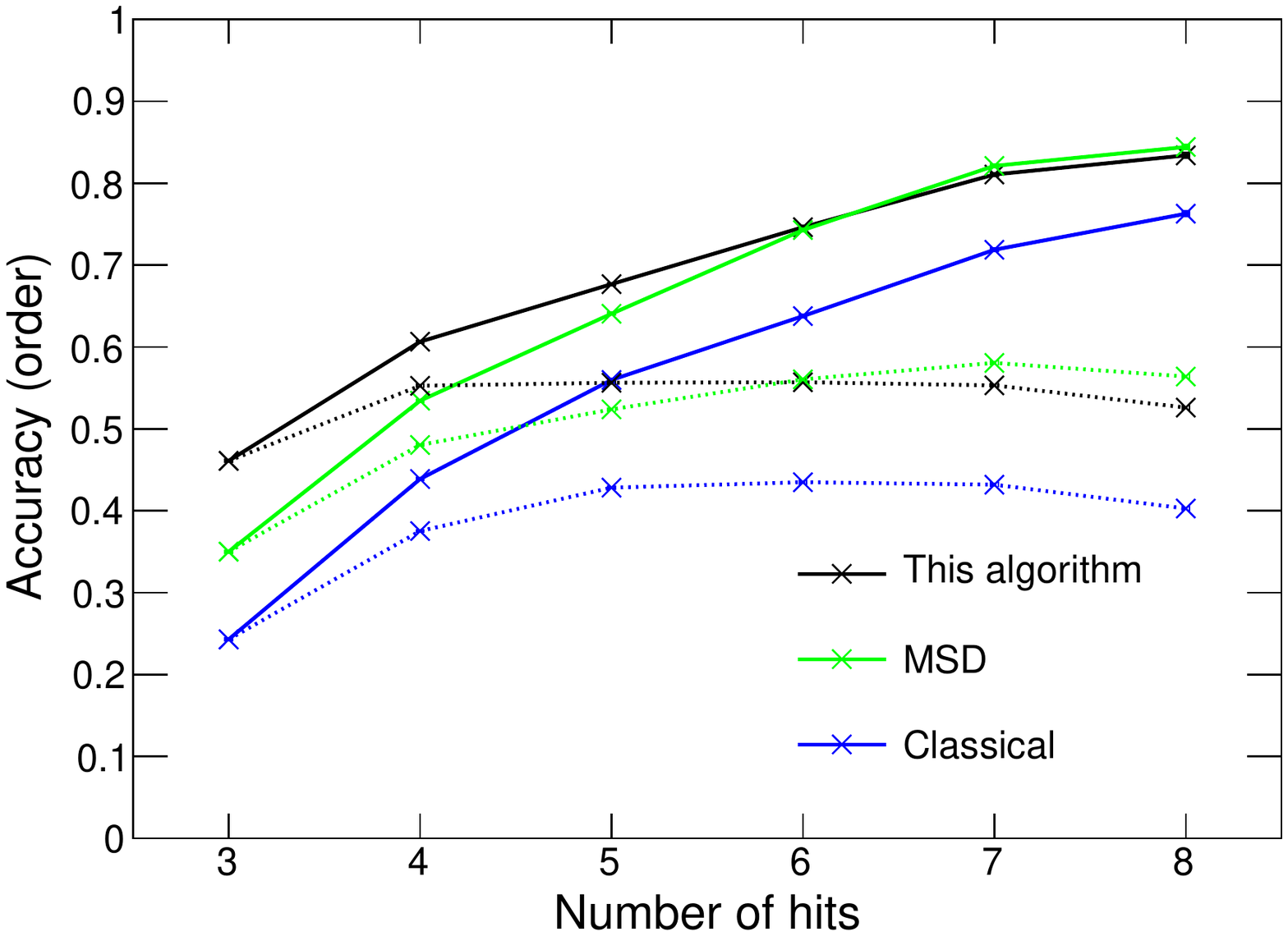}
\includegraphics[width = 8.0 cm]{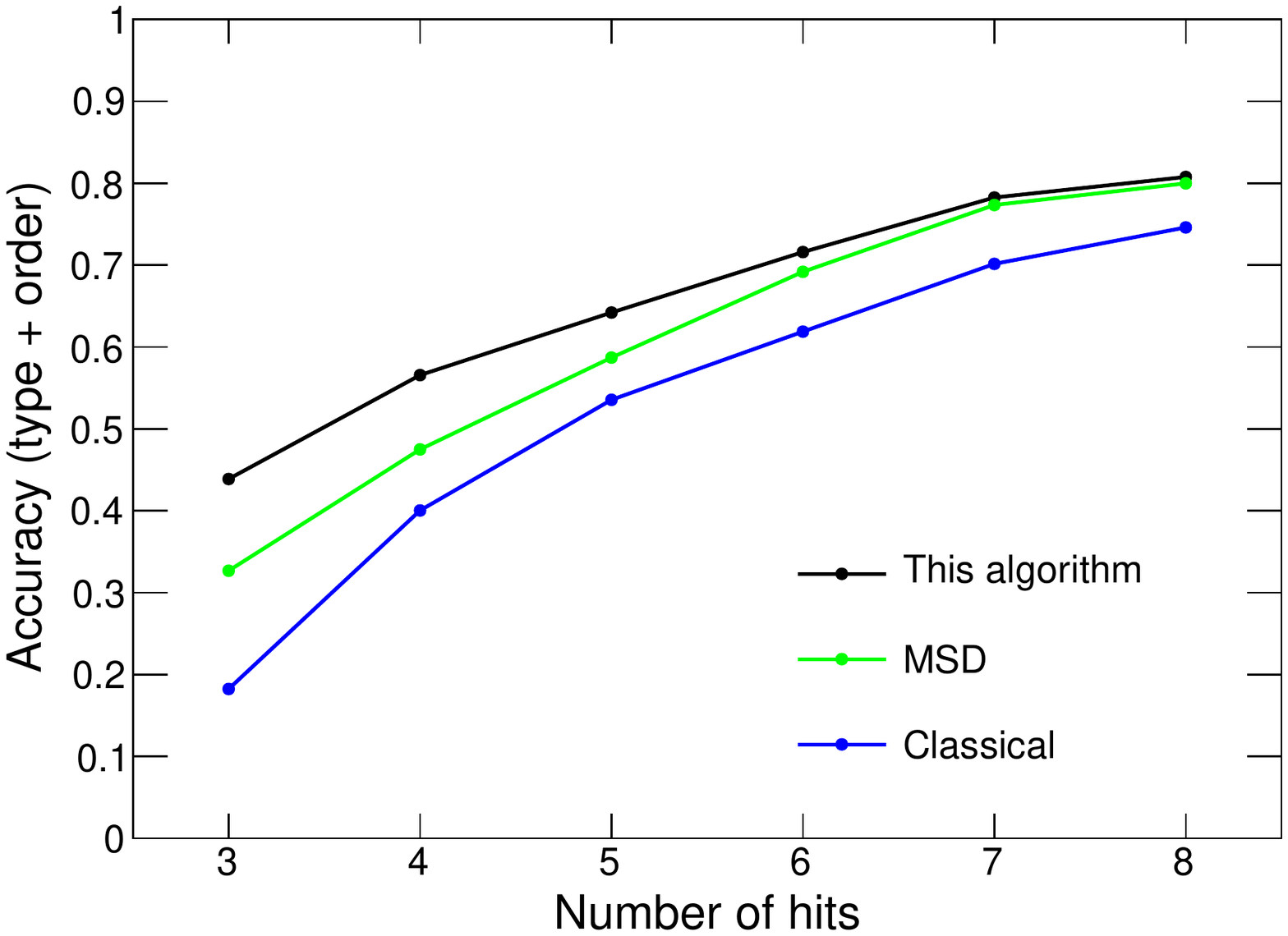}
\end{center}
\caption{Comparison of the performance with those of other algorithms. The black lines are the same as Figure~\ref{fig_accuray_at_1MeV}. The green and blue lines represent the modified TANGO algorithm \cite{TANGO2010} and the classical method \cite{oberlack2000Compton}.
The middle panel shows the accuracy of the scatter order of the first two interactions (solid), and all interactions (dotted).
The bottom one shows the accuracy for both event type and scattering order when focusing on the scattering order of the first two interactions for fully-absorbed events and the first three for escape events.}
\label{comp_TANGO_Oberlack}
\end{figure}

Figure~\ref{comp_TANGO_Oberlack} shows the accuracy of the reconstruction by our algorithm (this work), the modified TANGO algorithm (``MSD'' in the figure), and the classical algorithm when applying them to the 1 MeV gamma-ray simulation data.
The black points for the proposed algorithm are the same as Figure~\ref{fig_accuray_at_1MeV}.
This algorithm outperforms the others in the accuracy for the event type.
A plausible reason for this improvement is that the ratio of the figure-of-merits of the escapes to fully-absorbed events is calculated more accurately by formulating the probability model. 
Note that here the classical method reconstructs any event as a fully-absorbed event, and the accuracy for the event type of the classical method is equal to the ratio of truly fully-absorbed events to the total.

As for the accuracy for the scattering order,
the TANGO algorithm outperforms very slightly above 6 hits
because when events are dominated by fully-absorbed events the MSD algorithm tends to reconstruct small-scattering events with a slightly better accuracy \cite{Xu4pi}.
However, below 6 hits, our method has a better accuracy with a maximum improvement of about 11\% at 3-hit events.
When only considering the first two interactions for fully-absorbed events and the first three for escape events, our algorithm shows the best accuracy for both event type and scattering order (see the bottom in the figure).

\begin{figure}[thbp]
\begin{center}
\includegraphics[width = 8.0 cm]{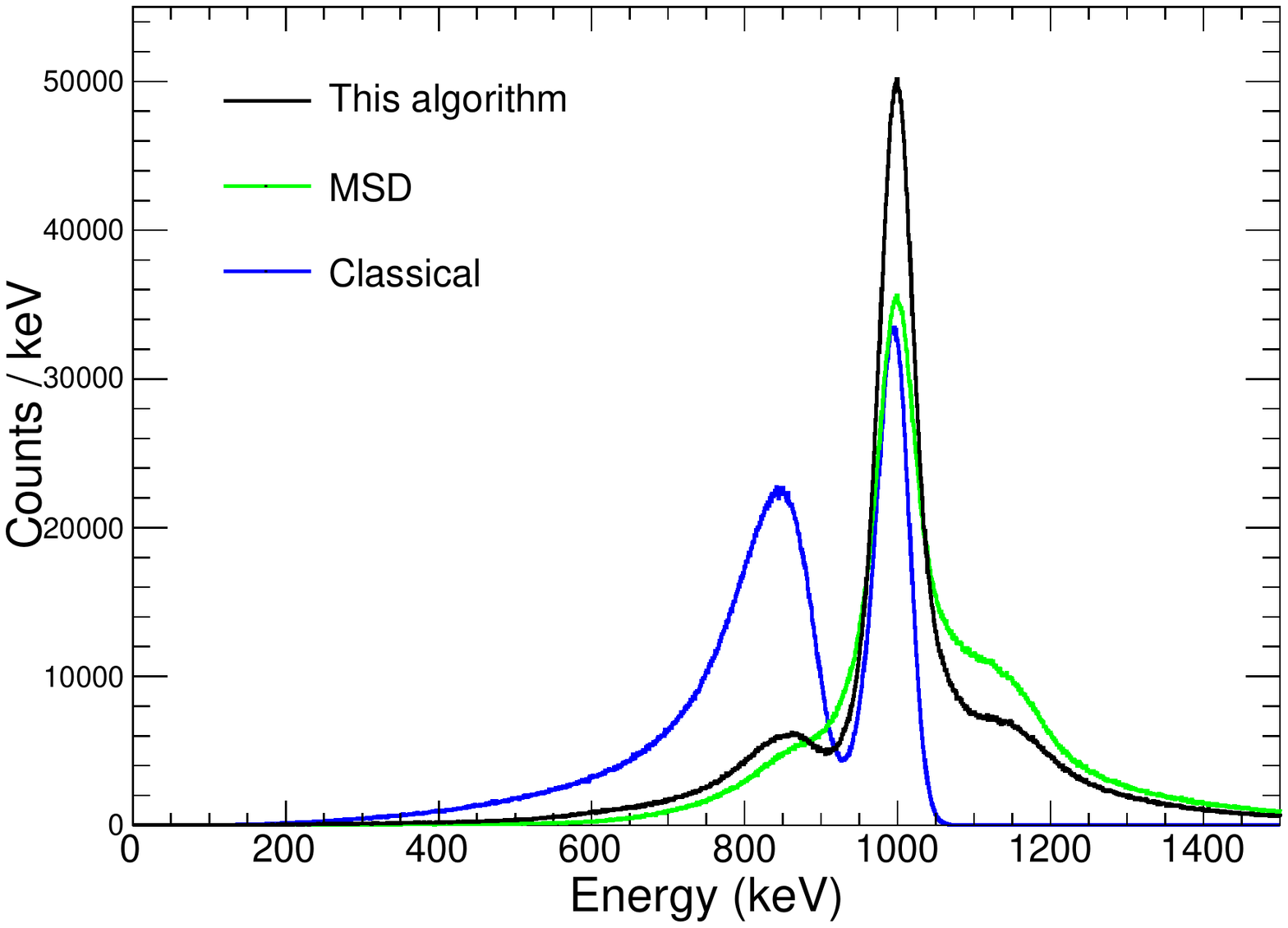}
\includegraphics[width = 8.0 cm]{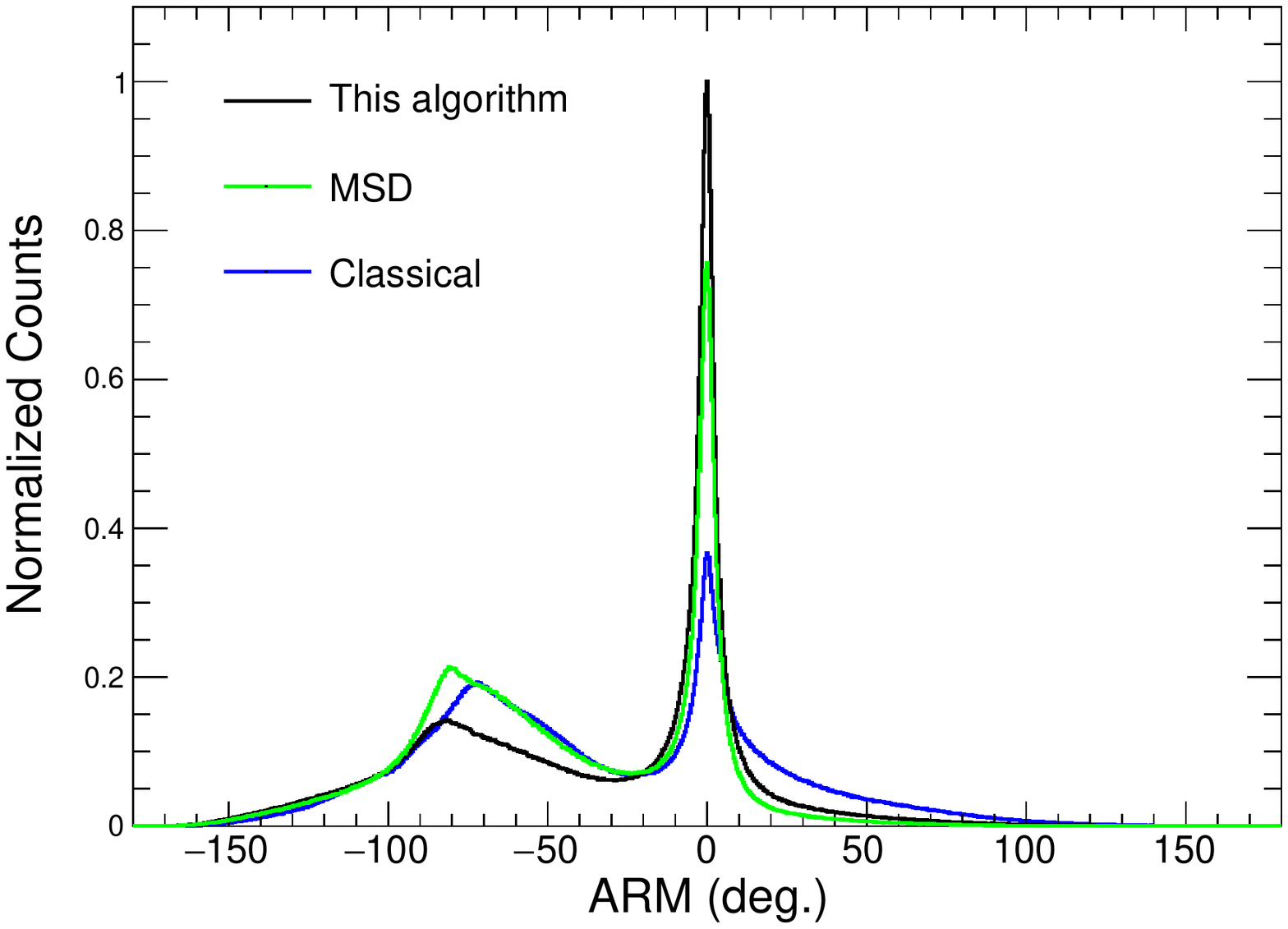}
\end{center}
\caption{The reconstructed energy spectra and ARM distributions of 3-hit events, obtained by this algorithm (black), the TANGO algorithm (green) and the classical method (blue). Here both fully-absorbed and escape events are used.}
\label{comp_TANGO_Oberlack_spectrum}
\end{figure}

This advantage is clear particularly for small-number-hit events, namely, 3-hit events.
Figures~\ref{comp_TANGO_Oberlack_spectrum} shows the reconstructed energy spectra and ARM distributions of 3-hit events.
The difference to the modified TANGO algorithm as well as the classical one is noticeable in the energy spectrum, where the peak height differs by about 40\%.
This improved performance is also observed in the ARM distribution as the sharpest peak around $0^\circ$.

\subsection{Performance in different gamma-ray energies}
\begin{figure}[thbp]
\begin{center}
\includegraphics[width = 8.0 cm]{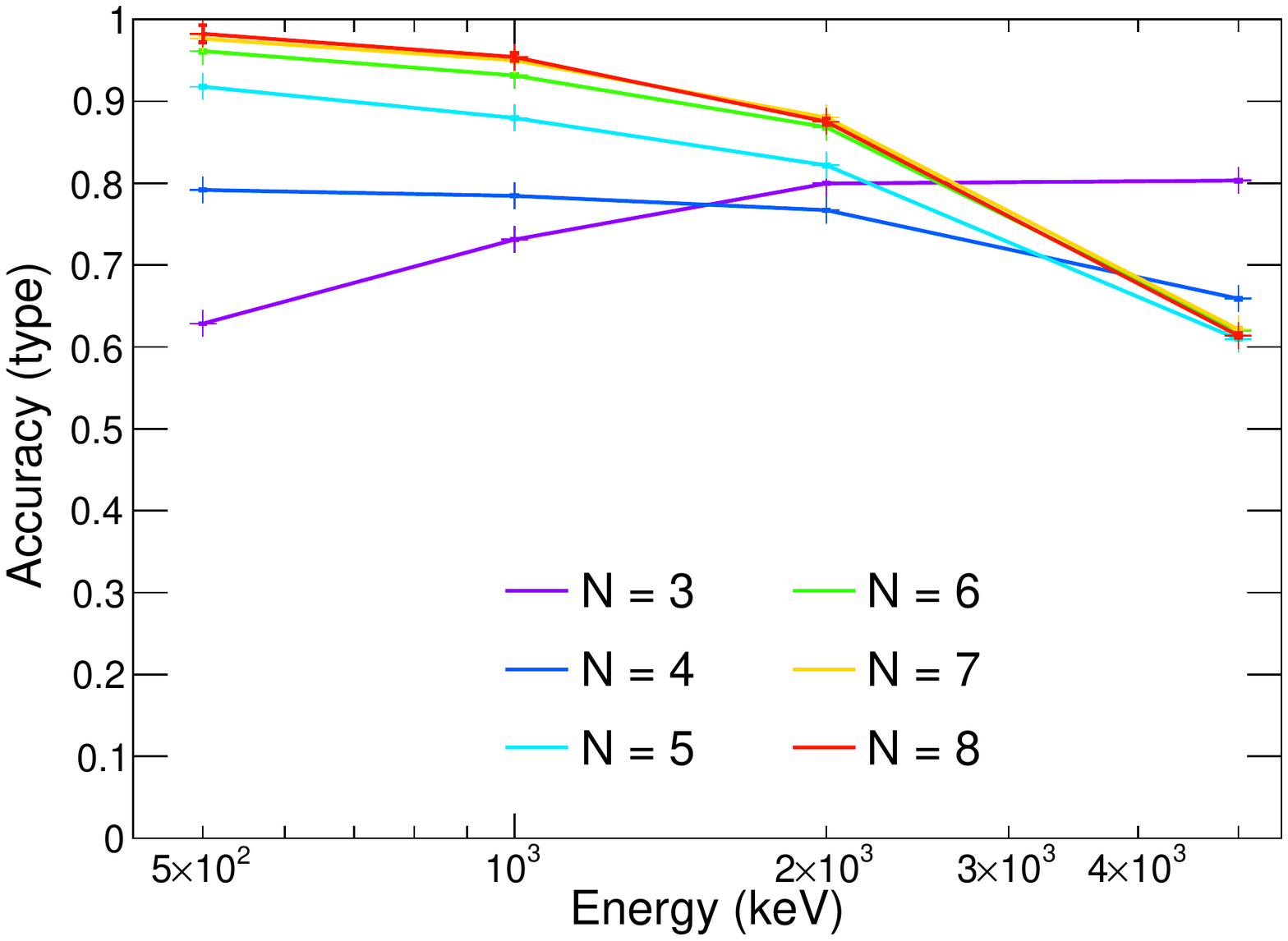}
\includegraphics[width = 8.0 cm]{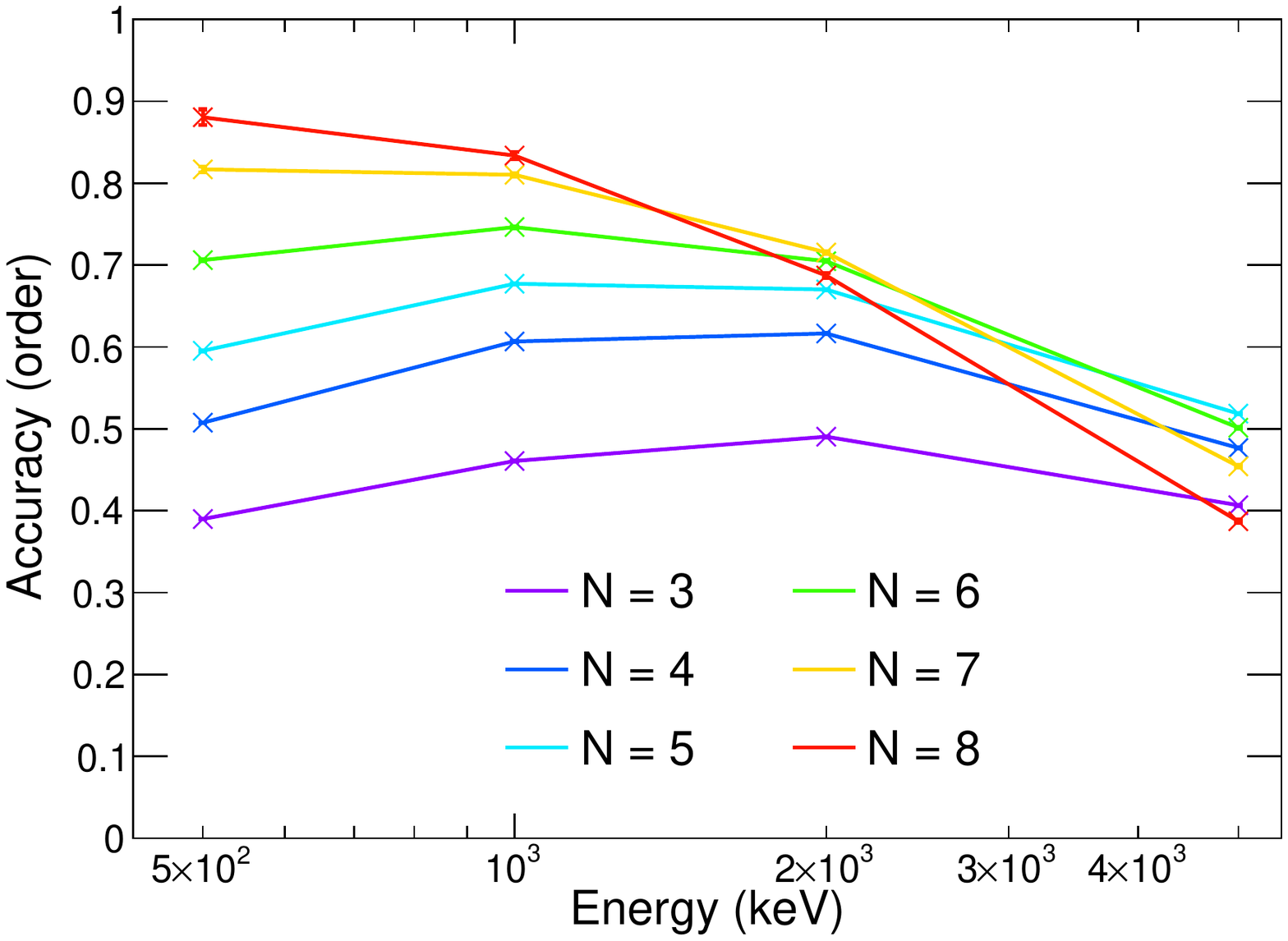}
\includegraphics[width = 8.0 cm]{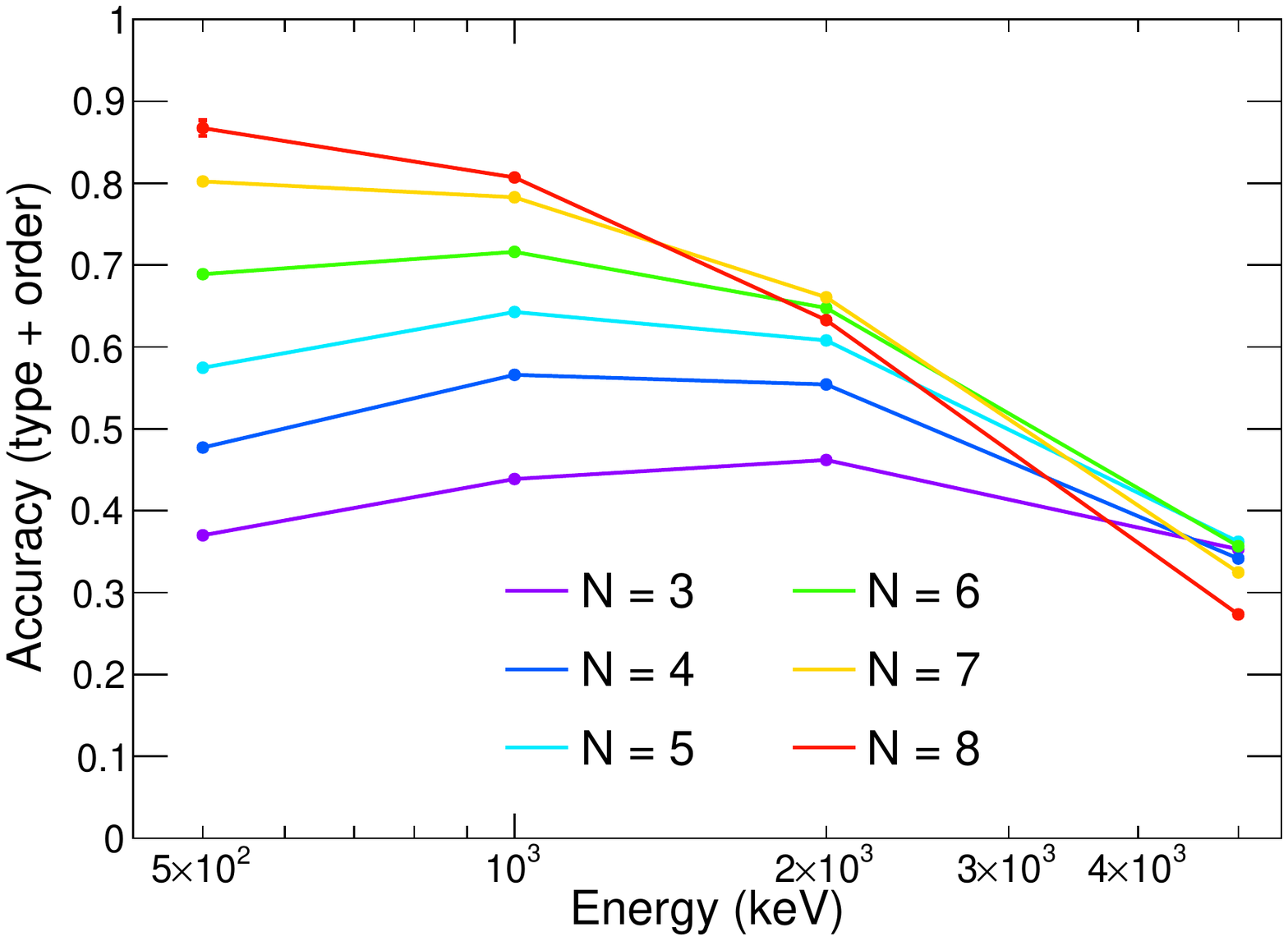}
\end{center}
\caption{The reconstruction accuracy against gamma rays from 0.5 to 5 MeV. The different colors represent the number of hits. The upper and middle panels show the accuracy for event type and for scattering order of the first two interactions, respectively.
The bottom panel shows the accuracy for both event type and scattering order when focusing on the scattering order of the first two interactions for fully-absorbed events and the first three for escape events.}
\label{fig_accuracy_wide_energy}
\end{figure}

To further investigate the proposed algorithm's performance for different incoming gamma-ray energies,
we also calculate the reconstruction accuracy of gamma rays from 500 keV to 5 MeV.
The results are shown in Figure~\ref{fig_accuracy_wide_energy}.
The accuracy gets the maximum at 1--2 MeV though it depends on the number of hits.
A factor limiting the performance at 5 MeV would be recoil electrons with energies above few MeV, which produce gamma rays via bremsstrahlung.
These gamma rays are not considered or filtered out in the algorithm.
Note that in this analysis we removed events that include pair creation to investigate the performance against truly Compton scattering events.
In the argon detector, at $\sim 12$ MeV the cross section of pair creation becomes comparable to Compton scattering.
The fraction of events that include pair creation is 2.3\% at 2 MeV and 21\% at 5 MeV.
In reality, the accuracy of removing such events also affects the performance.
We will discuss it in Section~\ref{sec_discussion}.

\section{Discussion and Conclusion}
\label{sec_discussion}

In this work, we have formulated probability functions of physical processes and measurements in Compton telescopes
and developed the reconstruction algorithm for the multiple Compton scattering events based on the probabilistic method.
This algorithm can treat both fully-absorbed and escape events in a unified framework.
The developed algorithm can be used in compact or large-volume Compton telescopes aiming at a few MeV.

A promising application is the GRAMS project which uses a large single-material detector with liquid argon.
We verified its performance using simulated data sets of a $140\times140\times20$ cm$^3$ liquid argon detector
and confirmed that the algorithm works very well for up to 8-hit events for 1 MeV gamma rays.
The accuracy for the event type is more than 73\%, with a maximum of 95\% at 8-hit events.
While the scattering order of the entire sequence is reconstructed with an accuracy of about 55\% at the maximum,
the accuracy gets much higher, e.g., $\sim$80\% for 8-hit events, when focusing on the essential part for the reconstruction, namely, the scattering order of the first two or three interactions.
For the application to the GRAMS project, we consider using the events from 3 to 8 hits to achieve a large effective area which is one of its advantages.
However, towards practical applications, we need further studies when considering the background, event selection, and some processes ignored in the proposed algorithm. 
In the following, we discuss several aspects for extending and improving the algorithm.

\subsection{The background reduction and measurement of gamma rays with known energy}
\label{sec_discussion_bkg_known_energy}

In astronomical observations, MeV gamma-ray signals are usually dominated by various background events, e.g., charged particles and neutrons of cosmic rays, radioactivation, albedo gamma-rays, and galactic/extra-galactic diffuse background.
Eliminating the background is critical for the achievement of high sensitivity.
For example, in the GRAMS project, the neutron background is removed by pulse shape discrimination, and surrounding plastic scintillators identify the charged particle events.
Towards the high sensitivity,
future works are needed to study how significantly the background contaminates signal events and whether the background events can be distinguished in some way based on background-included simulations.

\begin{figure}[!htbp]
\begin{center}
\includegraphics[width = 7.5 cm]{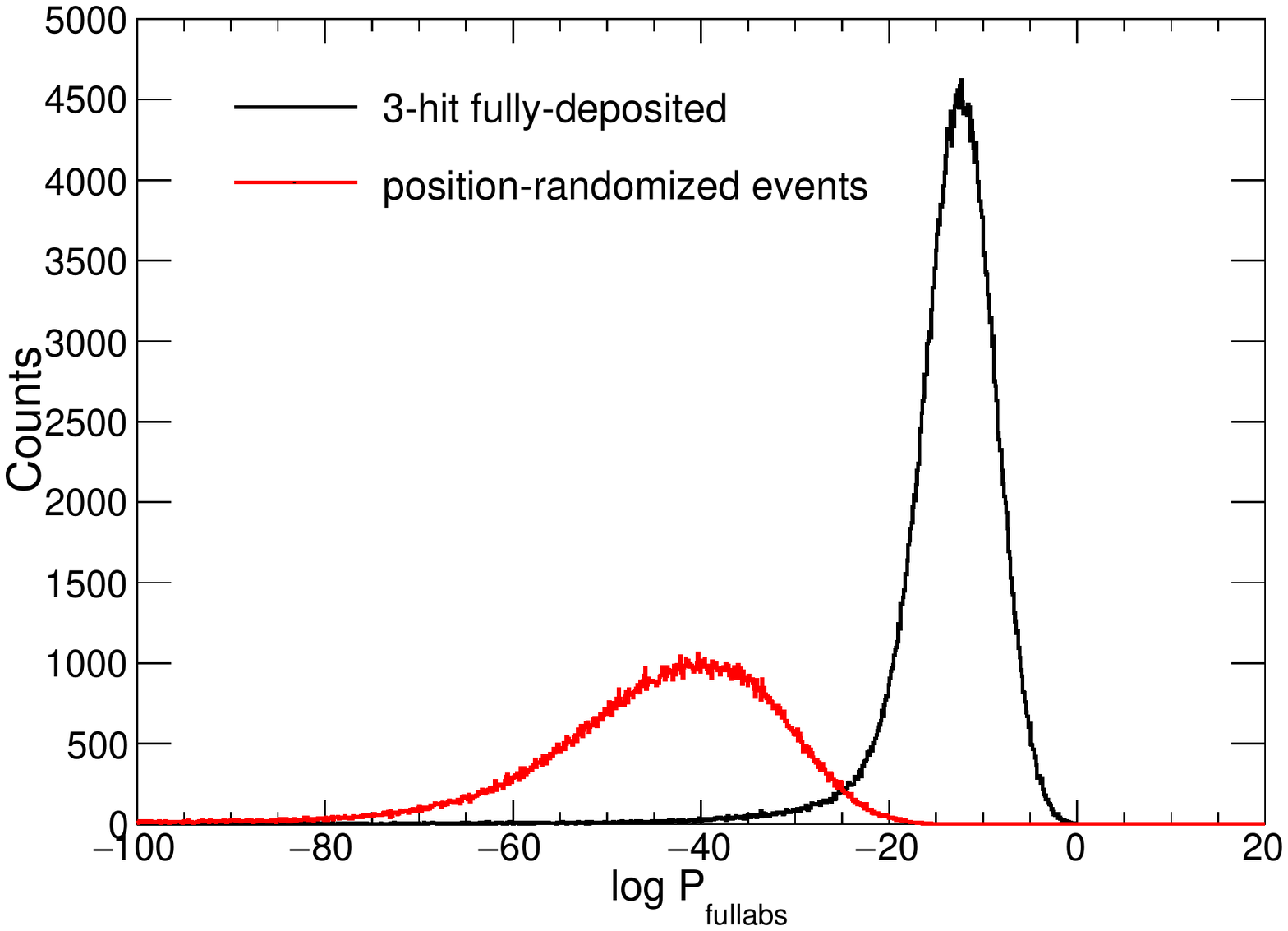}
\end{center}
\caption{The distribution of the conditional probability for 3-hit fully-absorbed events (black) and the position-randomized events (red). 
The incoming gamma-ray energy is 1 MeV.}
\label{fig_likelihood_distribution}
\end{figure}

As a possible idea, the conditional probability obtained from the algorithm ($\mathcal{P}_\mathrm{fullabs}$ or $\mathcal{P}_\mathrm{escape}$) could be used for background reduction.
The black line in Figure~\ref{fig_likelihood_distribution} shows the distribution of the conditional probability of 3-hit fully-absorbed events using the 1 MeV gamma-ray simulation data set in Section~\ref{sec_event_classification}. 
To generate events that do not originate from Compton scattering of gamma rays by a simple way,
we use the same data set, but for each hit in all events we re-sampled a position from a uniform distribution in the detector.
Then, we applied the reconstruction algorithm to these position-randomized events.
These events can be interpreted as a very simplified model of the background gamma rays, e.g., produced from nuclear interactions with cosmic rays \cite{SCHONFELDER2004Lessons}.
The red line in the figure corresponds to the distribution of the obtained conditional probability of the position-randomized events.
They have much smaller values than the gamma-ray events.
This result indicates that some background events can be removed by setting a threshold in the obtained value,
though a more realistic simulation is needed to validate this application.

\begin{figure}[!htbp]
\begin{center}
\includegraphics[width = 7.5 cm]{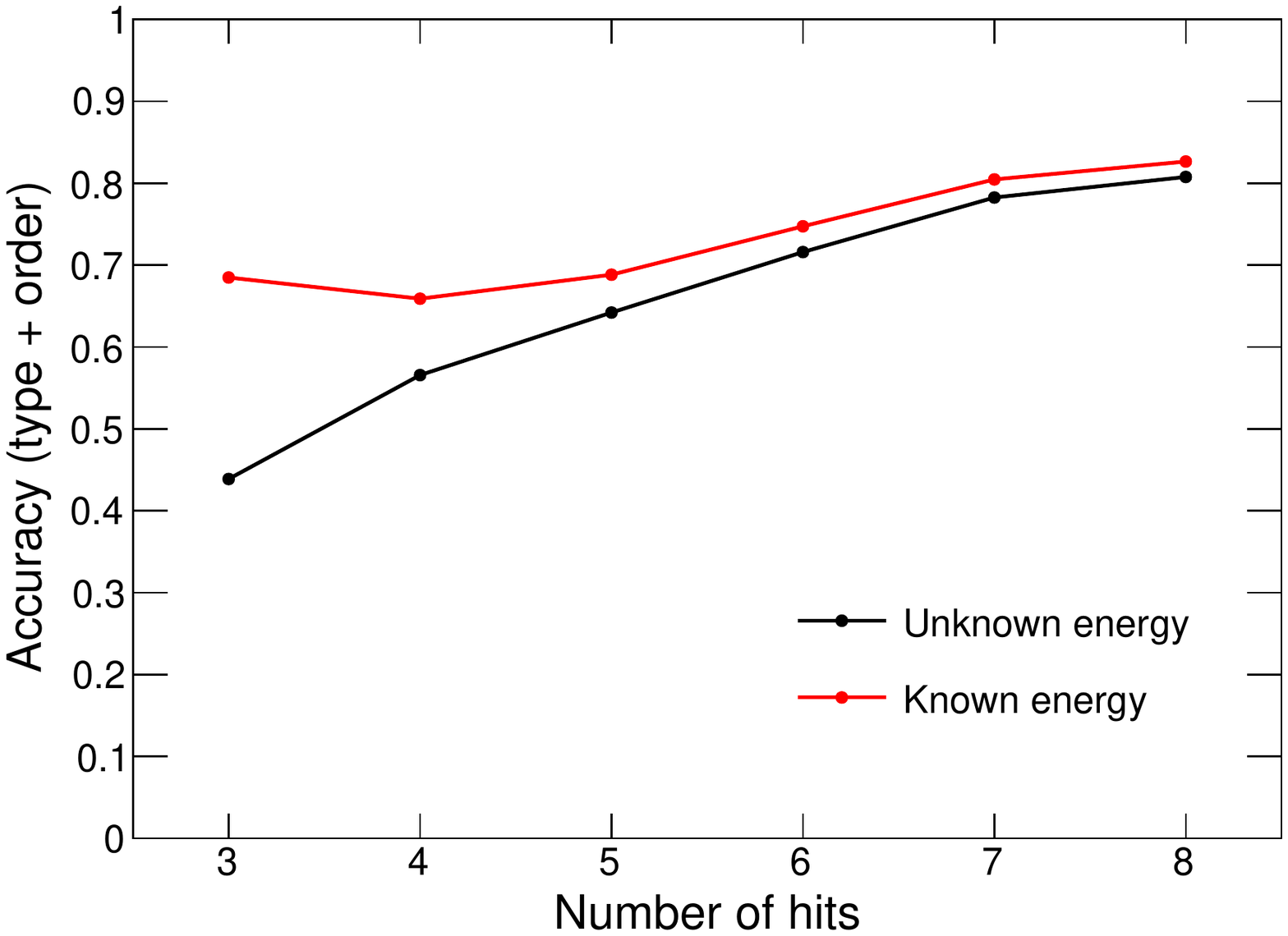}
\end{center}
\caption{The accuracy for both type and order, when the incident gamma-ray energy is known a priori (red). 
The scattering order of first two and three interactions are considered for fully-absorbed and escape events, respectively.
The black lines are the same as Figure~\ref{fig_accuray_at_1MeV}. The incoming gamma rays are 1 MeV.}
\label{fig_accuracy_at_known_energy}
\end{figure}

This work mainly focuses on astronomical application, but our algorithm could be used in other fields, e.g., imaging in nuclear medicine therapy or monitoring high-intensity radiation fields \cite{Takahashi2012,tomono2017First,yabu2021Tomographic}.
In these cases, incoming gamma-ray energy is often known a priori.
The reconstruction algorithm can be modified
by fixing the incoming energy instead of estimating it by Eq.~\ref{eq_e_ini_full_dep} or \ref{eq_esc_energy}.
This additional information can improve the reconstruction performance.
Figure~\ref{fig_accuracy_at_known_energy} shows the accuracy of 1 MeV gamma-ray event when fixing the incoming energy.
The reconstruction accuracy is improved, especially for events with a small number of hits by $\sim 30$\%.

\subsection{Possible extension of the algorithm for electron-tracking Compton telescopes}
\label{sec_discussion_etcc}
Though this paper focuses on the case that only the deposited energies and positions are measured,
our approach could be extensible even when a Compton telescope can also measure the trajectories of recoiled electrons and estimate their initial momentum directions \cite{tanimori2004MeV,vetter2011First,yoneda2018Development}.
In this case,
$\vector{D}_I$ is needed to be redefined as
\begin{align}
\vector{D}_I = \left( \vector{r}_{I}, \varepsilon_{I}, \vector{p}_{\mathrm{e},I} \right)~,
\end{align}
where $\vector{p}_{\mathrm{e},I}$ is the measured momentum direction of a recoiled electron.
Then, we modify the detector response function 
$P_\mathrm{det} \left( \vector{D}_{I} \mid \hat{\vector{q}}_{i-1}, \hat{\vector{q}}_{i} \right)$ by
including the electron trajectory information.
Here we introduce a function $P_{\mathrm{track}} (\cdot)$ that compares the expected direction of a recoiled electron 
$\left(\hat{\vector{p}}_{i} - \hat{\vector{p}}_{i-1}\right)$
and measured one $\left(\vector{p}_{\mathrm{e},I}\right)$.
The definition of $P_{\mathrm{track}} (\cdot)$ would depend on the configuration of Compton telescopes because the obtained electron images vary by the detectors, 
and in general $P_\mathrm{det}(\cdot)$ is reformulated as
\begin{align}
\label{eq_det_etrack}
\begin{aligned}
P_\mathrm{det} \left( \vector{D}_{I} \mid \hat{\vector{q}}_{i-1}, \hat{\vector{q}}_{i}\right) = & P_\mathrm{ene}( \varepsilon_I \mid \hat{\varepsilon}_{i})
\times P_\mathrm{pos}( \vector{r}_I \mid \hat{\vector{r}}_i) \\
& \times P_\mathrm{track}( \vector{p}_{\mathrm{e},I} \mid \hat{\vector{p}}_{i} - \hat{\vector{p}}_{i-1})~.
\end{aligned}
\end{align}

After modifying $P_\mathrm{det}(\cdot)$,
the scattering order can be determined in the same way as in Section~\ref{sec_algorithm} but with one exception.
In this case,
even if the incoming gamma-ray direction is constrained on a Compton circle,
the conditional probability depends on location in the circle 
because the direction of electron recoil varies depending on it.
Thus, with this additional information, one can constrain the incoming gamma-ray direction as a much narrower region, as the merit of electron-tracking Compton telescopes.

\subsection{For further improvement}
The developed algorithm successfully reconstructs multiple Compton scattering events with high accuracy. However, we should note that we ignored several important factors in the probabilistic model, e.g., interaction with passive materials, Doppler broadening effect, and bremsstrahlung. In the rest of this paper, we discuss them and a possibility for further improvement of this algorithm.

\subsubsection{Effect of passive materials on the algorithm performance}
\label{sec_passive_material}
The probabilistic model proposed in \S\ref{sec_def_likelihood_function} assumes that incoming gamma rays always interact with a sensitive part of the detector,
and interaction with passive materials is not considered.
However, the sensitive volume is surrounded by passive materials inevitably in a realistic detector configuration.
For example, the GRAMS project's liquid argon time projection chamber needs a cryostat.
Insensitive regions of liquid argon, photomultiplier tubes or Si-PMs (scintillation light detectors), pixels or wires with readout electronics for ionized electron detection can also act as passive materials.
These components may deteriorate the performance of the detector and hence that of the reconstruction algorithm since the energy deposits in them cannot be detected.
Part of this deterioration may be recovered by considering the interactions with passive materials in the algorithm.
Since the design of the GRAMS detector is still under discussion, 
here we summarize what kinds of scattering patterns occur with passive materials and discuss possible ways to consider them.

\begin{figure*}[thbp]
\begin{center}
\includegraphics[width = 13.5 cm]{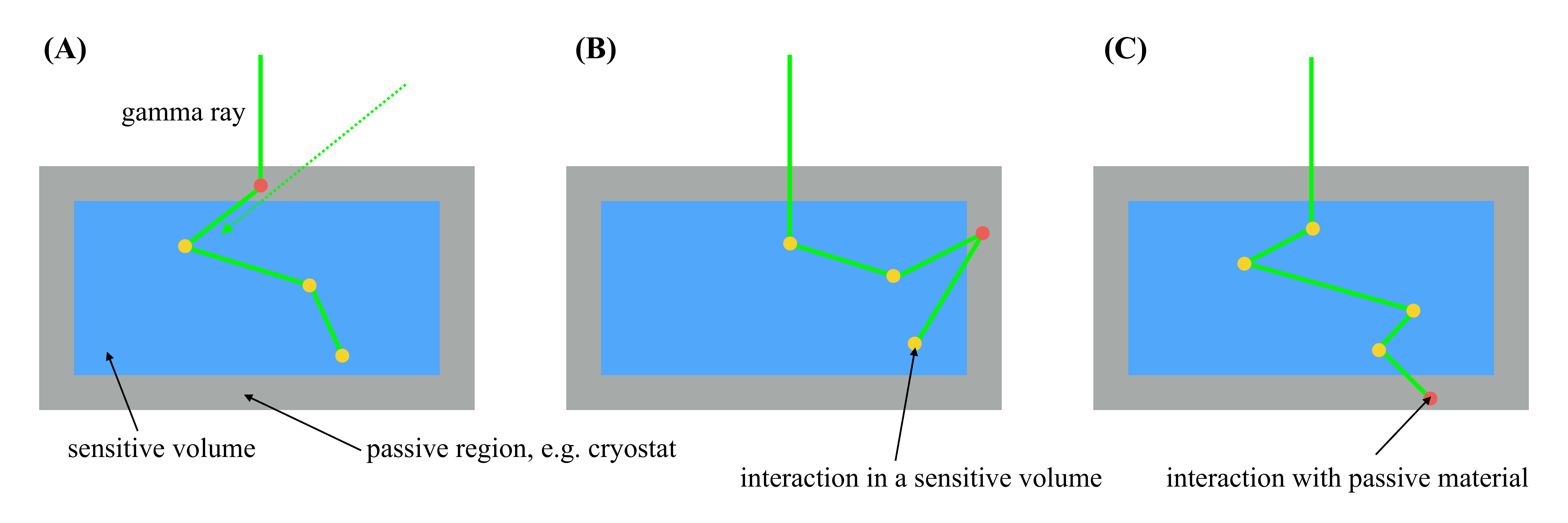}
\end{center}
\caption{Scattering patterns ignored in the proposed algorithm when passive materials are considered.
}
\label{fig_passive}
\end{figure*}

Figure~\ref{fig_passive} shows examples of the scattering patterns ignored in the probabilistic model when passive materials are considered.
In the pattern (A), an incoming gamma ray is scattered in a passive region as the first interaction. In this case, the detected hits are indistinguishable from an event with a gamma-ray incoming from the dotted arrow in Figure~\ref{fig_passive}.
Thus, it is difficult to identify and reconstruct an incident gamma ray by the algorithm, and this pattern should be treated as a background. In order to reduce these events, thin and low-density material is preferable for the top plate of the cryostat in the GRAMS detector.
Hereafter once the first interaction occurs in the passive region, we label the event as (A) no matter whether it is scattered in the passive region after that.

The pattern (B) in Figure~\ref{fig_passive} shows another example that a gamma ray is scattered in a sensitive region at first but suffers from scattering by passive materials as an intermediate interaction. There are two ways to handle such an event. One possibility is to use the fact that these events may yield a conditional probability smaller than events with interactions occurring only in a sensitive volume. Although the current implementation considers these events as fully-absorbed or escape events assuming no scattering in passive materials, the scattering angles calculated redundantly from kinematics and position information are hard to be matched due to the loss of information about the scattering in the passive region. It would result in that the conditional probability calculated by the algorithm becomes smaller. If so, these events could be distinguished and rejected by applying a threshold for the conditional probability. The other way is to include the scattering process with passive materials in the probabilistic model. While it may require complicated mathematical calculations, in principle, it is possible to include the scattering process in the model by marginalizing the scattering position and angle, similar to the marginalization introduced for the escape events in Eq.~\ref{eq_last_ene_esc}.

Finally, the pattern (C) in Figure~\ref{fig_passive} is a gamma-ray that is scattered in a sensitive region except for the last interaction. No matter whether the last interaction in the passive region is Compton scattering or photoabsorption, this pattern can be effectively regarded as an escape event considering only interactions in a sensitive volume. Thus, when the proposed algorithm works correctly, such an event is identified as an escape event, and the undetected gamma-ray energy can be estimated by Eq.~\ref{eq_esc_energy}.

\begin{figure}[thbp]
\begin{center}
\includegraphics[width = 8.0 cm]{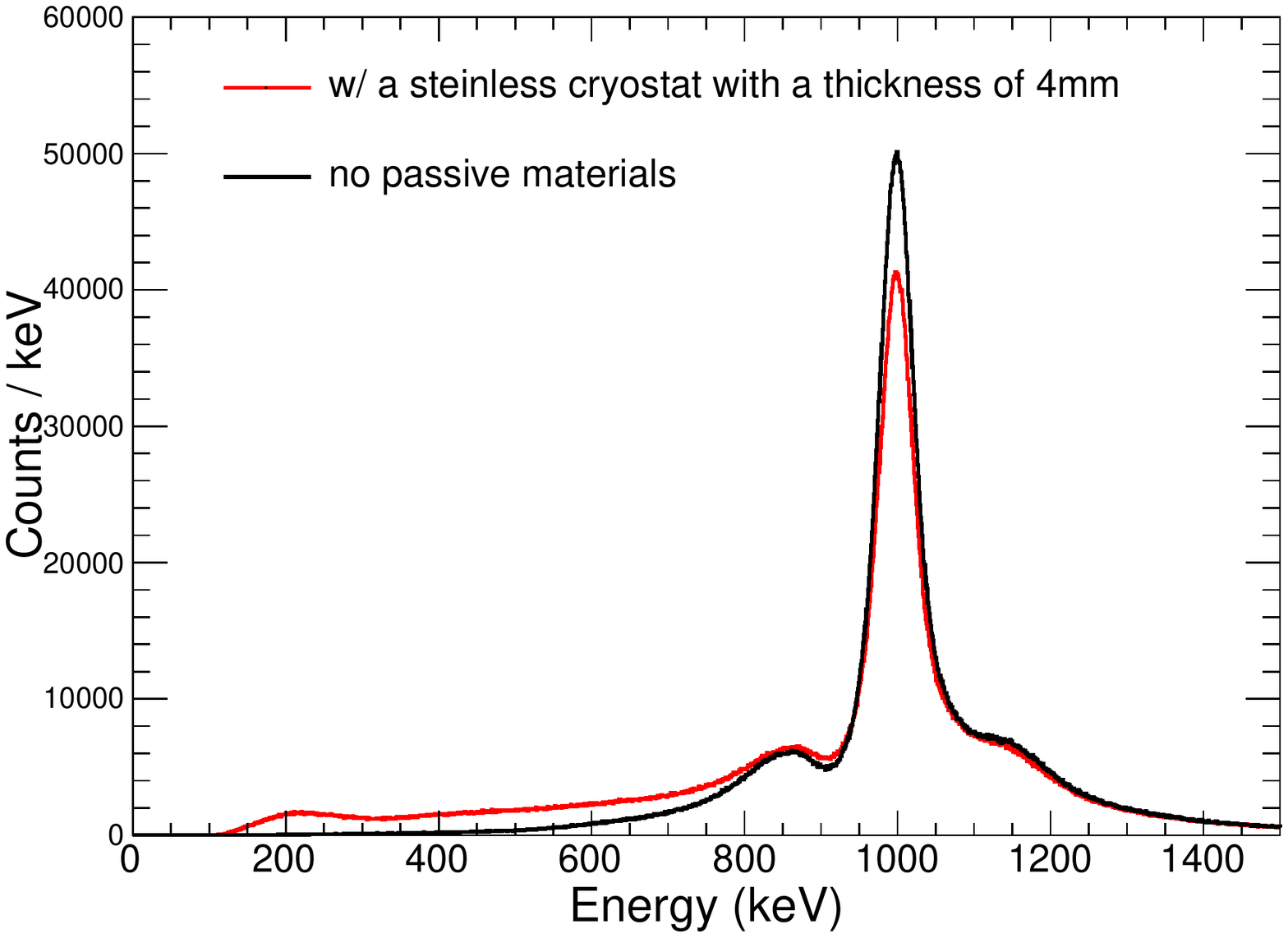}
\end{center}
\caption{The reconstructed energy spectrum of 3-hit events when including a stainless cryostat with a thickness of 4mm to the simulation in \S\ref{sec_event_classification}. Here both fully-absorbed and escape events are used.}
\label{fig_spectrum_passive}
\end{figure}

To consider these events more, we performed a simulation with 1 MeV gamma rays from the zenith the same as \S\ref{sec_event_classification}, but the sensitive volume is surrounded by a cryostat with a thickness of 4 mm. It consists of stainless steal (SUS304). 
Table~\ref{tab_ratio_passive_scattering} shows the ratio of the scattering patterns obtained from the simulation for different numbers of hits.
Only hits in the sensitive region are considered for calculating the number of hits, and the interaction in passive materials is not counted in it.
As the number of hits gets small, the ratio of events suffered from scattering in passive materials becomes large.
The dominant scattering patterns in this setup are (A) and (C).
The former can be reduced by using a thinner cryostat with other material, e.g., aluminum.
Also, we found that more than 80\% of the pattern (C) events are classified as escape events by the algorithm, and the energy deposit in the passive region is corrected.
Figure~\ref{fig_spectrum_passive} shows the reconstructed energy spectrum for 3-hit events comparing with the ideal simulation shown in \S\ref{sec_event_classification}.
We can see that the low energy tail appears mainly due to the scattering pattern (A).
Note that here we consider only the cryostat, but in reality, other passive materials exist as listed before. Thus, further considerations are essential, especially after the design of the GRAMS detector is finalized.

\begin{table*}[!htbp]
\caption{The ratio of the scattering pattern with passive materials. Here a cryostat with a thickness of 4 mm is considered.}
\label{tab_ratio_passive_scattering}
\centering
\begin{tabular}{ccccc}
the number of hits
& (A) & (B) & (C) 
& no scattering with passive materials \\ \hline
3 & 0.17 & 0.12 & 0.22 & 0.49 \\
4 & 0.15 & 0.10 & 0.13 & 0.62 \\
5 & 0.12 & 0.08 & 0.07 & 0.73 \\
6 & 0.11 & 0.06 & 0.04 & 0.79 \\
7 & 0.09 & 0.05 & 0.03 & 0.83 \\
8 & 0.09 & 0.05 & 0.02 & 0.84 \\ \hline
\end{tabular}
\end{table*}

\subsubsection{Doppler broadening effect and bremsstrahlung}
The proposed algorithm also ignores two important factor for physical processes.
One is the momentum distribution of electrons in the detector materials.
When the initial electron momentum is not zero,
the scattered gamma-ray energy differs from Eq.~\ref{eq_ene_scattering}.
This effect is known as the Doppler broadening effect \cite{zoglauer2003Doppler},
and limits the angular resolution.
Effectively, this effect could be considered by adding a function $\sigma_{\mathrm{Doppler}}$ in Eq.~\ref{eq_modification_of_energy_resolution},
which is an uncertainty in the energy determination at each interaction site due to the Doppler broadening effect.
Since it depends on the gamma-ray energy, the scattering angle, and detector materials,
the function $\sigma_{\mathrm{Doppler}}$ should be carefully modeled.
When the Doppler broadening effect is comparable to or dominates over the position and energy resolutions,
then such a modification would improve the reconstruction accuracy.
Note that this effect is not significant in the case of 1 MeV gamma-ray simulation shown in \S~\ref{sec_pv} \cite{zoglauer2003Doppler}.

The other factor is bremsstrahlung from scattered electrons in the detectors.
Bremsstrahlung becomes the dominant energy loss at high energies.
For example,
in the argon detector, it dominates over the ionization losses at a few MeV.
Gamma rays emitted by this process make additional signals, which makes the event reconstruction more complex.
Decreasing the accuracy above a few MeV is considered to be partially due to bremsstrahlung (see Figure~\ref{fig_accuracy_wide_energy}).
This process is not straightforward to treat because, unlike Compton scattering, it is challenging to identify the site where a bremsstrahlung photon is produced.
A possible way is to calculate the probability that a given signal is produced by a bremsstrahlung photon, considering 
energies and lengths to other interaction sites, and then merge it to corresponding signals if the calculated probability is large.

In a high energy band, typically above 5 MeV, it is also needed to distinguish pair creation events.
To consider these factors (the Doppler broadening, bremsstrahlung, and pair creation),
other statistical methods using large data sets might be effective, e.g., the deep neural network technique.
We expect that combining an analytical method like this work and simulation-based statistical methods can be a promising way to achieve even better reconstruction performance.
As a first step, we are also developing an event reconstruction algorithm using a multi-task neural network. 
In a subsequent paper \cite{Takashima2022}, we show that a neural network model improves the performance of the event reconstruction, especially for events with a small number of hits around 1 MeV.

\section*{Code availability}
The code for the reconstruction algorithm is available at \url{https://github.com/odakahirokazu/ComptonSoft}.

\section*{Acknowledgements}
We acknowledge support from JSPS KAKENHI grant numbers 18H05458, 19K14772, 20H00153, 20K14524, 20K20527, and 20K22355, and by RIKEN Incentive Research Projects, and by Toray Science and Technology Grant No.20-6104 (Toray Science Foundation). YI was supported by World Premier International Research Center Initiative (WPI), MEXT, Japan.

\appendix
\section{The effect of $L_0$ and $L_{\mathrm{esc}}$ on the algorithm performance}
\label{sec_effect_length_scale}

\begin{figure}[thbp]
\begin{center}
\includegraphics[width = 8.0 cm]{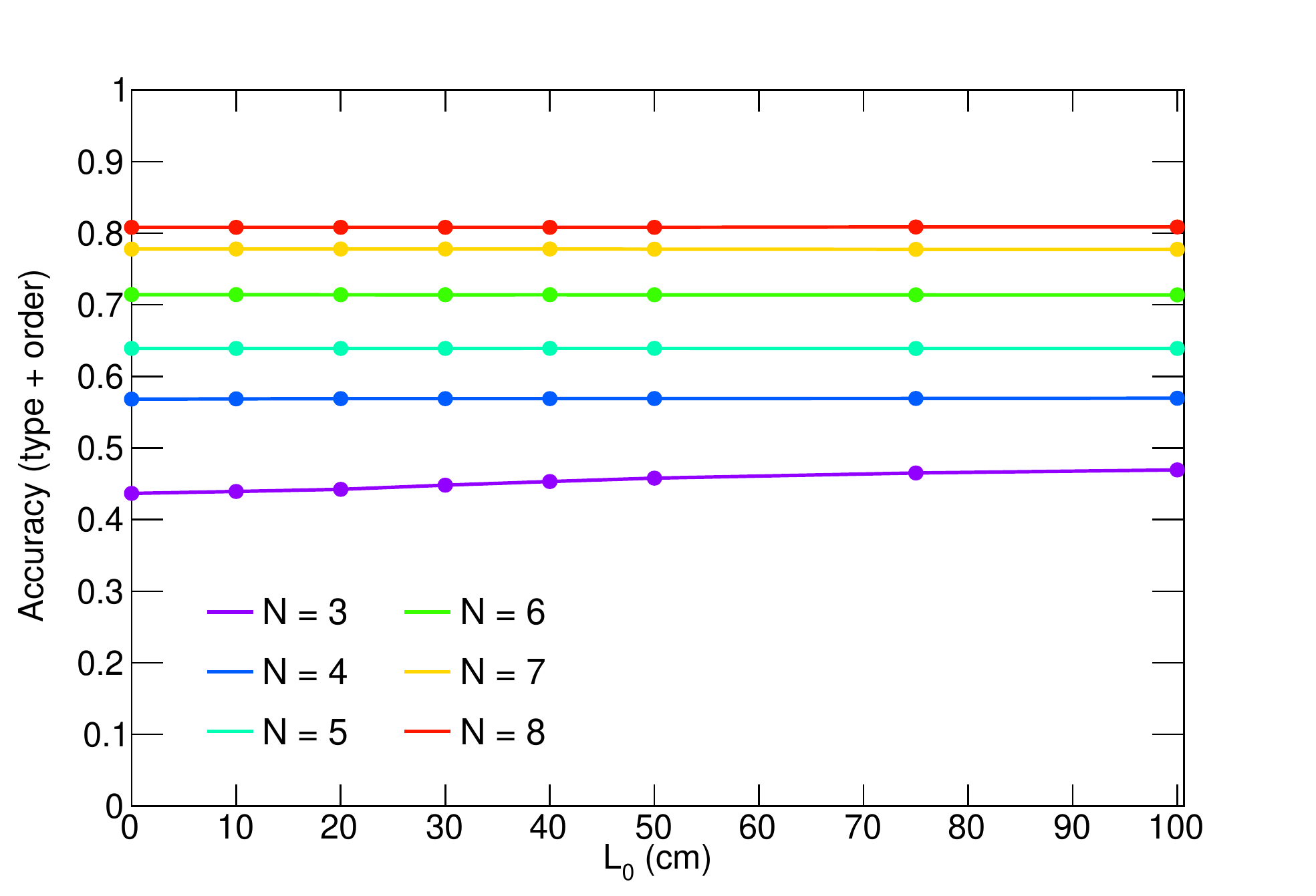}
\end{center}
\caption{The accuracy for both type and order, with different values of $L_0$.
The scattering order of first two and three interactions are considered for fully-absorbed and escape events, respectively.
}
\label{fig_L0_dep}
\end{figure}

\begin{figure}[thbp]
\begin{center}
\includegraphics[width = 8.0 cm]{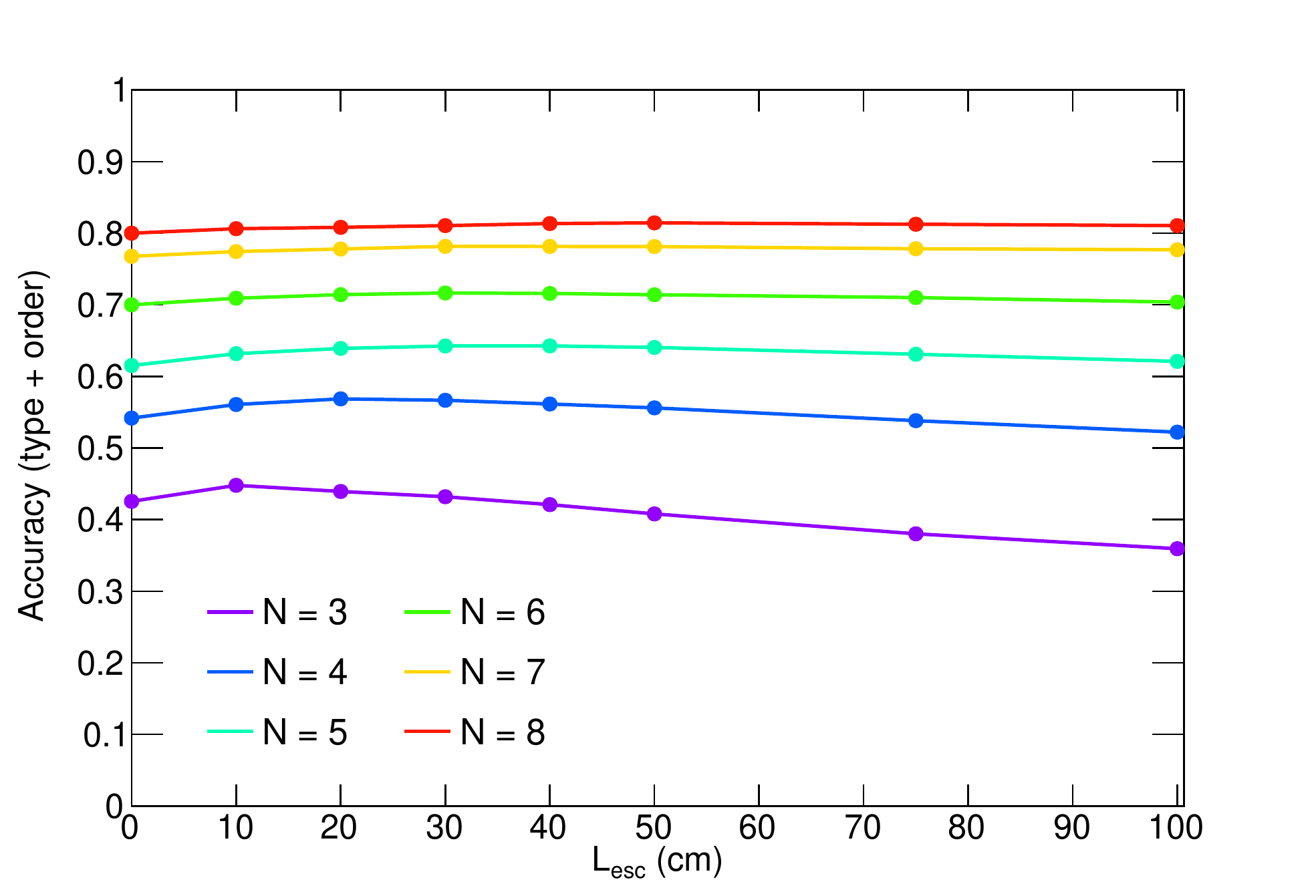}
\end{center}
\caption{The accuracy for both type and order, with different values of $L_{\mathrm{esc}}$.
The scattering order of first two and three interactions are considered for fully-absorbed and escape events, respectively.
}
\label{fig_Lesc_dep}
\end{figure}

Our algorithm assumes a constant value $L_0$ for the first interaction length, which is undetectable.
This approximation is very simple, and we checked how it affects the algorithm's performance.
Figure~\ref{fig_L0_dep} shows the accuracy for both type and order, with different values of $L_0$ from 0 cm to 100 cm.
The reconstruction accuracy varies by only 3\% at most.
We also checked the effect of $L_{\mathrm{esc}}$ on the performance.
Figure~\ref{fig_Lesc_dep} shows the accuracy for both type and order, with $L_\mathrm{esc}$ from 0 cm to 100 cm. 
In this case, the performance gets the best with $L_\mathrm{esc}$ of 10 cm for 3-hit events and 20 cm for 4-hits events, but as long as an extreme value like 0 or 100 cm is not used, the performance does not depend on $L_\mathrm{esc}$ so much.
Thus we conclude that $L_0$ and $L_{\mathrm{esc}}$ do not affect the performance so much, and this simple approximation is valid.
In \S\ref{sec_pv}, we set $L_0$ to 10 cm, a half of the detector thickness, and $L_{\mathrm{esc}}$ to 20 cm since it maximizes the performance at 4 hits.

\section{Performance with different energy/positional resolutions}
\label{sec_effect_energy_position_resolution}

\begin{figure}[thbp]
\begin{center}
\includegraphics[width = 8.0 cm]{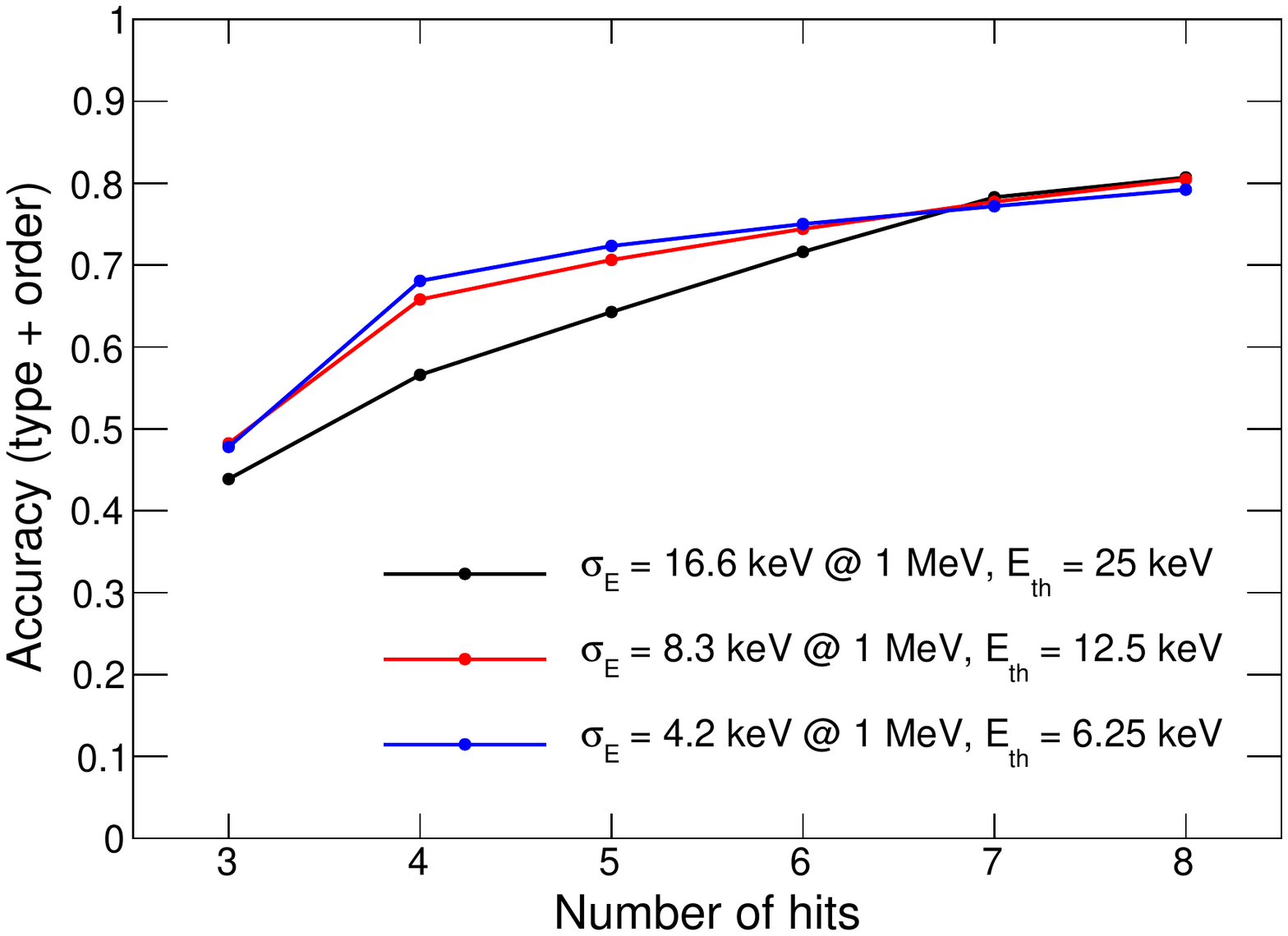}
\end{center}
\caption{The accuracy for both type and order, with different energy resolutions.
The scattering order of first two and three interactions are considered for fully-absorbed and escape events, respectively.
}
\label{fig_accuracy_with_different_energy_resol}
\end{figure}

\begin{figure}[thbp]
\begin{center}
\includegraphics[width = 8.0 cm]{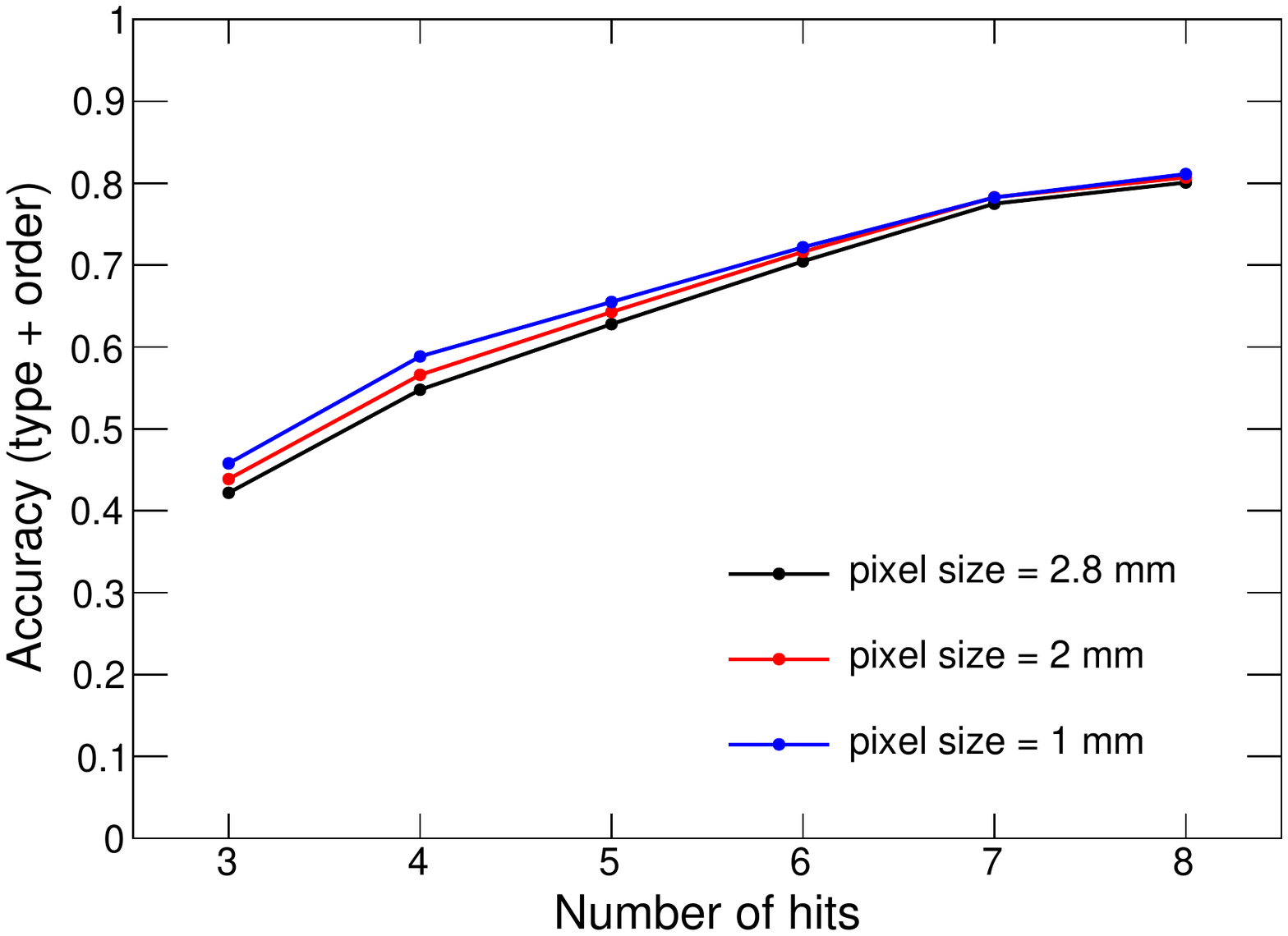}
\end{center}
\caption{The accuracy for both type and order, with different positional resolutions.
The scattering order of first two and three interactions are considered for fully-absorbed and escape events, respectively.
}
\label{fig_accuracy_with_different_position_resol}
\end{figure}

Here we examine how the reconstruction accuracy is affected by the assumed energy and positional resolutions.
In Figure~\ref{fig_accuracy_with_different_energy_resol}, we show the accuracy when the energy resolution is improved by a factor of 2 or 4.
In the latter case, the accuracy is increased by about 12\% at most.
Moreover, Figure~\ref{fig_accuracy_with_different_position_resol} shows the accuracy for different position resolutions.
Here we assume the pixel size of 1, 2 and 2.8mm. Respectively, $\sigma_{\hat{z}}$ is set to 0.5, 1.0, 1.4 mm.
In the best pixel resolution, the accuracy is improved by about 5\%.

\section{Performance with different viewing angles}
\label{sec_effect_viewing_angle}

\begin{figure}[thbp]
\begin{center}
\includegraphics[width = 8.0 cm]{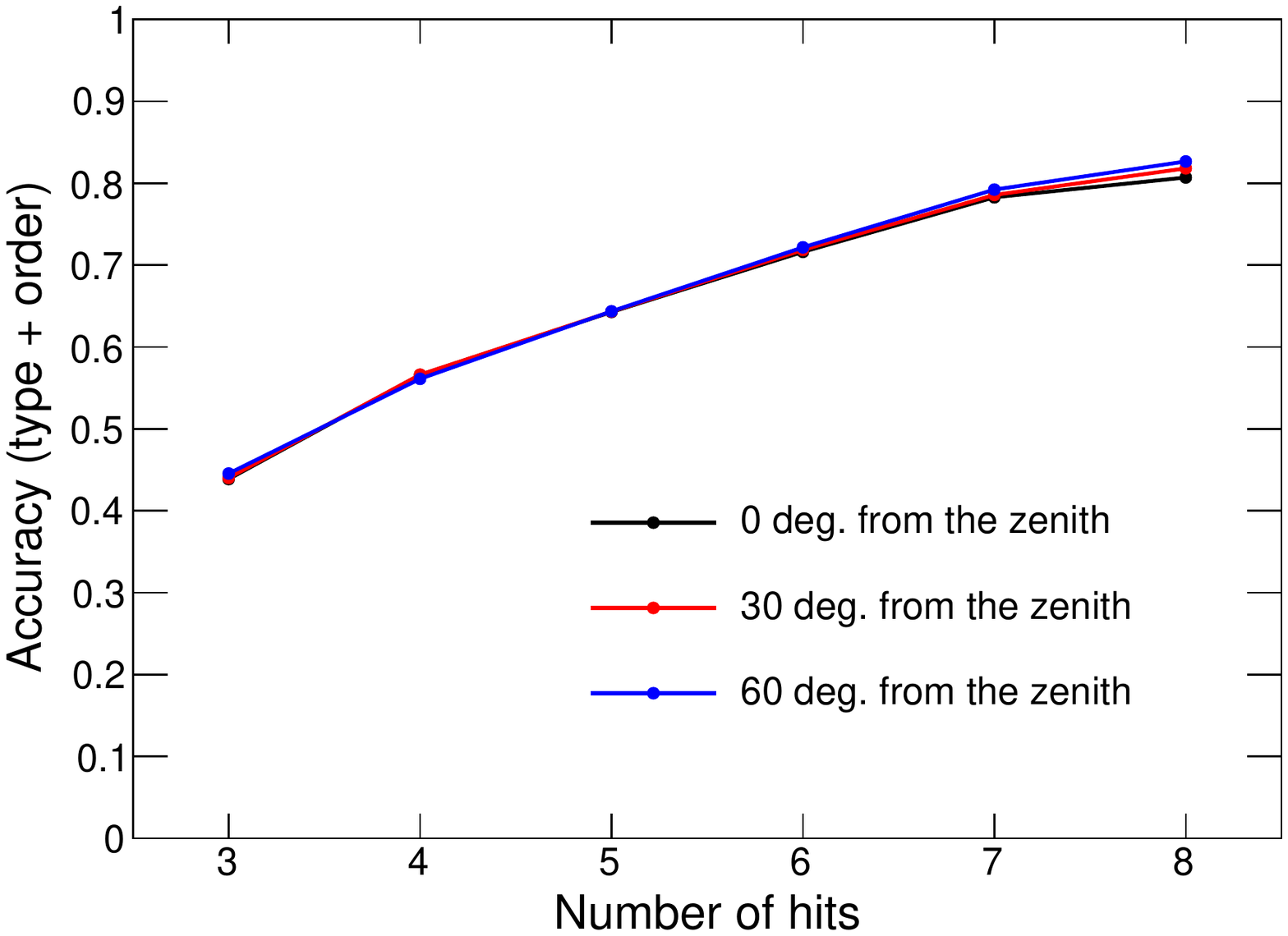}
\end{center}
\caption{The accuracy for both type and order, with different viewing angles.
The scattering order of first two and three interactions are considered for fully-absorbed and escape events, respectively.
}
\label{fig_accuracy_with_different_angle}
\end{figure}

We also examine how much the reconstruction accuracy depends on viewing angles of the source.
In Figure~\ref{fig_accuracy_with_different_angle}, we show the accuracy with different viewing angles of 0, 30, and 60 degrees from the zenith.
As a result, the viewing angle does not affect the performance significantly.

\end{document}